Malik Imran Daud

# Ontology-based Access Control in Open Scenarios

## Applications to Social Networks and the Cloud

DOCTORAL THESIS

2016

UNIVERSITAT ROVIRA i VIRGILI
Tarragona

Malik Imran Daud

# Ontology-based Access Control in Open Scenarios

## Applications to Social Networks and the Cloud

**DOCTORAL THESIS**

**Advisors**

Dr. Alexandre Viejo and Dr. David Sánchez

**Department of Computer Engineering and Mathematics (DEIM)**

UNIVERSITAT ROVIRA i VIRGILI
Tarragona
2016

UNIVERSITAT ROVIRA i VIRGILI
**Department of Computer Engineering and Mathematics**

*Av. Països Catalans*, 26,
43007 Tarragona
Tel. 977 55 82 70
Fax. 977 55 97 10

I STATE that the present study, entitled "*Ontology-based Access Control in Open Scenarios: Applications to Social Networks and the Cloud*", presented by *Malik Imran Daud* for the award of the degree of Doctor, has been carried out under my supervision at the Department of Computer Engineering and Mathematics of this university.

Tarragona, June 1, 2016

Doctoral Thesis Supervisor/s

Dr. Alexandre Viejo Dr. David Sánchez

*Dedicated to my mother and father*

# Acknowledgements

First of all, I would like to thank almighty for giving me strength to carry out my Ph.D research. Then, my special gratitude to my supervisors *Dr. Alexandre Viejo* and *Dr. David Sánchez* for their immense guidance, support, motivation and knowledge during my Ph.D research and in the writing of this thesis.

Then, I would like to thank my wife for her support during this period of time. I also acknowledge my brothers and sisters for their well wishes.

Finally, I would like to specially acknowledge my father and mother for their support and encouragement during my Ph.D studies.

# Table of Contents

# List of Figures

# List of Tables

# Chapter 1 **Introduction**

## 1.1 Motivation

Thanks to the advent of the Internet, it is now possible to easily share vast amounts of electronic information and computer resources (which include hardware, computer services, etc.) in open distributed environments. These environments serve as a common platform for heterogeneous users (e.g., corporate, individuals etc.) by hosting customized user applications and systems [Díaz-López, 2015], providing ubiquitous access to the shared resources and requiring less administrative efforts; as a result, they enable users and companies to increase their productivity.

Specifically, in recent years, open on-line environments such as Online Social Networks (OSNs) or the Cloud have attracted billions of users willing to share online resources and outsource data and computation [Boyd, 2007]. On the one hand, OSNs provide users with a common platform for social interaction based on their mutual interests and activities [Al-garadi, 2016], which is based onsharing digital information (e.g. photos, videos, text, profile data, etc). On the other hand, cloud computing has attracted many business organizations and end users due to its minimal management effort, maintenance cost and ubiquitous access of outsourced resources, which can be hardware or software [Younis, 2014]. These resources are shared in a large-scale distributed environment among the users of a computer network over the Internet [Bayramusta, 2016].

Unfortunately, sharing of resources in open environments has significantly increased the privacy threats to the users to whom the data refer to. Indeed, shared electronic data may be exploited by third parties, such as Data Brokers [Ramirez, 2014], which may aggregate, infer and redistribute (sensitive) personal features, thus potentially impairing the privacy of the individuals [Viejo, 2013]. Because of the potential confidentiality of many of the shared resources and data, ensuring the *right to privacy* of individuals has emerged as a main concern for data controllers, organizations and end users [Onn, 2005], and privacy protection in these environments has been considered as a challenging research problem [Ali, 2015, Bayramusta, 2016].

A way to palliate this problem consists on controlling the access of users over the potentially sensitive resources. Specifically, access control management regulates the access to the shared resources according to the credentials of the users, the type of resource and the privacy preferences of the resource/data owners [Ferraiolo, 2007]. Within this context, delegation of access control is also needed to transfer users' access rights to other entities on a particular resource in order to increase the flexibility of the access management and to reduce the administrative load in large scale scenarios.

The efficient management of access control is crucial in large and dynamic environments such as the ones described above. In the current literature, researchers have proposed several solutions [Blanc, 2013, Coyne, 2013, Zamite, 2013] that rely on basic access control models [Hu, 2014], such as discretionary access control (DAC), mandatory access control model (MAC), role-based access model (RBAC) or attribute access control model (ABAC). Most of these solutions are mainly based on priori and manually managed policies/rules specified for concrete entities and resources. These approaches are usually enough to manage users' access rights in closed environments, such as static organizations, which involve a limited number of entities and resources, and for which a manual management of access rules is feasible. However, manually managing privacy rules and access constraints in open environment, such as OSNs or the cloud, is not practical due to the following reasons:

(i) A large number of entities need to be managed. For example, Google Drive has billions of users and each user manages several types of resources.

(ii) The heterogeneous entities involved in these scenarios would likely have diverse privacy requirements. For example, for a cloud provider offering storage space, an organization involving employees, departments and resources would define significantly different privacy requirements than casual end-users.

(iii) The dynamicity and openness of such scenarios make the privacy requirements to change rapidly with respect to the type of services and users.

Moreover, many of the access control solutions proposed in the literature are ineffective for end users willing to manage the access to their sensitive information due to the following issues: i) the rigidness of the access control mechanism [Johnson, 2012], and ii) many users lack of technical knowledge about data privacy and access control [Liu, 2011]. Furthermore, in most access control mechanisms, there is a proportional decrease in performance as the number of entities involved in the process increase, which is problematic in large environments because of the number of entities to be managed. Thus, there is a need to develop generic access control solutions that scale well in large environments and that are able to overcome the above mentioned limitations.

## 1.2 Goals

To tackle the privacy challenges related to the exchange of data and resources in open environments via appropriate access control mechanisms, we define the following goals for this thesis:

- To study the state of the art on access control management, with a special focus on large and dynamic open environments. More specifically, we will study mechanisms to formally model the entities involved in access control management as a mean to alleviate the administrative efforts of manual management.

- To propose a generic access control mechanism that models the entities involved in access control and their interrelationships, which could be easily adapted to open environments. With this, we aim at providing the means to greatly simplify and speed up the definition of rules in complex scenarios and improve the interoperability between heterogeneous settings.

- To propose delegation enforcement mechanism for distributed open environments that automatically performs delegation, revocation and verification of access rights in an efficient manner.

- Access control should be driving the disclosure risks of the resources and this disclosure is caused by their (potentially sensitive) contents, we also aim at propose an automatic content and privacy-driven access control mechanism specially tailored to protect the sensitive resources published in open social environments. This will make the protection of sensitive resources transparent and alleviate the effort required to manually definite access control rules on specific resources.

## 1.3 Overview of this document

The rest of the document is organized as follows:

- **Chapter 2** presents the background of the work, discussing the need for privacy protection of sensitive resources and the main mechanisms available to enforce privacy, with a specific focus on access control. The main access control management models are also discussed. Finally, ontologies are introduced as knowledge representation formalisms suitable to model and manage access control entities and interrelationships in open environments.
- **Chapter 3** surveys and discusses works that aim at offering solutions to the problems faced by standard access control models in large-scale open environments. The state of the art is classified in ontology-based and non-ontology-based models and their main benefits and limitations are highlighted. Works are finally compared according to the desirable factors that contribute to improve the management of access control.
- **Chapter 4** presents a generic access control management ontology that can be easily extended for specific environments. The proposed solution simplifies the definition and enforcement of rules due to the automatic ontology-based inference of rules. To demonstrate its applicability and benefits, this ontology is applied to two large and open scenarios i.e., OSNs and the Cloud.
- **Chapter 5** elaborates important aspects of delegation of access control management in large and dynamic scenarios. It discusses the challenges of existing solutions and how they can be resolved through the ontology-based approach we propose. It also states how the processes of delegation, revocation and verification of access rights can be improved through the proposed algorithms.
- **Chapter 6** presents a content and privacy-driven access control mechanism that is able to automatically define access rules on sensitive resources according to disclosure risks inherent to their contents and the privacy requirements of the owner. This solution is especially suitable for open social media scenarios, such as OSNs.
- **Chapter 7** contains the summary of the work, highlights its main contributions, and depicts some lines for future work.

# Chapter 2 **Background**

## 2.1 Privacy

Privacy protection in an act that permit an individual or a group to define access boundaries around her private domain in order to manage public access according to the privacy requirements of their choice [Marcella, 2003]. This private domain may include information related to an individual/corporate or any tangible resource that needs protection from unwanted access. According to Solove [Solove, 2008], the concepts of privacy can be (i) access to the personal information, (ii) right to be alone, (iii) control on others to use one's information or (iv) option to conceal any information from others.

Several legislations regulate the notion of privacy of the individuals and the need to protect their private domain. In this respect, Article 12 of the Universal Declaration of Human Rights (1948) [Onn, 2005] proclaims that the *right to privacy* of individuals is essential and needs to be practiced. Likewise, current legislations on privacy, such as the EU Data Protection Act [EU, 1995], U.S. laws on medical data privacy [DoHNY, 2013] and the Health Insurance Portability and Accountability Act (HIPAA) [HIPAA, 1996], regulate which information is private and how it should be protected. Specifically, article 8 of the EU Data Protection Act [EU, 1995] declares personal data such as *medical health, religion, race, politics, membership of past organizations and sexuality* are sensitive and mandates that the information related to these topics cannot be uncontrollably released without the consent of the user. On the other hand, U.S. laws on medical data privacy [DoHNY, 2013] defines diseases such as *HIV, hepatitis, sexually transmitted diseases* as sensitive topics and potential sources of discrimination (e.g., in legal claims for workers' compensation claims) that should only be accessible to authorized parties. Moreover, the Health Insurance Portability and Accountability Act (HIPAA) [HIPAA, 1996] states safe harbor rules about the kind of personally identifiable information that should be removed in medical documents prior making them available to third parties; specifically it requires 18 data elements (called PHI: Protected Health Information), such as health information (which includes health status, provision of health care, or payment for health care) and census features (such as *names, locations, phone numbers,* etc.), to be protected.

The scientific community has proposed several methods to ensure the *right to privacy* of the individuals, which include, but are not limited to, the following methodologies: (i) cryptography, (ii) data anonymization, and (iii) access control management.

*Cryptography* is a mathematical technique that fully masks digital data to keep it secret, so that only authorized entities with the appropriate cryptographic material (i.e., decryption keys) may have access to the clear data. *Symmetric key cryptography* and *asymmetric key cryptography* are commonly used techniques. In *symmetric cryptography,* the key is common for both parties (i.e., sender and receiver) that is used for encryption and decryption of data. On the other hand, *asymmetric cryptography* operates on two keys, in which, one key (i.e., public key) is shared publically that is used for encryption and private key is used for decryption of data. Key management in the crypto-system is a real challenge, since the privacy of a set of individuals may be compromised if key is revealed [Rivest, 1978].

On the other hand, *data anonymization* seeks at removing or masking personal identifiable information related to the individuals from the publically shared data [Viejo, 2012, Martínez, 2013, Soria-Comas, 2015, Sánchez, 2016b]. In comparison to cryptography, data anonymization is irreversible and, thus, it does not require from managing secret materials (e.g., encryption/decryption keys). Moreover, contrary to cipher texts, anonymized data still retain some analytical utility, which is useful for research purposes, but that makes it impossible to guarantee a *perfect privacy* protection (i.e., some information is still being leaked). Some of the areas of research for data anonymization are statistical disclosure control [Hundepool, 2012] and privacy preserving data publishing [Domingo-Ferrer, 2016].

Finally, *access control management* regulates the authorization on shared sensitive resources according to the level of trust of the users. In access control management, access rights of the users are defined along with their level of access on the digital resources (see the next section for further details).

Encryption and anonymization are useful to protect data that is uncontrollably exchanged to third parties in open environments but, in environments with central authorities in which users are authenticated and associated to user credentials, access control is a more convenient alternative that does not modify data (so that information is perfectly preserved for authorized users), but just controls the access on them. Moreover, *cryptography* and *data anonymization* only deal with digital data, but *access control* can also manage the access to computing resources in addition to the digital content. Finally, access control management is an essential element to most information technologies in order to limit unwanted access to digital resources (e.g., business applications, social networks, cloud, etc). The following subsection details several aspects of access control and its applications.

## 2.2 Access Control

The management of access rights implies granting or denying access to a specific resource according to (i) the credentials of the users, (ii) the type of the resource and (iii) the privacy requirements of the resource owner [Imran-Daud, 2016a]. Access control systems rely on three basic building blocks in order to manage the access [Hu, 2006]:

(i) An *access control policy* that specifies how to manage access and who is eligible to access specific information/resource.

(ii) A *mechanism* that grants or denies access by examining the access request and enforces access policies accordingly.

(iii) A *model* that is enforced by the system and it is a formal presentation of the security policy that illustrates the methods to handle specific conditions when users try to access system resources.

Two well-known access control models that are used to manage security policies are: (i) discretionary access control (DAC) and (ii) mandatory access control (MAC). In DAC, access control is at the discretion of the resource owner or anyone who is authorized to control the access on the resources. In such model, access is restricted according to the identity of the users. On the other hand, MAC-based systems do not rely on the owner of the resource to take authorization decisions, but on the central authority that handles access requests and takes access decisions according to the rules that are defined for the entities of the system. These two models are the basis for several other models that have been proposed for different environments, which are discussed later in this section.

Access control management is essential for collaborative scenario dealing with sensitive data and/or resources, but it is more challenging in large scale open on-line environments due to large number of heterogeneous users and resources to manage. For example, in social networks, users publically share huge amount of content such as messages, profile data, or social apps that may contain sensitive information and may constitute a serious privacy issue, thus requiring proper access control management. To manage access control in a large scale environment, system designers have offered several models but two widely used generic models are: role-based access control (RBAC) and attribute-based access control (ABAC). These two models are more scalable (in terms of number of users), flexible (easy to implement in large environments) and convenient to manage than DAC and MAC models, because they refer to roles and attributes rather than relying on individual users to manage access control. Details of these models are explained in the following subsection.

### 2.2.1 RBAC and ABAC Models

In RBAC, access rights are managed through the roles of the users defined for specific resources, over which either MAC or DAC can be implemented to manage access rights. To model RBAC, Sandhu et al. [Sandhu, 1996] proposed four sub models:

(i) $RBAC_0$ (known as flat RBAC) specifies the *users* of a domain, their *permissions* to access resources and the *sessions*, and manipulates them through the *roles* of the users in order to manage access control.

(ii) $RBAC_1$ (known as hierarchical RBAC) introduces *role hierarchies* over $RBAC_0$. These hierarchies represent the organizational structure of the authorities, in which former roles manage their subsequent roles and authenticate them during an access request. For example, the project manager of any department in an organization is at the top of the role hierarchy with full access privileges over the subsequent project leaders that have limited access privileges inherited by the project leader. Engineers are at the last level of the hierarchy with the roles specific to their jobs that are allocated by the project leaders.

(iii) $RBAC_2$ (known as constrained RBAC) incorporates *constraints* in addition to $RBAC_0$. These constraints add conditions to the roles in order to limit privileges and to introduce the notion of separation of duties according to the users' roles. For example, in an account department, a constraint can be defined on the role of billing users to refrain her to access account receivable records.

(iv) Finally, $RBAC_3$ (known as symmetric RBAC) combines $RBAC_1$ and $RBAC_2$ capabilities, which provides *constrains* on the *role hierarchies*. In RBAC-based systems, access decisions are made according to the roles assigned beforehand to the users of a system. Roles represent the privileges the user have or tasks they are permitted to perform within the system.

On the other hand, the ABAC model [Hu, 2014], which can also be implemented with DAC or MAC, relies on the attributes of the entities of the system to manage the access rights (e.g., attributes of system users and resources). For example, project managers (defined as an attribute) can view records of employees of their respective departments. ABAC-based systems process access requests by examining attributes of the *subjects* (i.e., system users) and the *objects* (i.e., system resources) against the *policy* defined by the owner of resource. The attributes of the subject and object are managed locally in the repository that is managed by the system. The *policy* contains *rules* that are used to take authorization decisions and these rules are determined from the policy once the attributes are validated.

The core modules of the ABAC mechanism are the policy enforcement point (PEP) and policy decision point (PDP). PEP is responsible to process access requests of the user, invoke PDP and take authorization decisions based on the input of the PDP. On the other hand, the PDP takes authorization decisions based on the attributes of the subject, objects and rules defined for them.

In general, ABAC is better than RBAC in terms of scalability (i.e., number of users to be managed), flexibility (i.e., easier to implement in a large scale environment) and access control management (i.e., it is easier to associate attributes to other users or resources) [Priebe, 2006, Coyne, 2013]. Moreover, ABAC-based solutions are more efficient than RBAC-based solutions due to the following reasons: (i) ABAC does not require manual management of multiple roles (i.e., access roles and delegated roles) for entities of the system; and (ii) a single policy can be referred to multiple users in order to avoid the burden of multiple roles management for users sharing a common resource.

Researchers have also contributed to minimize the access control management burden by proposing several solutions relying on these models (i.e. RBAC and ABAC).These include classifying resources into categories [Cheng, 2012b], itemizing data into different elements [Aïmeur, 2010] or classifying users into lists (e.g., blacklist users) [Cramer, 2015]. However, these methods do not scale well in large and complex environments because of: (i) the growing privacy configuration requirements and the incapability of existing solutions to handle them in an efficient manner [Beato, 2009]; and (ii) the burden of the definition and management of rules and policies by users and administrators [Daud, 2015].

## 2.3 Delegation of Access control

In addition to standard access control tasks, delegation of access rights is another important aspect of access control management. By means of delegation, a user can delegate her access rights or privileges to other users, which is useful (i) for efficient management of access control in distributed environment (e.g., cloud, distributed databases etc), and (ii) to reduce the administrative burden of the owners of resource.

In the delegation process, a user is referred to as a *delegator* provided she holds access privileges on a specific resource and is legally empowered to exercise her right to delegate these privileges. On the other hand, the recipient of the delegation is referred to as *delegatee*. During the delegation process, the delegator defines scope and limitations of delegation that restrict the delegatee to exercise her inherited privileges under specific limitations. These terms are can be in the form of roles or policies depending on the access control model type.

Delegation is a proxy process that enables delegatee to perform specific tasks on behalf of the delegators [Pham, 2010]. For this purpose, it requires a reliable mechanism that supports these tasks and it must also address the following queries while processing a user's access request for a resource:

(i) Does a user has intended access privileges on a resource?
(ii) Is a delegator legitimate that granted access privileges to the user?
(iii) Does a delegator hold privileges that are delegated to the user?

To support such these tasks, researchers have proposed several mechanisms that rely on different standards(e.g., [Joshi, 2004]), being the XACML delegation profile [OASIS, 2013] [XACML-Profile, 2009] the most widely used standard. The advantages of the XACML are:(i) the policies can be written and analyzed independent of the specific environment [OASIS, 2013], (ii) administrators only need to describe access control policy once, (iii) it can accommodate dynamically changing access control policy requirements, (iv) a single policy can serve several entities at a time. This standard is briefly discussed in the following subsection.

### 2.3.1 XACML and the XACML Delegation Profile

XACML [OASIS, 2013] defines an access control policy language that is implemented in XML and provides an intuitive way to evaluate access requests according to the rules defined in the policies. Moreover, it provides a common terminology and enables interoperability among vendors that implement access control. XACML is an ABAC-based standard, even though it can also implement RBAC as a specialization of ABAC by using attributes of the entities for access control management.

XACML is structured into three basic elements: *policy set*, *policy* and *rule*. The owner of a resource manages the access control by specifying policies for several users. Policies that are defined for a common resource are managed in a policy set. Access control on a resource is managed by defining rules that are encompassed within a policy. Table 2.1 shows the actors modeled by XACML that are involved in the process of managing an access request for a resource and their architecture is shown in figure 2.1.

**Table 2.1** Actors of XACML

| Actors | Description |
|---|---|
| Policy Administration Point (PAP) | Repository for policies and serves policies to PDP |
| Policy Decision Point (PDP) | Takes access decision based on the access request and also collects related data from other actors. |
| Policy Enforcement Point (PEP) | It is interface to the requestor and the internal actors of the system. It processes request and response. |
| Policy Information Point (PIP) | Retrieves and evaluates attributes of the entities. |

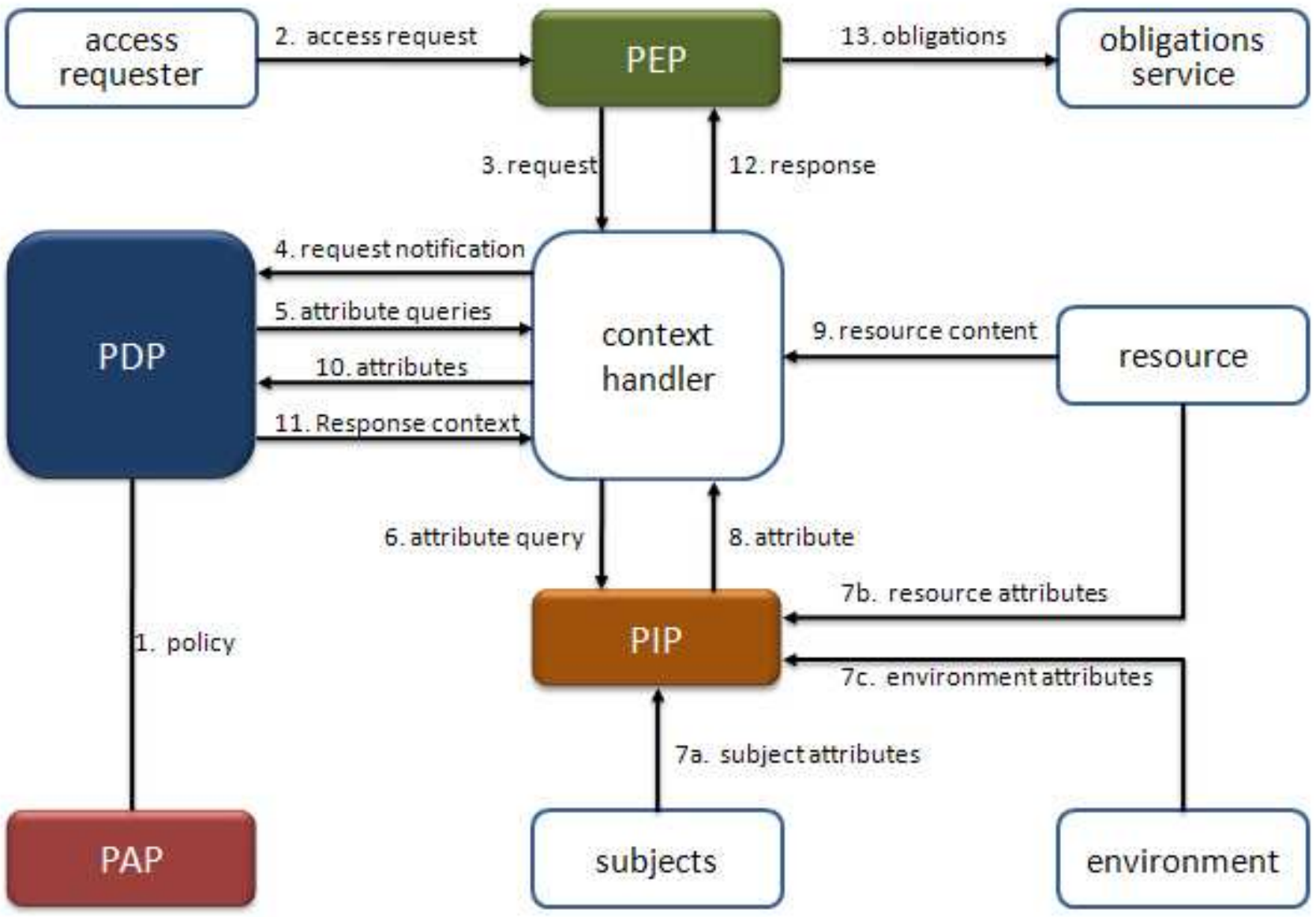

**Figure 2.1** Architecture of XACML actors [OASIS, 2013]

In addition to access control, delegation of access rights is also implemented in the XACML profile [XACML-Profile, 2009]. Delegation allows users to transfer their access rights to other entities on a particular resource [Wang, 2008]. In the XACML profile, access rights on the resources are delegated in the form of policies. Moreover, reduction is a process that is performed to validate the authenticity of the issuer of the policy. For this purpose, a graph of policies is generated as a result of each access request for a resource, which contains the hierarchy of the delegated policies. To generate a policy graph, the attributes of the *access request* are searched (i.e. the delegatee and requested resource) within the policy delegated to the requester and, then, edges between that policy and its delegated policy nodes are created by matching the attributes of the entities within the hierarchy of the delegated policies (attributes are the delegatee and its delegator). Then, the path of the graph, which connects the owner of a resource and the requester with all the intermediate delegators, is checked in order to verify the authenticity of the delegated policy of the requester. As a result, decisions are made in the form of permit or deny access to the resource.

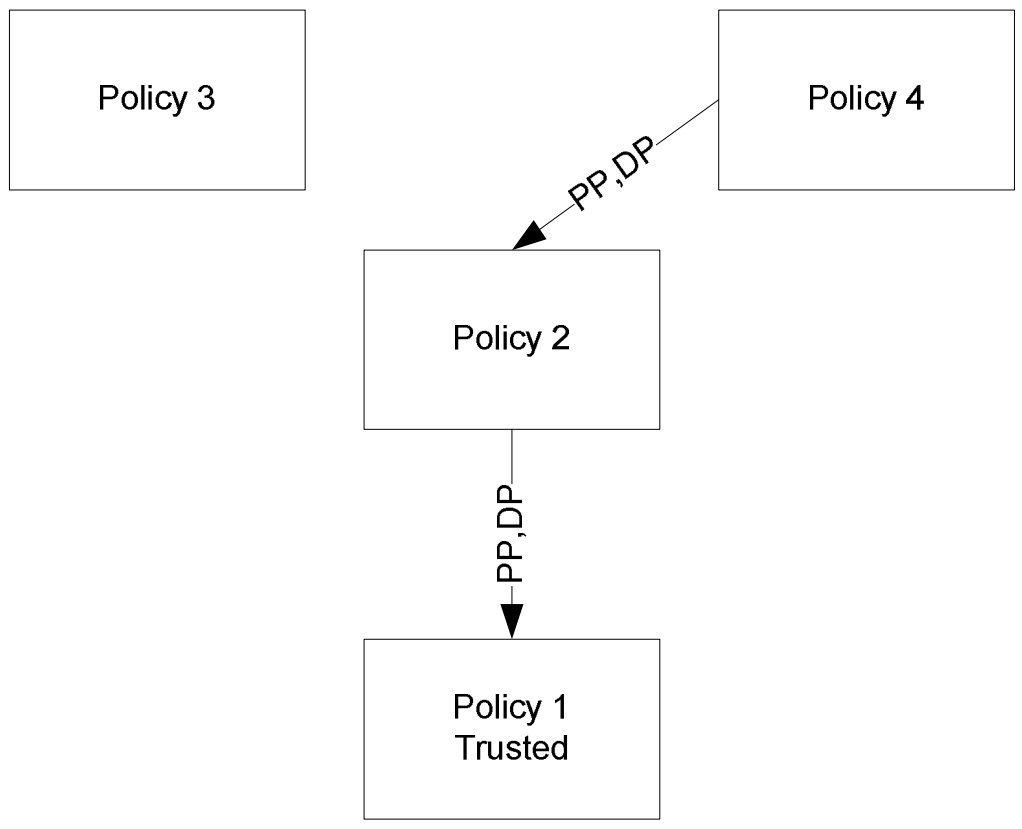

**Figure 2.2** XACML delegation graph[XACML-Profile, 2009]

In this approach, the method to evaluate an access request, that is, generating a policy graph and finding attributes within all policies for each *access request*, is a costly solution in terms of performance. Figure 2.2 illustrates a sample graph generated as a result of access request. The connection between the delegated policies represents the flow of the delegation, whereas the edges determine the delegation decision, that is, the policy permits (PP) the delegation or denies (DP) it to the other policy.

## 2.4 Ontologies and Access Control Management

Ontologies have gained a lot of attention in recent years due to their potential as tools to organize information and to limit the complexity of knowledge management. Neches [Neches, 1991] defines an ontology as:

> "*An ontology defines the basic terms and relations comprising the vocabulary of a topic area as well as the rules for combining terms and relations to define extensions to the vocabulary*."

Ontologies are useful to enable knowledge transfer and interoperability between heterogeneous entities due to the following reasons: (i) they simplify the knowledge sharing among the entities of the system; (ii) it is easier to reuse domain knowledge and (iii) they provide a convenient way to manage and manipulate domain entities and their interrelationships.

Ontologies are particularly helpful to formally specify the conceptualization and interrelations of a domain of knowledge [Mika, 2007] from which specific domain objects (e.g., users and resources) are defined as instances of this conceptualization. The components of an ontology are:

*classes*, *objects*, *relations* and *attributes*. *Classes* are the elements that conceptualize components of a domain. Classes are usually organized in taxonomies that are associated to each other through *relations*, which can be taxonomic (thus defining the type of inheritance among superclasses/subclasses) or non-taxonomic(which can define any other type of relationship, such as part-of, cause-effect, etc.). *Objects* are the specific entities of a domain and are represented as instances of classes; they may have specific properties that are represented by the *attributes* of these classes.

Ontologies are also useful to reduce users' inputs during a change in the modeled domain of knowledge [Zablith, 2008]. Nowadays, ontologies are extensively used in many fields of computer science, such as artificial intelligence [Batet, 2012, Isern, 2012], software engineering [Isern, 2011], knowledge engineering [Sánchez, 2010] and natural language processing [Gomez-Perez, 2007]. Ontologies have gained popularity in natural language processing because they provide conceptualization (senses) of linguistic entities and enable to properly understand and compare terms. Ontologies are thus useful to manage and understand the semantics of (i) textual documents [Sánchez, 2011], (ii) search engine queries [Sánchez, 2013c, Viejo, 2014], (iii) local databases [Martínez, 2012a, Martínez, 2012b], (iv) natural language translations, etc. For this purpose, several data sources are available that are being used by the system developers, which are Wikipedia, DBPedia[1], WordNet[2], etc.

The ontological modeling of a domain can be engineered through many languages (e.g., ontology inference layer OIL, simple HTML ontology extensions SHOE) but the Ontology Web Language (OWL) along with the Resource Description Framework (RDF) are the most commonly used ones. In practical terms, ontology engineering (i.e., the act of building ontologies representing a domain of knowledge) consists of [Gomez-Perez, 2007]:

- Define the classes (concepts) in the ontology.
- Taxonomically arrange the classes in a hierarchy.
- Define attributes, their allowed values and the relationships between classes.
- Define instances of the concepts and perform inferences.

In the context of our work, ontologies can be useful to manage access control. The modeled entities and their interrelations can be used to keep track of the owners of the resources, their type of relations with the resource requesters and with the resources. Through an ontology, the access privileges on the resources can be easily managed and enforced by following the interrelations of the ontological entities involved in access control. Specifically, during the access control management, *concepts* are the entities of a domain (e.g., *user* or *resource* that require access control management) and *objects* are the instances of those entities that are linked by means of

[1]http://wiki.dbpedia.org/
[2]https://wordnet.princeton.edu/

*relations*. For example: user *A can access* resource *R1*, where *A* and*R1* are the instances of the concepts/classes (*user* and *resource*) and *can access* is the relation between these instances. Properties of the *objects* are represented with *attributes* that are also used to state the nature of the relationship among the *concepts*. For example: only users of account *department* can access financial records of employees, where *department* is the attribute that signifies the employee's affiliation in an organization.

As discussed earlier in this section, ontologies are useful to study the semantics of textual documents, which can be also used to manage access control on the text published over the social web. For example: online public forums that allow people to interact with each other and share their personal issues to get solutions (e.g., medical advice). For this purpose, the lexical analysis of the knowledge source can be performed to identify the sensitive information (e.g., name of disease, personal data, etc), which leverages to regulate access control on the published data according to the privacy preferences of the users.

As we discuss in the next section, researchers have used ontologies in access control to alleviate some of the problems of RBAC and ABAC, specifically, the definition and management of rules and policies [Masoumzadeh, 2010b, Choi, 2014]. Modeling policies in an ontology can greatly increase the performance of the system because it can easily retrieve target policy from the workflow by following the interrelationship of the target entities instead of searching from the database.

# Chapter 3 **State of the Art**

As discussed in chapter 2, management of access control in large scale open environments (such as online social networks (OSNs) and the cloud) is a challenging task, which requires dealing with millions of users who share online resource, data and computation. Moreover, delegation of access rights in these environments has also been another focal area of research by the researchers. The scientific community has proposed several ontology-based and non-ontology-based solutions to manage access control in these environments. Some of these models and their limitations in such environments are discussed in following sections.

## 3.1 Non-Ontology-based Schemes

Researchers have proposed several solutions to manage access control management in OSNs. In due course, Ghazinour et al. [Ghazinour, 2013] proposed an artificial intelligence-based recommender system for OSNs (more specifically, for Facebook) that analyzes privacy requirements of the users and assist them by recommending solutions to improve their privacy settings. To do so, the system analyzes profile of users that have similar requirements in order to construct their data and later use them for recommendations. Moreover, the privacy configurations of the users are also analyzed in order to predict topics that are sensitive, and then privacy alerts are triggered based on this information.

In order to achieve the goals mentioned above, the recommender system first builds user data by analyzing attributes of the following categories of the OSN: (i) user's personal profile (e.g., user name, ID, locations etc.); (ii) user's interests (e.g., movies, books, music etc.); and (iii) user's privacy settings on photo album. The absence of value of any attribute is considered as the gesture that the user is reluctant to disclose this information, thus, it requires privacy protection. Moreover, the ratio of disclosed elements to the public or friend categories is calculated by the system and then used to predict the behavior of the users toward their privacy management and to compare them with other users.

In order to predict the users' behavior, system categorizes them into three types of people: (i) fundamentalists (those with high privacy concerns), (ii) pragmatics (users with medium privacy concerns); and (iii) unconcerned (low privacy concern users). The ratio of the profile attributes disclosure is calculated and then a decision tree is used to determine their type of category (e.g., disclosure ratio=0 then fundamentalist, if ratio > 50% then pragmatic else unconcerned).Based on this information, users who have less privacy configurations, but their privacy concerns are closer to the one that have more privacy configurations, receive suggestions to improve their privacy.

This system only recommends options to the users to improve their privacy, but it does not provide methods to ensure privacy of the users. Moreover, due to its probabilistic approach, the accuracy of recommended settings is a real concern about the reliability of the system.

In another scheme, Jung et al. [Jung, 2014] proposed property-based access control model for the purpose of secure information sharing over OSNs. In addition, they proposed administrative model that manages the activities of the OSN to employ access control model. This model resolves conflicts between the permissions of the OSN users through their proposed conflict handling techniques.

In this model, OSN entities are categorized into two categories: (i) user; (ii) community; and (iii) society. A SN *user* entity holds four type of properties, which are: (i) *contexts* that represent the attributes of a user like gender, age, date of birth etc.; (ii) *tasks* is the activity a user that she performs over OSN, e.g., publishing message; (iii) *resource* that is shared or managed by the user over OSN, e.g., photo, video, text messages etc.; and (iv) *policies* that are managed by the user for access control management. Similarly, a group of users form a *community* that cooperates with each other to retrieve information of the SN users. These communities have their own set of properties (i.e., *cooperation* property that maintains elements of cooperation among entities in addition to the formally mentioned properties of the *user* entity) that are used to collect related information to accomplish these tasks. Finally, the society represent the OSN that supports secure cooperation among the SN entities and it has properties similar to the user entity but used for different scenario (i.e., w.r.t community).

To manage access control, the related properties are analyzed in addition to the policies of the users in order to make an authorization decision within the given situations. This scheme maintains three types of administrators for each type of entity (i.e., *society* manager, *community* manager and *user* manager) that require special handling for the exchange of information between those parties, which represents an extra overhead for the system.

Access control management in the cloud is another area of research where researchers have contributed and proposed several solutions. In due course, Ngo et al. [Ngo, 2015] proposed a multi-tenant attribute-based access control management system for inter-cloud services. This model is inspired by the ABAC model that implies access control management through the attributes of the cloud entities (i.e., subjects, objects and resources). In this model, the access control is managed through policies and conflicts between the policies are managed through pre-defined constraints.

In this approach, ABAC subjects are decupled into three types, which are: (i) service providers; (ii) tenants (services of service providers); and (iii) users. These types of subject are modeled and their relationships are defined with each other along with the resources and their defined policy. To access a resource, authorization request is originated that contains information of the resource requester (i.e., user), requested resource and environment condition (e.g., time, location etc.). The system analyzes the request and determines the context relations from the model and the access to resource is granted according to the defined policy for the target user.

The proposed solution only focuses on the infrastructure resources of the cloud (i.e., IaaS) for the management of access control. Thus, it does not provide methods to manage access rights of the cloud services that may also need access on the resources. Moreover, it does not provide ways to manage access control on the distributed tenants or services.

In another approach, Habiba et al. [Habiba, 2013] proposed authorization-based access control model to manage access to the cloud data. For this purpose, a set of access rights (e.g., read, write, update etc.) are modeled in a tree that are used as a reference to manage access rights by the owner of data. To do so, the owner of a piece of data can transfer her access rights by simply associating a node of the tree to the intended user, thus, the following node and its child nodes in the branches of the tree that represent specialized access rights are automatically transferred to the user. These access rights are managed in the form of policies with several elements (e.g., rule, conditions etc.) that enforce access control management.

This model relies on a layered architecture to manage access control, in which, each layer is divided into several modules responsible to manage access rights. For example: *management layer* processes access request, *asset layer* deals with the cloud resources. This proposed scheme only deals with the data and it cannot be applied for other resources of the cloud. Moreover, to find target policy, the performance of the system declines due to increased number of policies stored in the database.

Delegation of access rights is another paradigm of access control management. In due course, Gusmeroli et al. [Gusmeroli, 2013] proposes a capability (ability of a user to perform specific operations) based delegation model in which tokens of authority (or capability tokens that grant access to the users to perform specific operations) are issued by the delegator to the delegatee. These tokens contain information about a resource and also the privileges that are being delegated along with the lifespan of this token. For this purpose, the capability tokens are created by the owner of the resource and distributed to the delegatees, who can further distribute to others (with same or different access rights) depending on the delegation depths restricted by the owner. To do so, the delegator attaches it's digitally singed capability token to the request and submits it to the *service manager* (locally maintained service for verification of authority). In turn, the *service manager* performs following two operations to make access decision, which are: (i) it formally validates capability from the authorization chain of tokens; and (ii) it validates requested set of operations.

This model is better in terms of verification of authority of the delegator (i.e., the relationship of the instances can provide information about a delegatee, delegator and a resource, instead of defining constraints/rules for verification). However, for multiple delegations, a delegator needs to maintain multiple tokens for each delegation, and this can represent a serious overhead during the verification of authority (e.g., in a department with a large number of employees). Furthermore, it does not provide any method to verify the authenticity of multiple delegations from a chain of delegations (i.e., delegation of different access rights from multiple users for a common delegatee).

In another scheme, Lui et al. [Lui, 2007] proposed a supervision-based delegation model that optimizes delegation certificate-based approach with the chained delegation of permission to enable users to further delegate access privileges. For this purpose, delegator appoints a supervising agent (e.g., it can be a third party user or a software agent) to monitor the activities of the delegatee and to grant authorization to the delegatee on her request to exercise delegated privileges. The delegation certificate is based on five elements, which are (i) delegator's public key; (ii) public key of delegatee; (iii) a boolean variable that indicates a delegator can further delegate access rights; (iv) authorization of rights; and (v) validity of the delegated rights. In this scheme, the authors proposed to add supervision as another element in the certificate that contains the public key of the supervisor agent.

In order to delegate privileges, the delegator signs the privileges being delegated (for example, read, delete etc.) to the delegatee and computes proxy authorization keys to be transferred to the supervisor agent. These proxy keys are used by the supervisor agent to verify the delegation on behalf of the delegator. This scheme involves administration burden for keys management and may not be suitable for large scale environment where a single supervisor agent need to monitor delegatees for their conduct and, at the same time, verify requests on behalf of delegator.

## 3.2 Ontology-based Schemes

In an ontology-based scheme, Carminati et al. [Carminati, 2009,2011] proposed an ontology-based model for social network in order to annotate OSN related publications. This ontology categorizes and models the following aspects of OSN: (i) users' personal information (i.e., profile data); (ii) users' relationships; (iii) OSN resources (e.g., pictures, videos, etc.); (iv) relationships among users and resources; and (v) actions performed by users (e.g., tag other users in photos, publish messages, etc.).

In this model, the personal information is modeled through the OWL-based ontology called Friend-of-a-Friend ontology (FOAF) [Dan Brickley, 2014]. As FOAF models only the basic profile information (e.g., name, groups, documents, etc.), thus, this model extends such information through the RDF/OWL language that cannot be modeled by the FOAF (e.g., profile identities, etc.). Moreover, users' relations are modeled using a n-ary relationship that connects an individual with several others. For this purpose, related classes and their interrelationships were defined. A data property *TrustValue* is defined in order to quantify the level of trust of the users and to measure the strength of their relationships. Similarly, in the case of OSN resources, their relationships with users and users' actions are also modeled through the RDF/OWL language using several classes with their data properties.

In these schemes, access control is enforced through the *access control policies* that authorize users according to the type of relationship, relationship depth and the level of trust among these users. Authorization of access is regulated through the policy rule. Moreover, users can control inappropriate published content through the use of *filtering policies*. For this purpose, the user defines filtering preferences in order to block unwanted content. *Admin policies* are defined to regulate *access control* and *filtering policies* that keep record of the users that are authorized to define such policies.

In another ontology-based scheme, Masoumzadeh et al. [Masoumzadeh, 2010b] proposed a*social network ontology* (*SNO*) and an *ontology-based OSN model* (*OSNAC*) that enforce access rights of the users by means of defined rules.*SNO* models key entities of the OSN into 14concepts and 10object properties that represent the relationship among these entities. The ontology was built using the OWL web language and RDF was used to model the knowledge of the domain. On the other hand, the *OSNAC* model relies on the *SNO* for access control management. More specifically, it manages two types of access control rules, which are: (i) user rules (it maintains personal authorization rules of the users, for example, delegation or authorization of access to the personal resources/data); and (ii) system rules (it maintains the rules of the system to manage system policies in order to keep a track on the resources and regulate access on them according to the users' rules).

In order to regulate the authorization, the authors proposed the use of an *access control ontology* (*ACO*) in order to model the actions of the users, their privileges on the resources and the type of relationships among them. This ontology models the knowledge through the basic ontology of OSN (i.e., *SNO*) and the privacy policy that contains the privacy preferences of the users.

The above mentioned schemes have some common limitations. The proposed schemes are not flexible to modifications in the ontology, profiles or other contents, because they should undergo with a lot of manual changes in the ontology and also in the annotation of resources. Moreover, there is no mechanism defined to evaluate the sensitiveness of the resources, which leads the system to provide a coarse grained access control; this implies that an access to a resource is binary, this is, it generates a full access decision or a full deny decision. Besides, a lot of manual management by the users and the social network administrator is required in order to configure policies for each type of users and OSN resource.

Bourimi et al. [Bourimi, 2012] proposed an interesting approach to address privacy issues related to the linkability of profile data and information sharing by the users on various online social networks (OSN) within the context of the European project *Digital.Me*. In this approach, the profile data from two or more OSNs is analyzed and privacy warnings or recommendations are triggered to alert the users to apply common privacy settings for similar profiles for her contacts. Similarly, the messages published by the users are also semantically analyzed to examine the content that may raise privacy concerns for the users. For this purpose, they rely on the ontologies that are already defined by them in their prior research work, which are: (i) *contact ontology* (that models user's contacts); (ii) *personal information ontology* (to link profile of the users); and (iii) *live post ontology* (to model information content).

In order to detect similar profiles, natural language processing libraries (NLP) are used and these profiles are linked together through the ontology for the management of common privacy settings. To do so, the profile data from OSNs is retrieved through the contact APIs (for similar attribute detection) and then modeled through the defined *contact ontology*. For this purpose, NLP libraries evaluate profile attributes and compare them syntactically and semantically. Finally, the profiles that match with each other are linked together through the *personal information ontology*. On the other hand, the published content that may hamper privacy of the users is analyzed. To do so, the content to be published is analyzed semantically by using NLP libraries to identify information content and related resources (e.g., actions, user names etc.), which are then modeled through the *live post ontology*.

This scheme only models users that have a similar profile and cannot model other entities; as a result, it may require separate access control management for the rest of entities. Moreover, the proposed solution does not specify any method to enforce privacy.

Choi et al. [Choi, 2014] proposed ontology-based access control model that manages access rights in the cloud environment. They proposed a content-aware approach that determines the type of users (i.e., service provider or normal user), their context information from the ontology (i.e., relationship type of the user with the resource) and their access rules from the policies that are managed locally in a repository. This model is composed of the following two basic modules: (i) *context analysis engine*; and (ii) *access control module*, which are discussed in the following paragraph.

The *context analysis engine* is responsible for selection and integration of context information of users and resources, and then renders this information to the *access control module*. The *access control module* manages this information by means of an ontology (i.e., context ontology), moreover, it identifies and authenticates an access requester based on her context information and security policy, which is kept locally by this module. Finally, the authentication information is rendered to the context module that, in turn, regulates the access to the requested resource.

The proposed ontology only models the context information of users and resources (i.e., type of users and their relationship with the resources); it does not model the policies defined for these entities. Thus, the system needs to map the context information with the policy database in order to get the appropriate policy, which represents an extra overhead and makes the processing of any access request more complex.

In another ontology-based approach, Liu [Liu, 2014] modeled a set of operations of cloud business services: (i) payment status (to keep record of users' payment to access cloud resources); (ii) service level agreement (the level of access on the resource) to manage access control of the users on cloud resources; and (iii) role of users (to distinguish valued users from regular ones). For this purpose, they designed an access control model (i.e., CSAC) that regulates access based on the ontological relationship between users and resources. In this model, the central server manages access control by processing an access request of the users and it regulates access based on the ontological relationships and the policy that is managed by the administrator of the system. In addition, several rules are specified to tackle policy conflicts and to manage unauthorized access of the users.

Again, this ontology is not generic and it is limited to model specific cloud services (i.e., payment status only), thus, it provides ad-hoc inference system for rules. Moreover, administrator of the system needs to manage ontological relationships and other operations of the system that may affect the performance of the system, which is also not feasible for the distributed scenarios.

Xu et al. [Xu, 2009] proposed a delegation model that extends XACML delegation profile by incorporating roles as the attributes of the users that are stored within the policies. These roles are delegated to the users in order to grant access privileges that can be further delegated to other users and represented in role hierarchy. In order to manage access control, following roles are defined for the actors of the system, which are: (i) regular roles (used for normal access rights); (ii) delegable roles (roles that can be delegated to other users); (iii) delegated roles (delegable role that has been delegated to users); and (iv) administrative roles (assigned to administrators to manage roles that are delegable). They further enhanced XACML profile by providing delegation and administration enforcement mechanism in order to tackle policy conflicts (i.e., read and write operations occurring simultaneously).

In this approach, the authority of the delegators is verified from the role hierarchy that is generated as a result of each resource request. This can result in a serious performance overhead in a large scale environment dealing with a large number of delegatees during the following actions: (i) searching within a large number of delegated policies on an access request; and (ii) generation of the graph on each access request. Moreover, management and enforcement of multiple roles is an extra overhead to the administrator of the system.

In another XACML based approach [Seitz, 2005], a delegator can delegate either access level authorization or administrative level authorization. Access level authorization is a four-way tuple that contains information about the subject (delegatee), object (resource to be shared), method (corresponds to action) and time interval (timestamp until the authorization is valid). Whereas, administrative level authorization allows users to delegate their access privileges to other users through the policy rules.

In this approach, access privileges are defined through the policy and system is capable to create a new policy by updating an existing policy of a delegatee (in the case if administrative privileges are begin granted to the normal access users). In order to enforce delegation, a locally maintained system performs the authorization of delegated policies. To process an access request, they used a tool XPath (which has XML-like structure) that matches the attributes of the request with the attributes of the policy (i.e., subject and object).

This system is customized for a specific application (i.e., account management) and cannot serve large scale cloud operations (i.e., federated cloud environments). Moreover, the delegation records are maintained in the local repository of the policy that can be expensive solution in terms of performance in order to search a required policy from this repository.

In another approach, Wainer et al. [Wainer, 2005, Wainer, 2007] proposed a user-to-user privilege delegation and revocation mechanism by extending the RBAC model. They implemented this model for workflow management systems (i.e., a system designed to manage users' workflows within the business processes). In this model, a delegator can delegate specific access rights rather than the entire role (which contain all the access rights) to a delegatee. For this purpose, a generic algorithm for delegation is proposed and authority of the delegators is verified by generating a delegation graph. In order to enforce delegation, authorization rules are specified (in the form of constraints) that overrule the roles already assigned to a delegatee in order to implement the delegated privileges. Furthermore, the verification of the delegator's authority to delegate privileges is assessed by the administrative roles (roles that acknowledge her privileges as delegator) assigned to them. These delegation and authorization processes are applied by means of specially designed services and protocols.

In general, RBAC-based mechanisms may not perform efficiently in cloud environments due to the following reasons: (i) the manual management of multiple roles (i.e., access roles and delegated roles) for each delegator and delegate; and (ii) the burden on a single entity for the enforcement of these roles for multiple tenants (e.g., users and CSPs). Moreover, it is hard to verify the authenticity of the delegators from the role hierarchy due to the following reasons: (i) it is difficult to determine who is delegating and what resources does she own to delegate; and (ii) the role hierarchy can be forged to gain unauthorized access on given resources.

## 3.3 Comparative analysis of State of the Art

In table 3.1, we have analyzed the aforementioned approaches according to the several elements that are required for any efficient solution, which are:

(i) **Adaptability:** Due to the dynamic nature of open environments, ontology-based approaches must support additional entities without requiring to change the structure of the ontology.

(ii) **Automatic Configuration:** Systems must provide automatic configuration management of users' privacy requirements and it must have less manual management by the administrators or users of the system.

(iii) **Scalability:** the proposed approach must be suitable for large scale and distributed environments and it must be capable to serve large number of users in an efficient manner.

(iv) **Fine grained access control:** The proposed solutions must provide a fine grained access control mechanism that could support instance level access instead of a whole class.

(v) **Policy Modeling / Policy-based:** Proposed systems must model policies within the ontology to avoid policy search overhead (as happens in the repository-based systems).

(vi) **Generic ontology:** Finally, the approach needs to be generic that it can be transformed to any environment that requires access control management with fewer modifications.

These elements are derived based on the limitations of the existing solutions (discussed in sections 3.1 and 3.2); thus, they serve as essential elements for any reliable/efficient access control mechanism. The elements such as *Adaptability*, *policy modeling* and *generic ontology* are the crucial elements for the success of any ontology-based scheme that helps a mechanism to be integrated in any large and dynamic open scenarios. The remaining elements enhance the efficiency of the mechanism that is required for any productive access control mechanism.

In the following analysis, there is no such solution that fulfills the above mentioned criteria and they have some limitations. Though, the solution proposed by Bourimi et al. [Bourimi, 2012] is better than rest of the solution that assures most of the analytical elements. Though, this solution is not a generic solution and it may require major structural changes during the course of transformation into another environment. Moreover, it has overhead of policy search from database while processing resource access request.

**Table 3.1** Comparative analysis of ontology and non-ontology-based schemes

| **Approaches** | **Adaptability** | **Automatic Configuration** | **Scalability** | **Fine grained access control** | **Policy Modeling / Policy-based** | **Generic ontology** |
|---|---|---|---|---|---|---|
| Carminati et al. [Carminati, 2009,2011] | ✗ | ✗ | ✓ | ✗ | ✓ | ✓ |
| Masoumzadeh et al. [Masoumzadeh, 2010b] | ✗ | ✗ | ✓ | ✗ | ✓ | ✗ |
| Bourimi et al. [Bourimi, 2012] | ✓ | ✓ | ✓ | ✓ | ✗ | ✗ |
| Choi et al. [Choi, 2014] | ✗ | ✓ | ✗ | ✓ | ✗ | ✗ |
| Liu [Liu, 2014] | ✗ | ✗ | ✗ | ✗ | ✗ | ✗ |
| Xu et al. [Xu, 2009] | ✓ | ✗ | ✗ | ✓ | ✓ | ✗ |
| Seitz et al. [Seitz, 2005] | ✗ | ✓ | ✗ | ✓ | ✗ | ✗ |
| Wainer et al. [Wainer, 2005, Wainer, 2007] | ✗ | ✗ | ✗ | ✓ | ✗ | ✗ |
| Ghazinour et al. [Ghazinour, 2013] | N/A | ✓ | ✗ | ✗ | ✗ | N/A |
| Jung et al. [Jung, 2014] | N/A | ✓ | ✓ | ✓ | ✓ | N/A |
| Ngo et al. [Ngo, 2015] | N/A | ✗ | ✓ | ✓ | ✓ | N/A |
| Habiba et al. [Habiba, 2013] | ✗ | ✓ | ✗ | ✗ | ✓ | ✗ |
| Gusmeroli et al. [Gusmeroli, 2013] | N/A | ✓ | ✗ | ✗ | ✗ | N/A |
| Lui et al. [Lui, 2007] | N/A | ✓ | ✓ | ✓ | ✗ | N/A |

# Chapter 4 **Ontology-based Access Control Management**

## 4.1 Introduction

As mentioned in the chapter 2 (section 2.2), the access control management on the resource is regulated based on the privacy requirements of the owner. This greatly depends on the openness of the mechanism how efficiently it deals with such requirements and provides options/solutions that fulfill them. To achieve this goal, system designers have offered several solutions that are based on either RBAC or ABAC as generic models to manage access control. These include: classifying resources into categories [Cheng, 2012b], itemizing profile data into different elements [Aïmeur, 2010] or classifying users into lists (e.g., blacklist users) [Cramer, 2015]. However, these methods do not scale well in large and complex environments because of: (i) the growing privacy configuration requirements and the incapability of existing solutions to handle them in an efficient manner [Beato, 2009]; and (ii) the burden of the definition and management of rules by users and administrators [Daud, 2015].

To overcome these shortcomings, the scientific community has proposed solutions to manage access control that model entity types as graphs [Pang, 2014, Cramer, 2015]; within ontologies [Carminati, 2009, Masoumzadeh, 2010b, Choi, 2014]; for role-based access control [Ben-Fadhel, 2015]; or for attribute-based access control [Smari, 2014]. Ontologies are particularly helpful to formally specify the conceptualization and interrelations of a domain [Mika, 2007], so that specific entities (e.g., users and resources) can be defined as instances of this conceptualization.

Then, access control can be easily managed according to the (privacy-oriented) interrelations defined in the ontology for the involved entities. Usually, ontology-based approaches define ad-hoc ontologies for concrete scenarios, which limit their generality and hamper the interoperability between heterogeneous settings (i.e., each one is based on a different ontological backbone) (e.g., see [Pang, 2014]).

To tackle these limitations, we present a generic ontology-based solution inspired in the Attribute-based Access Control (ABAC) paradigm that models entities and their access policies. This system provides the following benefits: (i) a generic ontology that can be easily extended for specific environments, so that access control can be defined at different levels of granularity; and (ii) it simplifies the definition and enforcement of rules, thanks to the automatic ontology-based inference of rules. In order to demonstrate its applicability and benefits, we have applied it to two large and open scenarios: OSNs and the Cloud.

The rest of the chapter is organized as follows. In section 4.2, we present general ontology for access control management. In section 4.3 and 4.4, we extend our general ontology for open scenarios, OSNs and cloud respectively, in order to show its applicability and access control rules management. Finally, in section 4.5, we provide the conclusions.

*The content of this chapter has been published in:*

- *Imran-Daud M, Sánchez D, Viejo A, (2016), Ontology-based Access Control Management: Two Use Cases, in: Proceedings of the 8th International Conference on Agents and Artificial Intelligence, Rome, Italy, pp. 244-249.*

## 4.2 A general ontology for access control management

The backbone of our ontology (which is shown in Figure 4.1) is inspired in the ABAC model. It models the three basic (ABAC) entities required to manage access control: *subject*, *object* and *policy*. *Subjects* can be the owners of the resources that define access rights for other users or they can be the target users over whom the access control should be enforced. *Objects* are the resources (e.g., services, files, messages, etc.) that require protection from unauthorized access; they are protected by defining *policies* that contain access rules. The access *rule* is represented by the following tuple.

$$rule \equiv < s_i, o_j, a >$$

where $s_i$ is the subject target user, $o_j$ represents the object resource and the element $a$ is the action that holds access decision (e.g., allow, deny).

The ontological property (i.e., *access rights on*) between the *subject* and the *object* determines the role of the user w.r.t. the resource (i.e., owner of the resource or the one who requests access to the resource). Likewise, the *defines* property between the *subject* and the *policy* indicates the relationship of policy maker with the policy, whereas, the *written for* property shows the relationship between the target user and the policy itself. Finally, each *resource* is associated with the *policy* through the *has* property.

The generic design of the ontology allows us to define general rules that refer to the abstract classes (i.e., *subject*, *object* and *policy*) rather than to specific entities. Then, entities involved in the specific scenario (i.e., concrete users and resources) can be represented as instances of ontological classes and, thus, access control over these entities can be enforced on the basis of general rules by relying on the ontological structure (i.e., specific rules at an entity level can be automatically derived from the general rules defined at a class level). Moreover, the generic ontology can be specialized with more specific classes that are appropriate for a concrete scenario and, accordingly, more specific rules can be tailored (in any case, without require to define them on entity-basis).

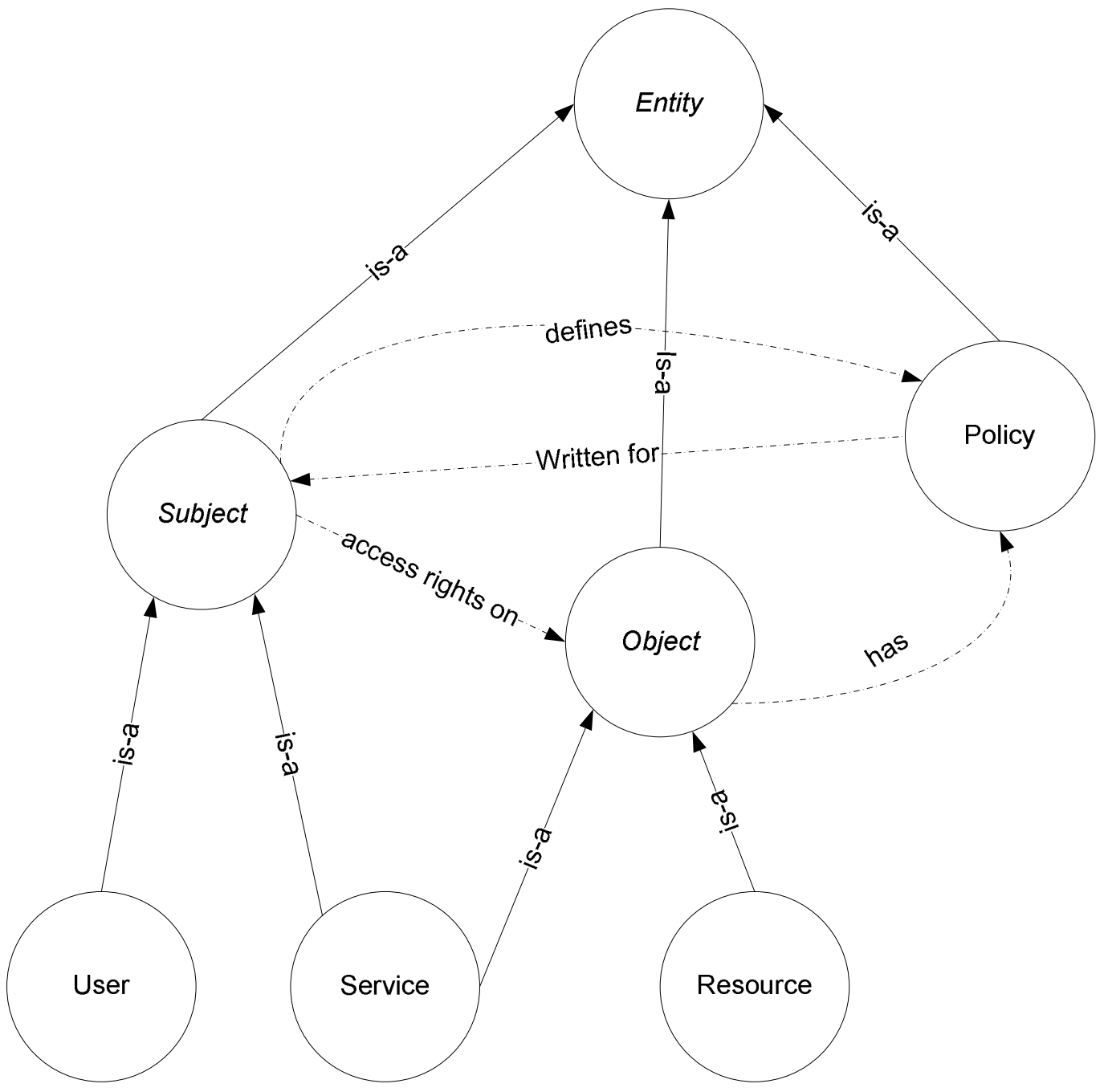

**Figure 4.1** Access control ontology

In order to take authorization decisions, the access control mechanism evaluates the interrelationship and the attributes of *subject*, *object* and *policy*, as stated in the ABAC model. Specifically, the system determines the following information from the ontology: (i) the resource requestor, (ii) the owner of the resource, (iii) the resource itself, and (iv) the policy defined by the owner of resource.

In the following subsections, we show how our generic ontology can be extended to model the entities involved in two widespread scenarios: OSNs and the Cloud.

## 4.3 OSNs Use Case

Online social networks (OSN) such as Twitter, Facebook, Google+, Myspace, etc., are platforms where people interact with each other by publishing messages. In these platforms, users can build their own social circles of friends and join social groups or communities. In these groups and communities, strangers may connect with each other according to their common interests, views or activities [Boyd, 2007]. In social networks, users spend most of their time in publishing or accessing information about such activities. Very frequently, the published content may contain sensitive data such as date of birth, political views, religious views, medical-related information or others.

Publicly shared content containing that sensitive information can be easily revealed by means of messages, profile data or social apps (like games). This data may portray a person's social or inner life [Gross, 2005], which constitutes a serious privacy issue. Social networks such as Facebook[3] and Twitter[4] consider trusted and non-trusted users as friends [Shehab, 2012], but the trust of such friends cannot be measured [Boyd, 2007]. As a result, sensitive information may be revealed to non-trusted users. Moreover, Johnson's analysis [Johnson, 2012] concludes that the majority of users are more concerned with internal threats of privacy (i.e., from friends) rather than strangers. For this reason, most of the OSN friends are considered untrustworthy to share sensitive information. On the other hand, according to the *European Union Agency for Fundamental Rights & Council of Europe* [Europe], this sensitive information needs to be protected from untrusted third parties, because it can be exploited by such parties for their own benefit [Viejo, 2013].

For this purpose, we present a mechanism to ensure privacy of the OSN users by managing access rights on the shared resource. Figure 4.2 presents the extension of our general ontology that models OSN entities and their interrelationships. In this scenario, the *subject* entities of the OSN (i.e., owners of the resources) manage their access rights on *objects* (e.g., photos, text messages, videos, etc.) by defining access control *policies* over other *subjects* (i.e., other users with whom the owners are in contact). The *rules* are the attributes of these *policies* that hold access right decisions (i.e., allow or deny access to a resource uploaded by the owner). Since OSNs allow users to classify their contacts into different categories (e.g., close friends, family friends, strangers, etc.), the *subject* class has been specialized with a *contact* subclass that encompasses the contact types of the users. This specialization is also helpful for the users to define different access rules according to the contact category of the users. Finally, *user* is modeled in a subclass of the *subject* class; their membership to a certain contact type of the owner of a resource is represented with the *has* property.

[3] http://www.facebook.com
[4] https://twitter.com/

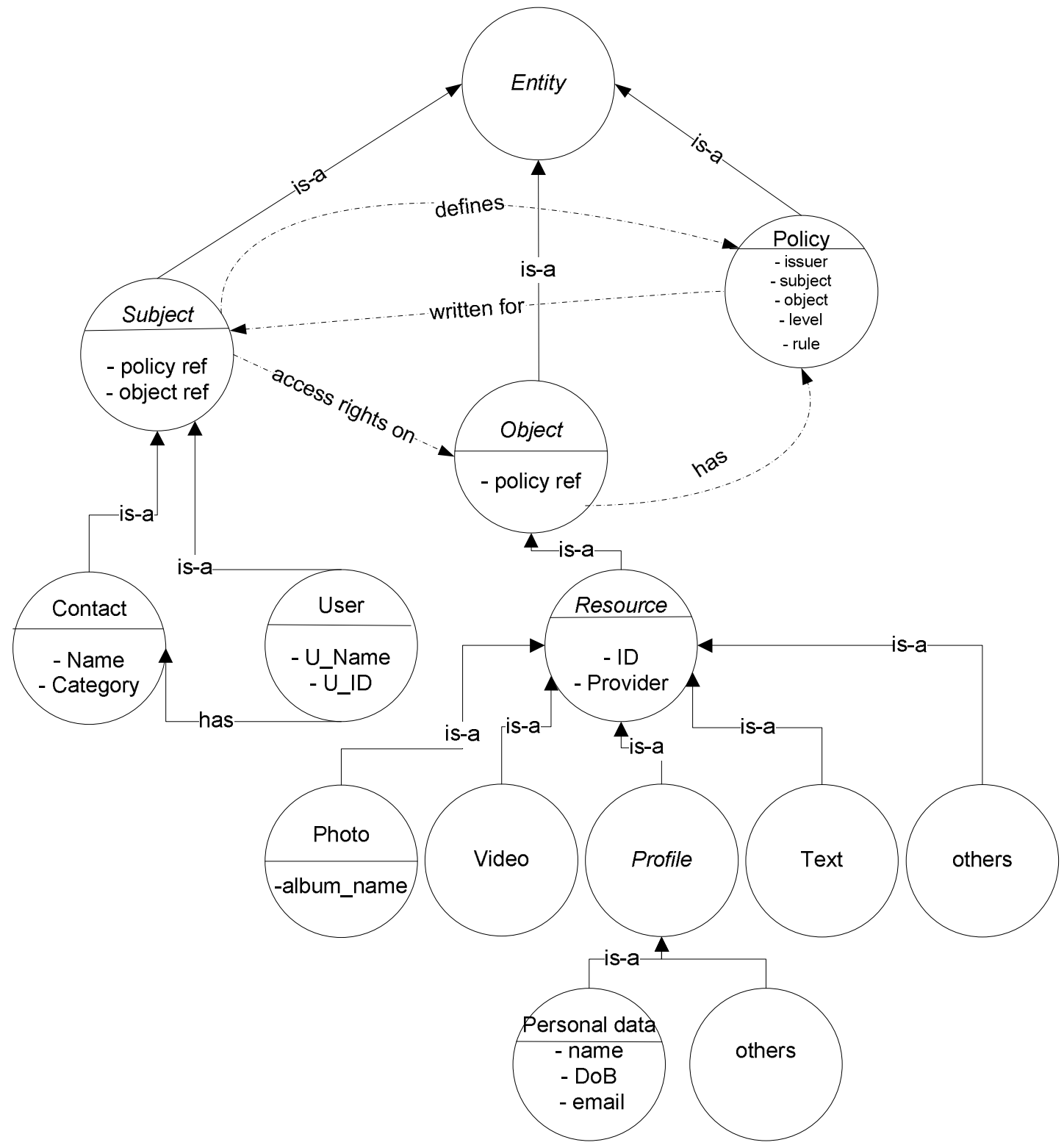

**Figure 4.2** Extended access control ontology for OSNs

The *object* class constitutes the resources that require protection from unauthorized access. In the context of OSNs, objects are specialized in specific *resource* types (i.e., *photo, video, profile, text,* etc.) so that a more fine-grained access control can be enforced; that is, managing access control on each resource type rather than applying the same rule for all the resources. The *profile* class is further classified into two subclasses: (i) *profile data*, which details the identity of the users, and whose access could be protected in order to avoid identity disclosure and (ii) other related information (e.g., interests of the users), which may refer to confidential information.

Even though this ontology represents the entity types involved in an OSN, it can be further extended to accommodate the specificities of a particular vendor (e.g., Facebook), such as predefined contact types or more specific resource types.

## 4.3.1 Access control management and enforcement

As discussed in section 4.2, a user may limit the access to her resources by defining an access rule for a set of target users. With our ontology-based approach, the rule can be defined for ontological classes at any level of abstraction so that it would be automatically enforced for the corresponding subclasses and, finally, instances (entities) of such classes. Within the OSNs scenario, the default rule for all the resources is *deny* access, so that the user only needs to define *allow* permissions. The following example illustrates the extension and instantiation of the OSN ontology for a specific scenario and the automatic inference of rules and their enforcement.

*Example 1*: Figure 4.3 illustrates the ontological specialization and instantiation of social network entities associated to the *Alice*'s social account (e.g. Facebook). As privacy preferences, she defines a *rule* to allow her *family friends* to access her *resources* (i.e., rule$_{\text{Alice}}$≡ < *family friends*, *resource*, '*allow*'>). This *rule* is encompassed in the *policy* instance that is linked with the instance of the *resource* being referenced and the instance of the target *subject* (i.e., contact type *family friends*). Since, this rule is defined at a class level (i.e., *family friends* in *contact* and *resource* as a whole), by ontological inference, it will be automatically enforced on all the subsequent entities. Since *Bob* is a *family friend* of *Alice* and by the inference of generic rule, the system grants full access to *Bob* on *photo* and *video* instances. Specifically, the following *rules* are generated for the instances of the *user* that are *family friends* of *Alice* (only *Bob* in the given case).

rule$_{\text{Alice}}$ ≡ < *Bob*, *"college.jpg"*, *'allow'* >
rule$_{\text{Alice}}$ ≡ < *Bob*, *"family.jpg"*, *'allow'* >
rule$_{\text{Alice}}$ ≡ < *Bob*, *"party.avi"*, *'allow'* >
rule$_{\text{Alice}}$ ≡ < *Bob*, *"festival.avi"*, *'allow'* >

In any case, *Alice* can also define rules for specific instances of the *user* class. For example: *Alice* may only allow *Alex*, from *close friends* contacts, to access all of her photos (i.e., rule$_{\text{Alice}}$≡ < *Alex*, *photos*, '*allow*'>). Thus, the following rules are inferred from this generic rule.

rule$_{\text{Alice}}$ ≡ < *Alex*, *"college.jpg"*, *'allow'* >
rule$_{\text{Alice}}$ ≡ < *Alex*, *"family.jpg"*, *'allow'* >

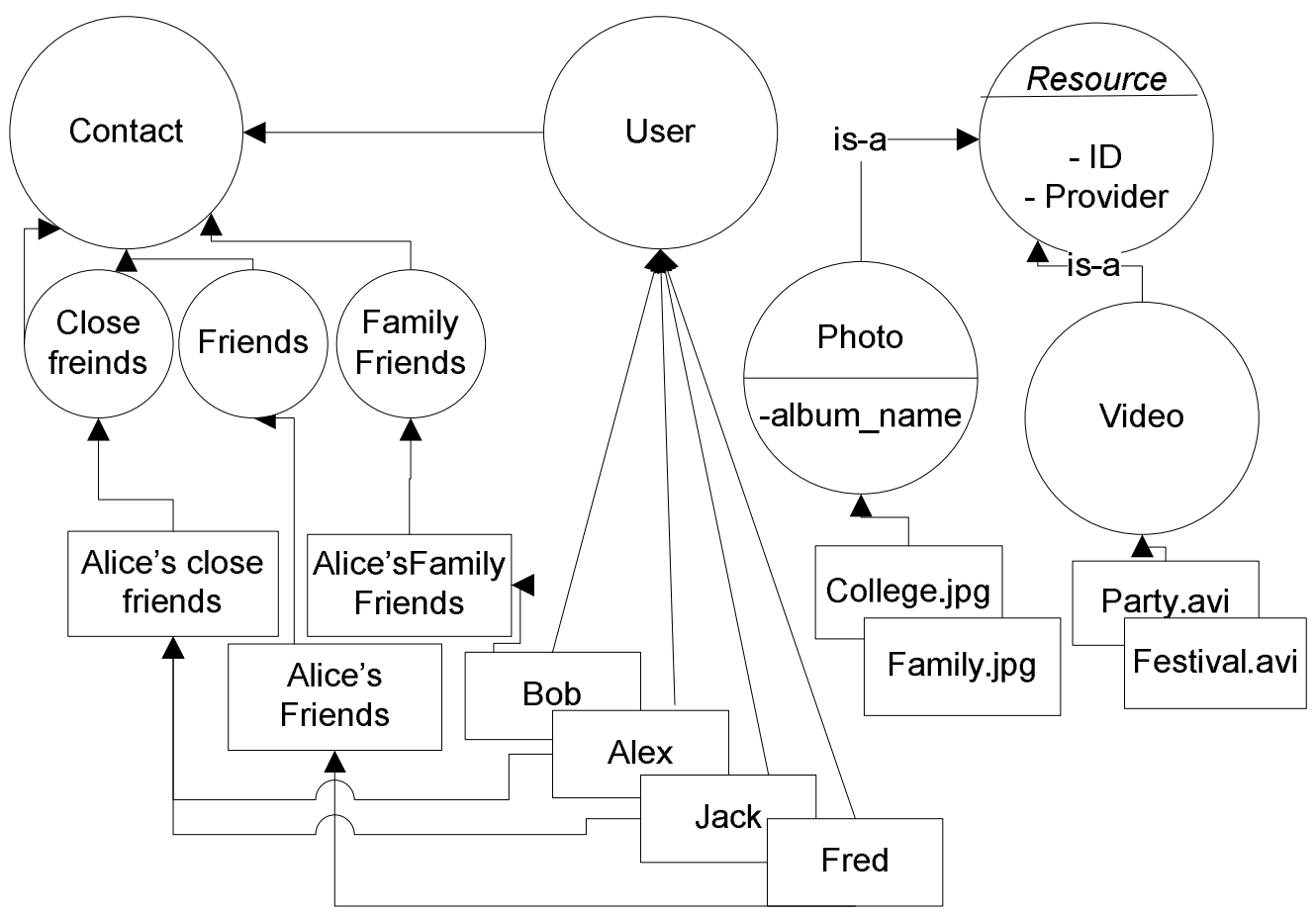

**Figure 4.3** Instantiation of the OSN ontology for user *Alice*

## 4.4 Cloud Use Case

Cloud computing provides a ubiquitous platform to share resources and to provide cloud services to tenants. Because of its open nature, it requires a scalable mechanism that manages access control on the shared resources. Usually, a cloud relies on the following multi-level service models [Rountree, 2014] that are used to deliver cloud services: (i) Infrastructure as a Service (IaaS), where physical computing resources are shared among cloud users; (ii) Platform as a Service (PaaS), which provides databases and operating systems; and (iii) Software as a Service (SaaS), which provides software applications. To manage access rights in cloud, we extend our general ontology to incorporate the cloud entities and the attributes that are relevant to manage access rights in the cloud environment.

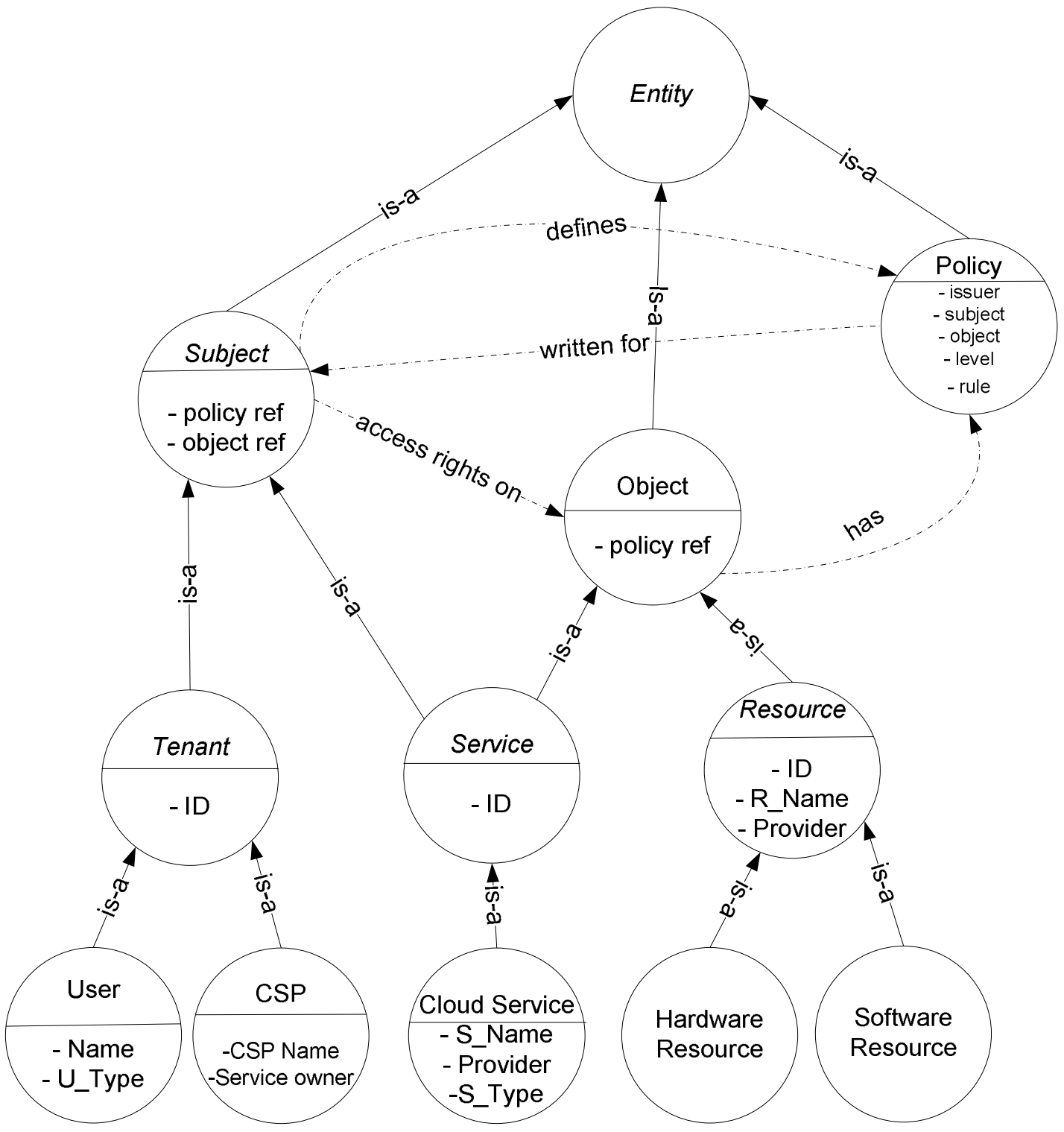

**Figure 4.4** Extended access control ontology for the cloud

Figure 4.4 illustrates the extended ontology that lists cloud entities (*tenant*, cloud *service* and cloud *resource*) and their interrelationships. In this illustration, *tenant* is a subclass of *subject* that holds cloud actors, which are: (i) *user* (which use cloud services) and (ii) *cloud service provider* (*CSP*) (which provides and shares cloud services). Likewise, *service* is a subclass of *subject* that represents the services provided by the CSPs, these services may require access to shared resources to accomplish their tasks. On the other hand, *service* is also a subclass of *object* because the tenants may access them as *cloud service*. Finally, cloud resources can be *hardware resource* (e.g., servers, storage space, etc) or *software resource* (e.g., web application, web services, etc.). Cloud service providers can manage the access to their shared resources and services by defining a *rule* that is encompassed within a *policy*, as explained for OSNs in the previous section. The following example illustrates the enforcement of rules in the cloud scenario.

### 4.4.1 Access control management and enforcement

*Example 2*: Figure 4.5 illustrates the extension and instantiation of the cloud ontology for CSP (*Google*) that offers its *services* and *resources* to the *users*. In this example, Google offers different cloud services at different service levels (i.e., SaaS, PaaS and IaaS) for standard users and educational institutions (e.g., educational institutions are offered more space on Google drive and a professional domain for email). *Google* configures the access to its resources and services with the following two rules: (i) it allows *SaaS* services to access all the *resources* (hardware and software); and (ii) it grants *users* belonging to any educational institution with special access to its *Cloud services* that are meant for an educational purpose. In this last case, and in coherency with the ABAC model, we can rely on the attributes defined for the ontological classes and instances. Thus, the following generic rules are defined:

$\text{rule}_{\text{Google-R1}}$ ≡ < *SaaS*, *resource*, *'allow'* >

$\text{rule}_{\text{Google-R2}}$ ≡ < *Users <U_Type="Education">*, *Cloud Services <S_Type="Education">*, *'allow'* >

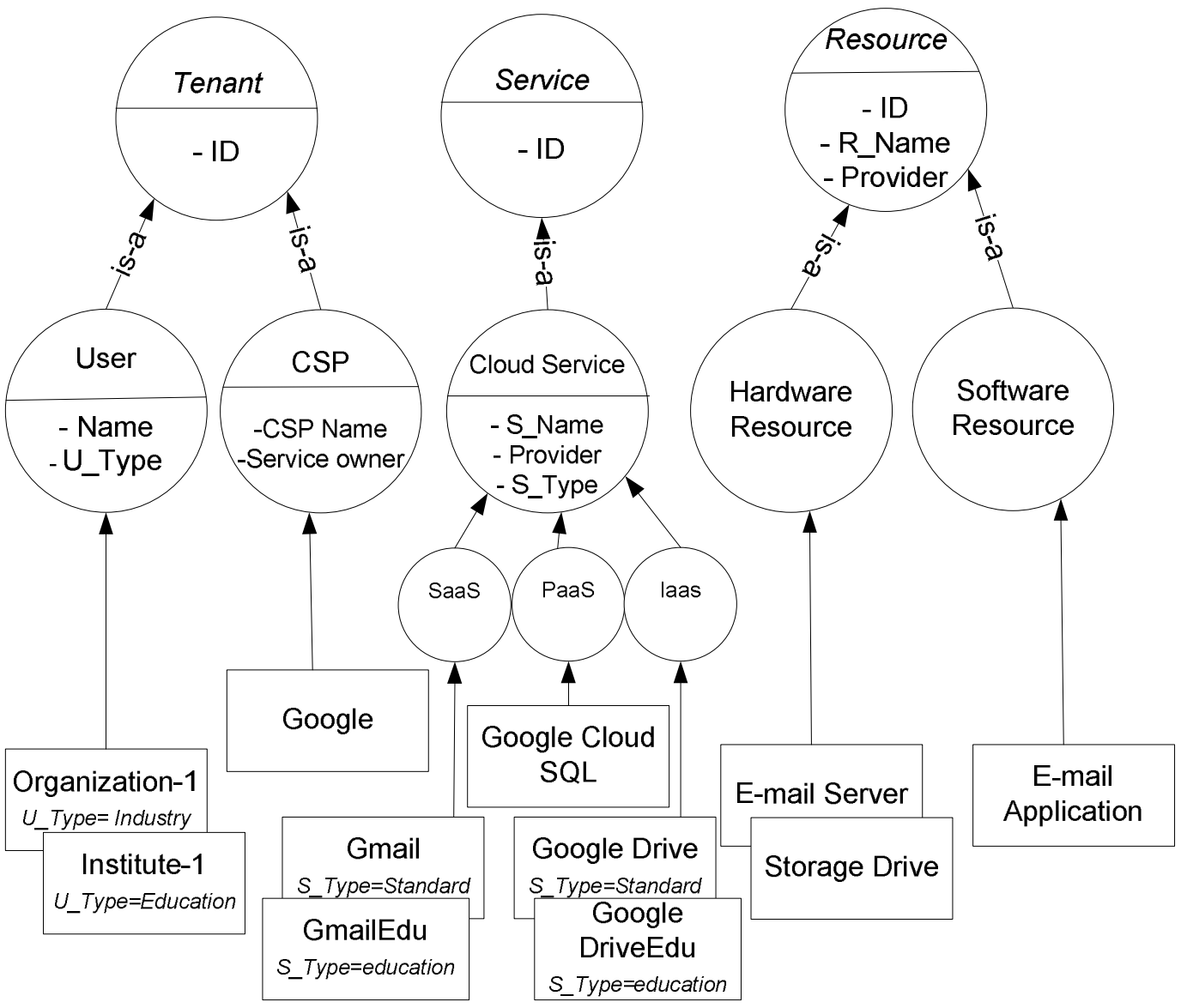

**Figure 4.5** Instantiation of the Cloud ontology for *Google*

The rule$_{\text{Google-R1}}$ is defined at the conceptual level (i.e., at *resource* and *SaaS* classes) of the ontology and, thus, it covers all the entities below the *hardware resource* and *software resource* classes. By inferring specific rules at the instance level, we obtain the following ones:

rule$_{\text{Google-R1}}$ ≡ < *Gmail*, *e-mail server*, *'allow'* >
rule$_{\text{Google-R1}}$ ≡ < *Gmail*, *storage drive*, *'allow'* >
rule$_{\text{Google-R1}}$ ≡ < *Gmail*, *e-mail applications*, *'allow'*>
rule$_{\text{Google-R1}}$ ≡ < *GmailEdu*, *e-mail server*, *'allow'* >
rule$_{\text{Google-R1}}$ ≡ < *GmailEdu*, *storage drive*, *'allow'* >
rule$_{\text{Google-R1}}$ ≡ <*GmailEdu*, *e-mail applications*, *'allow'* >

On the other hand, rule$_{\text{Google-R2}}$ grants access to cloud services that are specifically allocated to educational institutions. To manage this, the type of *users* is determined through the value of the *U_Type* attribute of the entities, whereas the educational services are determined by the value of the *S_Type* attribute. As a result, the educational instances of the *user* class are distinguished and granted access to all cloud services that are allocated for educational institutions. The following rules are, thus, generated due to the inference of this generic rule.

rule$_{\text{Google-R2}}$ ≡< *Institute-1*,*GmailEdu*, *'allow'* >
rule$_{\text{Google-R2}}$≡<*Institute-1*,*Google DriveEdu*, *'allow'* >

## 4.5 Conclusions

In this chapter, we proposed a generic ontology that models entities, their interrelationships and access control policies, and it can be easily extended for specific environments. To show its applicability, we extended it for two large and open scenarios: OSNs and the cloud. We also illustrated through examples how the definition of rules and the management of access control are greatly simplified for system administrators, because they can be intuitively made at a conceptual class-level. Then, specific (and dynamic) rules can be automatically inferred according to the specific entities, which would also be likely dynamic in open scenarios such as those tackled in the chapter.

# Chapter 5 **Ontology-based Access Control Delegation Enforcement**

## 5.1 Introduction

As introduced in chapter 2 (section 2.3), delegation mechanism enable users to transfer their access rights on a particular resource to other in order to lower their administration burden and to manage access control. Most of the delegation mechanisms are based on the role-based access control (RBAC) model [Sandhu, 1996], where access rights are delegated in the form of roles. Few models rely on the attribute-based access control (ABAC) model [Hu, 2014] for delegation, where delegation is managed by using policies instead of roles.

Delegation of access control is particularly challenging in dynamic and distributed scenarios due to large number of entities with heterogeneous requirements. Therefore, delegation enforcement in these scenarios require extensive administrative burden on the users/administrators. In the following we illustrate this kind of scenarios with a paradigmatic example (i.e., cloud computing).

Cloud service providers (CSPs) share these services among distinct organizations or end users that are called *tenants*. In a multi-tenant environment (i.e., a single resource shared with multiple tenants), access control management on shared services is a serious concern for CSPs [Manvi, 2014]. Ideally, the shared services need to be accessed only by the tenants that are authorized by the pertinent CSPs. On the other hand, CSPs are burdened to efficiently manage access control due to the diversity of cloud services and the diverse security requirements of the tenants. For example, an organization may have a large number of users with different types of roles, and the access to shared services must be distributed according to these roles. To handle this situation, one of the solutions implemented by CSPs is the *delegation* of access control. Delegation brings a number of advantages that includes decentralization of access control, better scalability for large organizations, and seamless management of role-change requests. With delegation, a CSP acting as the delegator can transfer its administrative privileges on a particular service to other tenants that are called delegatees. In turn, these tenants can manage the access control on the delegated services for other users according to their own requirements.

In a cloud environment, delegators delegate their access rights to tenants on the same or different level of the service model in a hierarchal way (i.e., IaaS, PaaS and SaaS). However, due to the heterogeneity of the cloud federation [Kurze, 2011] (a cloud environment where CSPs share cloud resources to other service providers) and the lack of trust, the delegation of access rights can be a serious concern for end-to-end authorization and verification of the delegators [Nahrstedt, 2012]. *Authorization* refers to the process of specifying and transferring access rights from a delegator to a delegatee within the cloud. In contrast, *verification* validates the privileges of a delegator that allow her to manipulate the authorized resources. However, the identity of the delegators could be forged in order to gain or delegate unauthorized privileges (which is a spoofing attack) [Wang, 2008]. In this respect, existing cloud services (e.g., Microsoft Azure) implement conventional and simple methods of delegation; for example, the delegation mechanism of Azure is based on the credentials of the user (i.e., username and password). This approach requires a centralized management of the user credentials that can overhead the administrator and may not be feasible for an organization that requires further delegations to employees with limited access.

In this chapter, we propose an ontology-based delegation framework (with cloud-centered improvement) to the XACML delegation profile [XACML-Profile, 2009] (i.e., XACML extended with delegation concepts), wherein the access rights are delegated in the form of policies. In contrast to the methods [Sohr, 2012, Ruan, 2014], our approach does not require specifying constraints or rules for enforcing delegation; instead, the delegation enforcement is managed by means of automatic algorithms. To attain this, we use our already proposed access control ontology (in chapter 4, section 4.2) that can be instantiated with the cloud entities in order to demonstrate delegation workflow. Consequently, our scheme overcomes the need of generating a delegation graph as a result of each resource request by the user; instead, the interrelations of the entities are used to validate the users' access request. Moreover, trust is built through the trusted policy that originates delegation workflow, and it is digitally signed by the resource provider authority that initiates the delegation process, which also owns the resource. In the following, the main contributions of our research work are summarized:

- We use our already proposed ontology in chapter 4 (section 4.4) to model the entities involved in the delegation process (with cloud as scenario), which includes subjects (i.e., delegators or delegatees), objects (resources or services), policies (document that translate delegated privileges) and their interrelations. Through that ontology, we can keep a track of who is delegating, what privileges on a resource are being delegated and it also provides an intuitive solution to verify the attributes of the actors involved in any delegation.

- We present a distributed delegation model for the cloud by classifying its main actors (e.g., cloud providers, CSPs, organizations and users) into different delegation levels, wherein the access rights on the resources can be delegated in a distributed way. In contrast to other solutions ( e.g., [XACML-Profile, 2009]), the delegation of access rights are managed in a distributed way and the authenticity of the delegation policy is verified with the trusted policy that is written and signed by the CSP (which is the owner of the resource).

- In contrast to the method that verify the delegator's authority through roles [Ahn, 2007], our system automatically verifies the authority through the attributes of the entities and the policy of the delegator by following the interrelations of the entities (represented as instances of the ontology), which leads to the trusted policy.

- Contrary to studies [Wainer, 2005, Wainer, 2007], our system does not require the specification of rules for delegation enforcement, but it automatically enforces delegation, verifies the delegated authority and also revokes the delegated privileges by using simple algorithms. In addition, we do not require any additional rules to implement delegated policies; instead, it automatically implements a delegated policy by applying a policy combining algorithm that combines the normal access policy and the delegated policy. Moreover, this algorithm resolves possible policy conflicts within the entities.

The rest of the chapter is organized as follows. In section 5.2, we present the cloud-centered delegation ontology. In section 5.3, we detail the enforcement of the delegation by means of several algorithms. Section 5.4 discusses and evaluates the performance of the proposed system. Finally, in section 5.5, we provide the conclusions.

*The content of this chapter has been published in / submitted to:*

- *Imran-Daud M, Sánchez D, Viejo A, (2015), Ontology-Based Delegation of Access Control: An Enhancement to the XACML Delegation Profile, in: Fischer-Hübner S, Lambrinoudakis C, López J (Eds.) Trust, Privacy and Security in Digital Business: 12th International Conference, TrustBus 2015, Valencia, Spain, September 1-2, 2015, Proceedings, Springer International Publishing, Cham, 2015, pp. 18-29.*
- *Imran-Daud M, Sánchez D, Viejo A, Ontology-based Access Control Delegation Enforcement for the Cloud, in: Security and Communication Networks, (Under Review)*

## 5.2 Ontology-based Delegation

Our model is based on the XACML delegation profile [XACML-Profile, 2009]. In this profile, the authority of the policy issuer is validated through the delegation graph that is generated on each user's request for a resource. To do so, the attributes of an access request (which includes the attributes of the user and the requested resource) are matched with the attributes of the delegated policies stored in the database. As a result, the delegated policies related to the requested resource are determined and represented as a graph by creating edges between the delegated policy nodes. This graph plots the hierarchy of the delegated policies of the delegators of the same resource. Finally, the authority of the delegator that issued the policy to the requester is verified through this graph by verifying the hierarchy of all preceding delegators. As a result, the access to the resource is managed according to the flow of the graph (i.e., each predecessor delegator in the hierarchy grants access privileges to the successors). We can see that, due to the overhead of searching policies and generating the policy graph for each access request, this approach may not scale well in the cloud scenario due to: (i) the large number and heterogeneity of entities (users, resources, policies) that could be involved in a delegation; and (ii) the large number of access request that should be managed by the CSP.

Our model is extension of the XACML delegation profile that incorporates an ontology and models the semantics of (attribute-based) access control and its delegation in the cloud scenario. The use of an ontological paradigm [Guarino, 1998] to model the delegation workflow has the following benefits: (i) it is simple and the delegation process can be implemented easily; and (ii) it automatically defines and interprets the relations between the entities [Carminati, 2009]. Our ontology models the entities (i.e. delegator, delegatee and resource types) and their interrelations involved in the delegation process in cloud scenarios. In this ontology modeling, the entities correspond to the instances of the classes, and the delegation workflow is represented through their interrelations. Similarly, to process access request of the tenant, the delegator's authority (e.g., a CSP)) is evaluated and verified through the interrelationship of the instances of the ontology (it include hierarchy of the delegators', the delegatee and resource that is delegated) in spite of searching entities within the policies In this solution, the attributes of an access request are determined and compared with the instances that represent cloud entities; as a result, their related policy is determined for from the database in order to get the rules defined for these entities. Contrary to the XACML profile that maintains policies of all delegators sharing a common resource within a policy set, in our system, a *policy set* only contains the policies issued by a single subject for a single resource and each delegator maintains its own *policy set*. By doing so, we reduce the overhead of (i) searching policy issuer from a policy set in a large scale environment where delegators are large in number, and (ii) generate policy graph to validate delegation authority for access request.

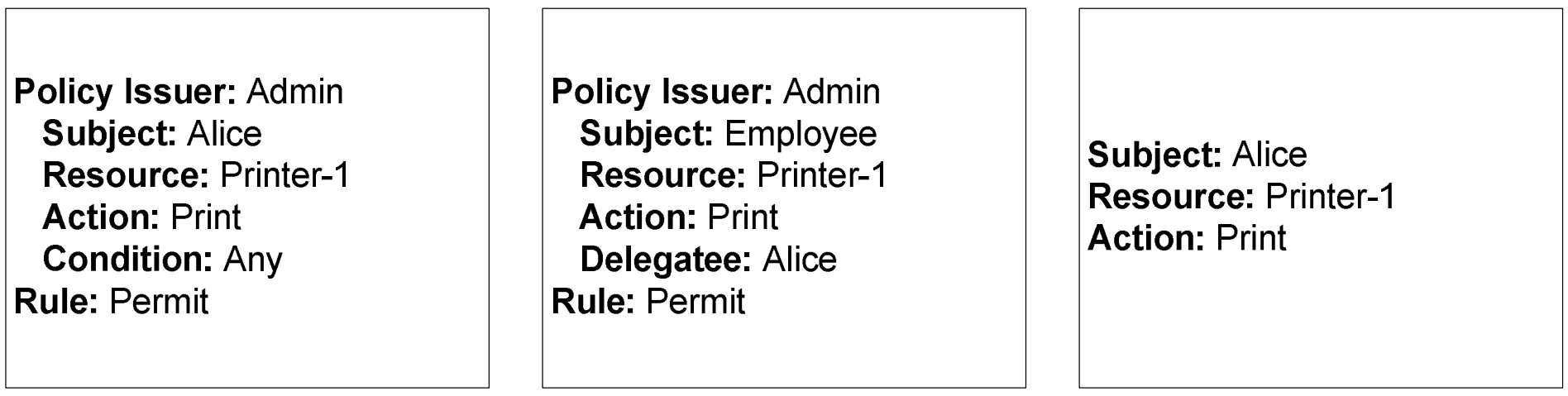

a) Access Policy b) Administrative Policy c) Access Request

**Figure 5.1** Structure of policies and access request

In this model, a cloud user can manage two set of policies (listed in Figure 5.1): (i) an *access policy* (which permits or denies access on the resource); and (ii) an *administrative policy* (which gives privileges to the tenant to issue policies to other tenants). For example, in Figure 5.1(a), a CSP manages the access to a resource (i.e., printer-1) by allowing *Alice* to use the printing facility. On the other hand, in Figure 5.1(b) the CSP delegates the authority to *Alice* (of the *employee* department) to issue policies (*access* or *administrative policies*) on the same resource. To grant a limited access to the resource, the delegator can limit the privileges through the rule attribute of the policy.

Contrary to the XACML profile, each delegator only maintains those policies in a *policy set* that are defined for a common resource. Whereas, in the XACML profile, a policy set holds policies of all delegators that share privileges on a common resource. As mentioned in the introduction, the cloud relies on a multi-level service model, wherein, services and resources are managed in a decentralized way. Therefore, due to its central management of the policies, this profile is not feasible in a cloud environment due to its distributed management of users and policies. Moreover, there is a significant overhead while searching some entity (as a result of *access request*) from a given *policy set* due to: (i) the usually large number of users within the cloud; and (ii) the need for searching all the policies to get the related policy from a given *policy set*. Thus, this approach renders following benefits: (i) decentralized management of *policy sets*; (ii) feasible to delegate in the distributed environment; and (iii) it improves the methods of delegator's authority verification.

## 5.2.1 Ontology Representation of ABAC and Cloud Entities

The entities involved in the ABAC model are *subjects*, *objects* and *policies*. Subjects usually own or delegate privileges on resource objects (e.g., services, network resources, data sets, or applications) by managing the access control as policies. That is achieved by defining access control policy rules for the resource objects and the target subjects. In order to enforce access rights, the subject and the object attributes are evaluated to take authorization decisions. For authorization, these attributes are evaluated against the policy rules and, based on this evaluation, access decisions are taken. To do so, a module of XACML (i.e., PIP, the policy information point) manipulates the attributes of the entities. In the same way, access rights can be delegated to other subjects through delegation policies, which contain the attributes and the rules of the entities defined for the delegated resources.

To manage the delegation workflow in an automatic way, the cloud entities modeled in an ontology (chapter 4, section 4.4) are represented as classes. In this manner, the policy attributes (those in figure 5.1(a) and 5.1(b)) can be manipulated in order to model interrelations of the entities as ontological properties. The directed edges of this ontology represent the relationships among the entities, which depict a taxonomic dependency within classes or subclasses (e.g., types of subjects or objects) or associative properties between classes (e.g., access right of a subject over an object).

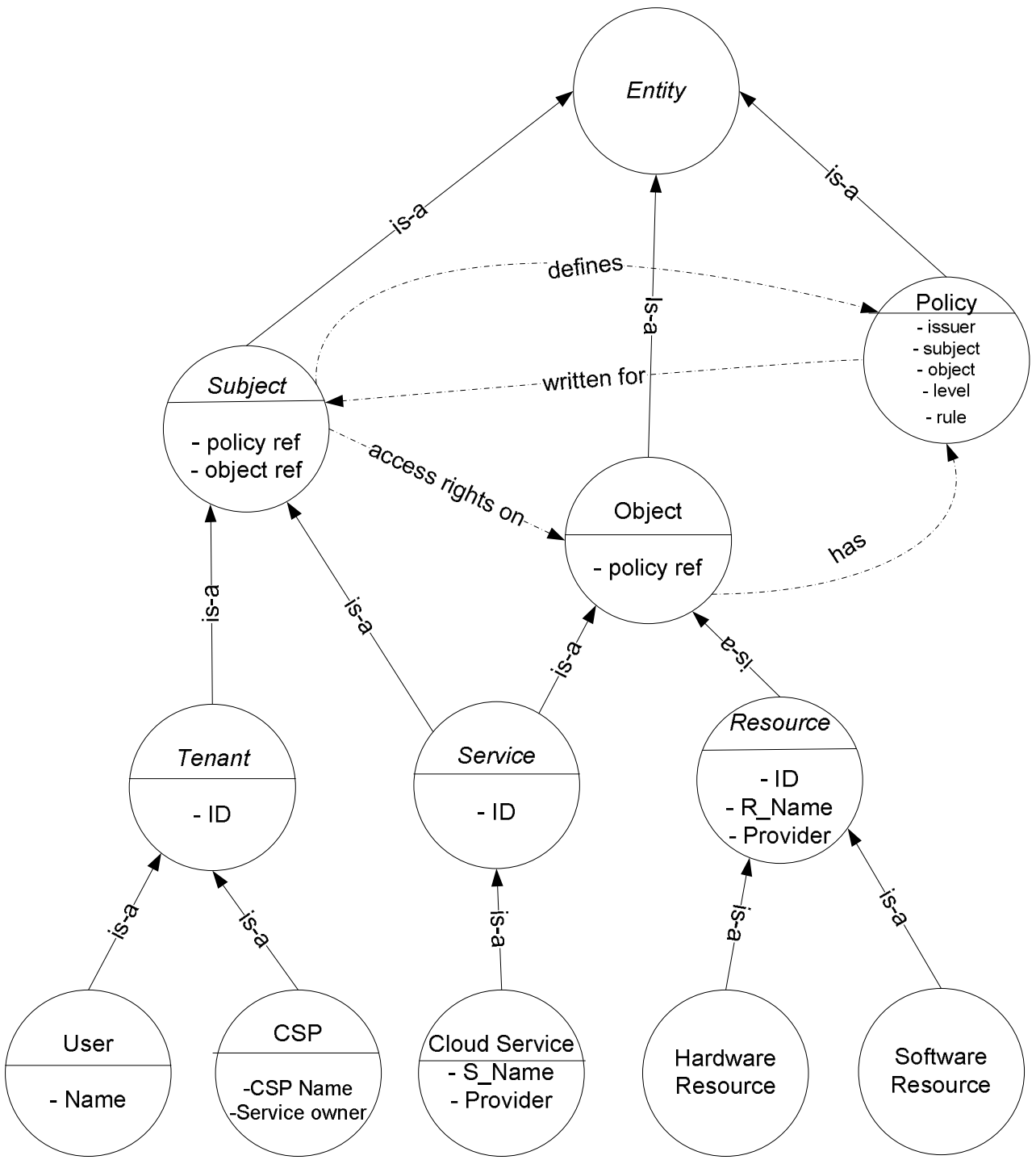

**Figure 5.2** Ontology modelling delegation process for cloud entities

For our reference, we redraw our proposed ontology presented in chapter 4 (section4.4) in figure 5.2 that depicts the cloud entities (as ontology classes) and the logics of the delegation process (as class interrelationships). This cloud ontology enables *users* and *CSPs* to delegate access rights on the *cloud services* or *resources* by managing delegated *policies* in order to manage access control.

### 5.2.2 Workflow Delegation Representation

In this section, the proposed ontology (shown in Figure 5.2) is instantiated in order to show the delegation workflow and its applicability in the cloud environment. For this propose, the subjects, objects and policies entities are instantiated and interrelated according to ontological properties. In this modeling, the delegator *subject* maintains its *policy set* in the local repository that lists the details of the delegatees authorized on a certain resource. The instances of the delegator and delegatee are linked together through the policy instance, whereas, the delegated resource has a direct interrelation with the owner and the delegated policy.

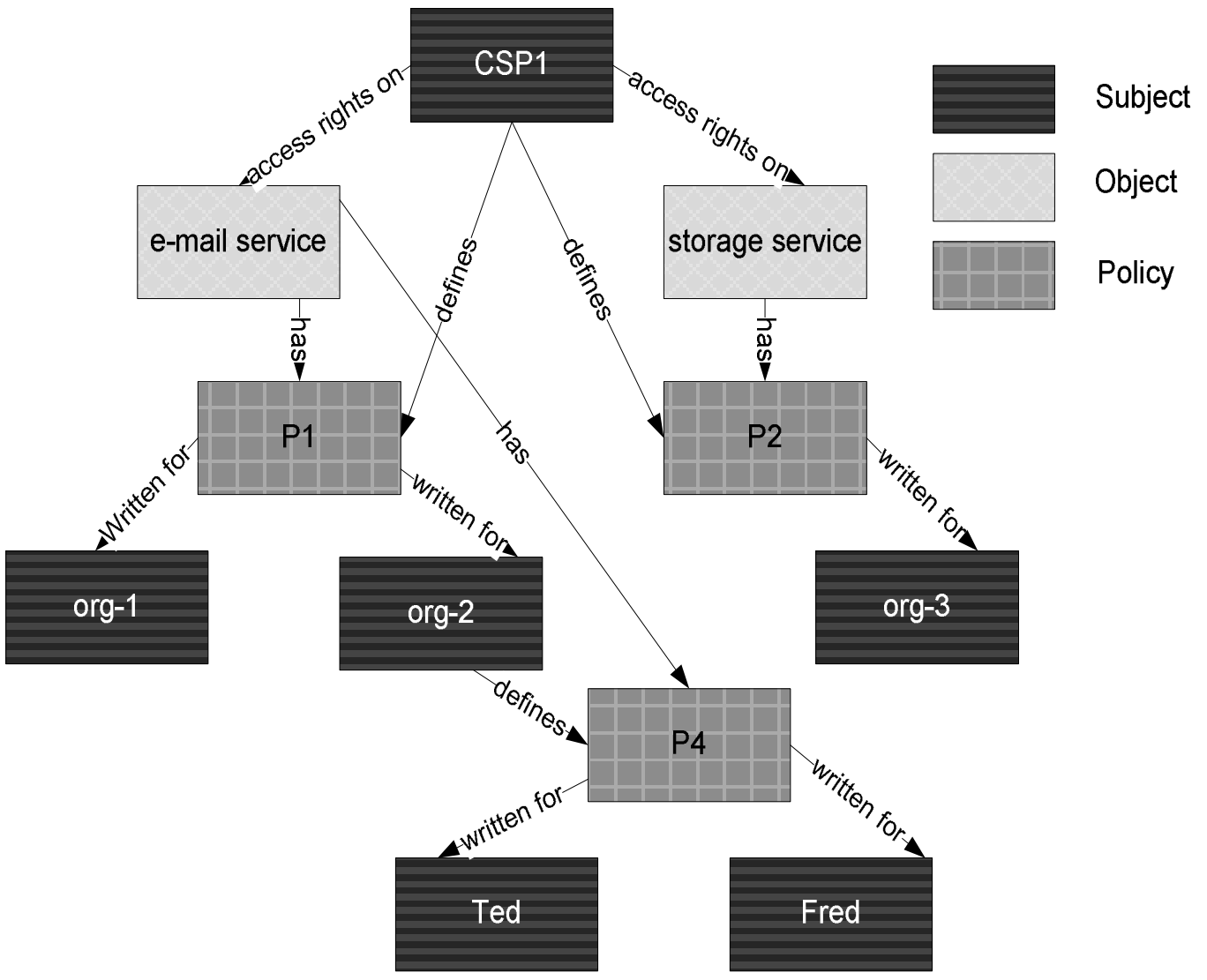

**Figure 5.3** Delegation workflow representation

Figure 5.3 presents the delegation workflow instantiated from the basic ontology proposed for the cloud scenario (presented in section 5.2.1). In this workflow, *e-mail service* and s*torage service* (instances of the *Cloud service* class) are offered by the cloud service provider *CSP1*, who is the owner of these services. In the next level of this workflow, the privileges on these services are delegated to other organizations by maintaining separate policies for each service. Thus, the policy set *P1* maintains the access privileges delegated to the *org-1* and *org-2*, whereas, *P2* maintains the access privileges delegated to *org-3*. *Org-1* and *org-3* are granted only with an *access policy* (i.e., they can access *e-mail* and *storage services* respectively, but they cannot further delegate); whereas, *org-2* holds an *administrative policy* and it can further delegate access privileges of the *e-mail service*. In the next level of the delegation workflow, the access privileges on the *e-mail service* are further delegated by *org-2* to users *Alex* and *Ted* by maintaining a policy set *P4*, which only contains *access policies*.

### 5.2.3 Workflow of the system

When a cloud user delegates her access privileges, the system automatically determines the instances and generates the interrelations of the entities (i.e., delegator, delegatee, resource delegated and policy set) involved in the delegation process according the ontological structure; this process is algorithmically formalized in Section 5.3. As a result, and on contrary to the XACML profile, we do not need to generate a delegation graph on each request. Instead, our workflow is updated on each delegation request, thus rendering the delegation verification more efficient, as it will be discussed in section 5.4.

In order to process an access request, first, the system verifies the access privileges of the user from the policy linked to her within the delegation workflow. Then, the system determines the policy issuer (i.e., the one who issued policy to the access requester) from the delegation workflow, and verifies her authenticity; this process is formalized in section 5.3.2. For this purpose, the attributes of the policy instances in the delegation workflow that are connected to the access requester are examined and cross matched with the attributes of the access request (as listed in Figure 5.1(c)). In the case that the required delegated policy is found, the system verifies the authenticity of the policy issuer by generating an *administrative request*. For this purpose, the delegator's instance linked to the target *access policy* is determined, and the delegator's attributes are examined and verified. In this way, the authority of the delegators who are in the chain of delegation to the target delegator (until the policy of the resource owner) are examined and verified by generating a series of *administrative requests* at each level of the delegation. Finally, a user is granted access to the resource provided that (i) the delegation hierarchy is legitimate (i.e., it originates from the trusted policy that is digitally signed by the resource owner), and (ii) a user is granted access privileges; otherwise, system denies the access to the resource.

Similarly, a *revoke request* is generated by the delegators in order to revoke access privileges of the tenants. As a result, the target policy and its chain of ascending policies in the hierarchy are verified. In case the request is genuine, the system revokes further delegations, and deletes the instances from the ontology-based workflow.

### 5.2.4 Policy Conflicts

Within a delegation workflow (especially if it is defined within a distributed environment involving heterogeneous entities such as the cloud), there could be cases in which the rules of the policies contradict each other:

1. A user has access policies issued by two different subjects on a common resource, where one policy permits the access to the resource while the other one denies it. For example, in an organization, an employee is working in two different departments, where one department denies access to a particular resource and the other department grants access to the same resource.

2. A user has two policies issued by two different subjects on a common resource, where one policy contains an *access policy* that only allows her to access to the resource, whereas the other policy contains an *administrative policy* that delegates access rights to issue policies on the same resource. For example, in an organization, an employee is working on two different departments, where one department allows the user to only access a particular resource and the other department grants access rights to further delegate them to other users on the delegated resource.

The above mentioned policy conflicts are automatically handled by the policy combining algorithm (which is formalized in Section 5.3.4), and it does not require any other rule or constraints specifications for these situations (as practiced in [Sohr, 2012, Ruan, 2014]).

Our system handles policy conflicts between two or more delegators according to their positions in the delegation workflow; that is, preceding delegators are priority, because they can revoke privileges to their subsequent delegators. If delegators have the same level of delegation, the policy that holds the strictest rule is implemented.

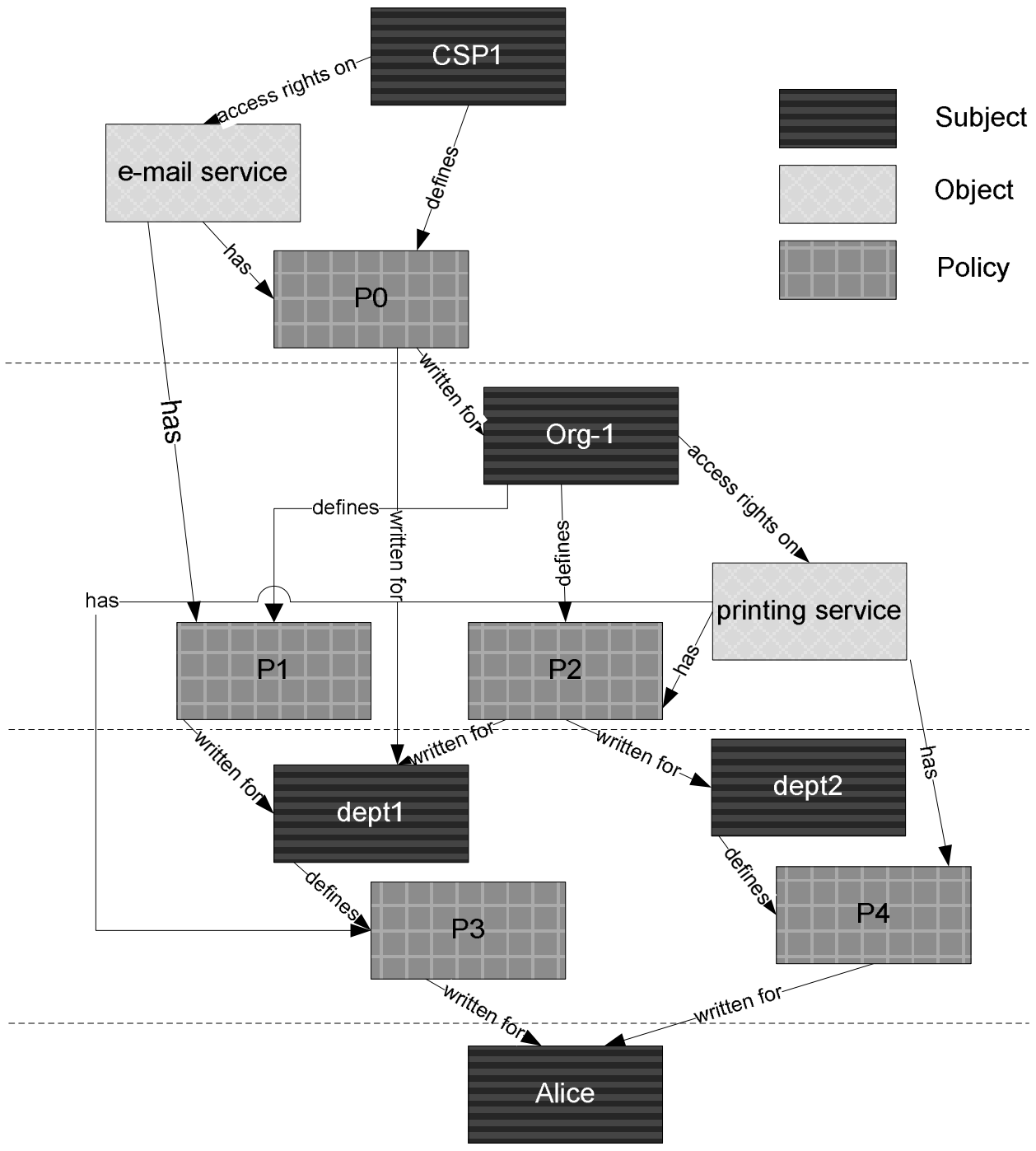

**Figure 5.4** Example of policy conflict

The above mentioned conflict scenarios are illustrated in Figure 5.4, where several policies contradict the rules of each other. Regarding conflict '1', *dept1* (an instance of object class) *has* two policies (in policy sets *P0* and *P1*) on a common resource *e-mail service* by the delegators *org-1* and *CSP1*. Let say that the policy in *P0* grants access to *dept1* on resource *e-mail service* and *org-1* denies access to the same resource in *P1*. When the employees of *dept1* perform an access request, the conflict between policies in *P0* and *P1* is resolved by comparing the precedence of the delegators. In a given scenario, the policy in *P0* is given priority over policy in *P1* due to its precedence over *P1* and the employee of *dept1* is given access to the desired resource.

Regarding conflict '2', *dept1* grants access to *Alice* on the resource *printing service* (by defining *P3*), and *dept2* denies access on the same resource in policy *P4*. During the *access request* by the *Alice*, the precedence of the issuing authorities of *P3* and *P4* are checked (i.e., *dept1* and *dept2*, respectively). In this case, the precedence of *dept1* and *dept2* is the same; therefore, the policy that has the strictest rule, that is *P3* (deny), will be implemented.

# 5.3 Delegation Enforcement in the Cloud

In order to support access control delegation in cloud scenarios, cloud entities need to be categorized according to their roles and the type of services they provide. The purpose of this categorization is to manage the delegation within the cloud service levels (i.e., IaaS, PaaS and SaaS) and to provide an infrastructure for the management of the policies within these levels. Accordingly, there are three main entities that manage cloud resources: (i) the cloud provider (e.g., Google); (ii) cloud service providers (e.g., Gmail, Google drive, etc.); and (iii) tenants (e.g., organizations or local users using the cloud services). Cloud providers provide infrastructure (e.g., servers, storage, etc.) to the CSPs. Likewise, CSPs provide platforms or services/applications to tenants. Finally, tenants, who are the actual consumers of the cloud services, can also manage shared services with other users (e.g., users/departments within the organization).

In order to provide a decentralized delegation, these entities must have distributed local repositories to store their policies so that they manage their resources independently. The purpose of this distribution is to provide a mechanism for decentralizing the authority to reduce the overhead of the policy management by a single entity. The flow of the delegation and the management of the policies within the cloud entities are shown in Figure 5.5:

- In the first layer of the delegation, the cloud provider (CP) can delegate its resources to the CSPs by storing their policies in its local repository. In this repository, the CP manages *policy sets* for each shared resource and each *policy set* contains the policies (i.e., *access policies* or *administrative policies*) of all the entities that share a common resource.

- Similarly, in the second layer, the entities manage access rights on resources (e.g., the services owned by the CSPs or the delegated resources by the CP) by maintaining policies in the local repository managed by the CSPs. The CSPs can delegate resources to other CSPs (on the same level e.g., in a federated cloud scenario) or to the tenants of the cloud (i.e., third level entities) by maintaining *policy sets*.

- The third layer of tenants manages resources that are delegated by the CSPs, which can further delegate (provided the delegated policies allow them to do so) to other tenants by maintaining their own policies.

As the storage space is managed and shared by the CP, the entities (at each level of delegation) can use the same space to store their policies but they can independently manage their local repositories. Even though the repositories are managed in a distributed way, the system can couple all the entities through a common delegation workflow. Thus, as a result of an access request generated at any level by the entities, the delegator's authority can easily be verified by following the common delegation workflow.

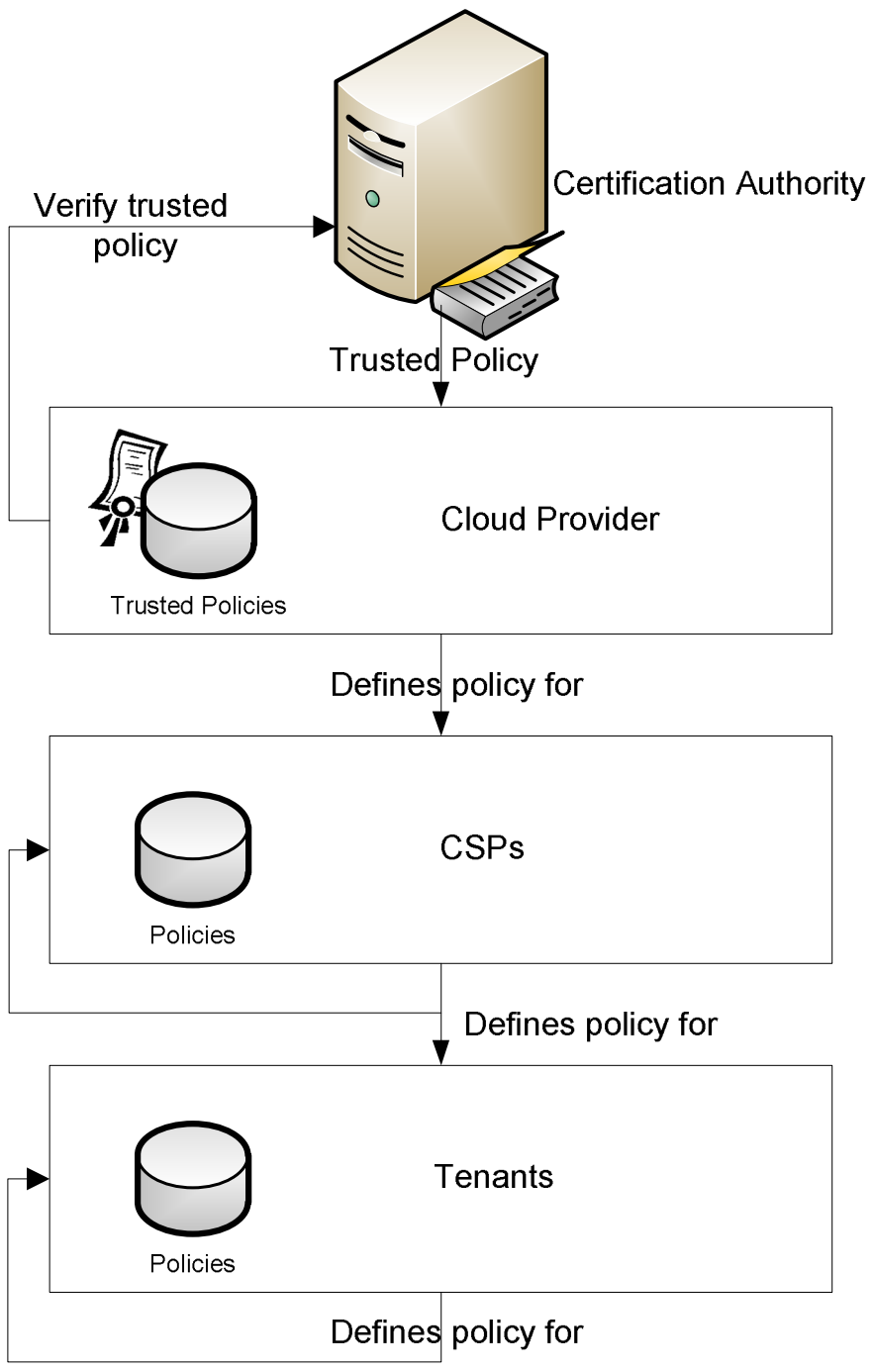

**Figure 5.5** Delegation and management of policies within cloud entities

For this purpose, the workflow generated as a result of the delegation process needs to be trusted to avoid the forgery of the identities of the delegators. To develop this trust, the owner of the resources digitally signs its policies, which are referred as trusted policies. The purpose of the trusted policy is to provide a reliable mechanism to validate the delegators' authority on each *access request* to a resource. To do so, the delegation workflow needs to be originated from the trusted policy that is digitally signed by the owner of the resource.

In the cloud, the CP and the CSPs can be the owners of the cloud resources/services, thus, the policies written by them are the trusted policies (i.e., all the policies of a *policy set*). A CP delegates infrastructure to the CSPs and the CSPs, in turn, offer infrastructure-based services to other CSPs or tenants. In this situation, we consider CSPs as the owners of the services; thus, the trusted policies of the CSPs are considered. The trusted policies of the CP are only considered during the verification of the authority of the *access request* by the CSPs. We assume that the resources and their owners are registered in a trusted third party (i.e., certification authority), which provides verified information on request (e.g., keys, certificates, etc). In the cloud, this certification authority can reside with the CP or it can be any other reliable trusted third party certificate provider (e.g., Verisign). The CP is responsible to acquire certificates for CSPs and it would originate revoke requests in case the delegation needs to be revoked.

In a trusted policy, the identity of the owner (which can be a name or an ID) is digitally signed (using a private key) by the owner and stored as an attribute within the trusted policy; this process is only performed by the owners. Within this scenario it is not possible for any attacker to forge identities to perform unauthorized delegations. Moreover, due to ontological relationships between the users, resources and their policies, it will be difficult for other users to forge their identities as delegatees. As a result of the *access request*, the authority of the delegator that issued the *access policy* to the requester is validated by following the flow of the delegated policies that leads to the owner of the policy (i.e., from the requester's access policy with all intermediate delegator's policies). Then, it is verified by using a public key of the owner of the policy that is provided by the certification authority. As a result, the requester is granted or denied access to the resource if the delegation is genuine..The mechanism we propose is more reliable than the one implemented in the current XACML profile, in which a policy is considered trusted if it does not contain a policy issuer element in it, which can be can be forged easily (i.e., a user can spoof its identity to gain access to the cloud resources).

The following subsections detail the processes of delegation of authority, verification of authority and revocation of authority through several proposed algorithms. Moreover, the policy combining algorithm depicts the method to resolve policy conflicts in case of multiple delegations by different delegators.

### 5.3.1 Delegation of Authority

Algorithm 1 details the process of delegation of authority from a delegator to a delegatee on a given resource. The request for the delegation is received by the system with the instances of the delegator, the delegatee and the resource that is being delegated. To process a delegation request, in line 1, the delegator's authority is verified in order to check its privileges as a delegator (discussed in Section 5.3.2). In case that the delegator has enough delegation rights (line 2), the delegation process starts by establishing a connection with the local repository of policies managed by the delegator (line 3). At this point, a distinction has to be made about the type of the delegator (i.e., owner or delegatee) in order to write the corresponding policy, that is, a trusted policy or a delegated policy. For this purpose, the type of the subject delegator is checked in line 4. If the delegator is the owner of the target resource, the parameters for the trusted policy are prepared. To do so, the attribute of the delegation level is set (in line 5) by the owner to define the maximum number of delegations on the target resource. In line 6, the identity of the owner is digitally signed by using its private key and it is stored in the attributes of the trusted policy. In the case the delegator is not the owner of the resource, the delegation policy is updated by decreasing the delegation level of the resource for new policies (line 7-8). The system will not allow the delegation in case that the delegation level reaches the threshold level (i.e., the maximum level of delegation that was set by the owner) in order to limit number of delegations (line 19). In addition, the *delegation policy* is updated by defining a policy rule according to the access right i.e., intended *action* (e.g., read, write, etc) to be performed (line 11-18).

**Algorithm 1:** *DelegationOfAuthority (subject_delegator, subject_delegatee, object_resource, action)*

```
1:  authority ← VerifyDelegatorAuthority(subject_delegator,object_resource, action);
2:  if authority is valid then
3:      establish connection with local database of subject_delegator;
4:      if subject_delegator is the owner then
5:          set delegation_level; // e.g. default value is '5';
6:          prepare trusted policy by signing an attribute of the owner;
7:      else   prepare delegated policy;
8:              delegation_level = Delegation_level -1;
9:      end if
10:     if  delegation_level is greater than zero then
11:         create new policy_rule for the subject_delegatee;
12:         define new policy by adding policy_rule and delegation_level;
13:         if subject_delegator has policySet for the object_resource then
14:             add policy to the existing policySet;
15:         else    create new policySet;
16:                 add policy to the policySet; end if
17:         store policySet in the local database of the subject delegator;
18:         CreateInstanceRelations (subject_delegator, subject_delegatee, object_resource,
            policySet);
19:     else object_resource reached maximum limit of delegation; end if
20: else   not a valid delegator;  end if
```

After that, the policy is created by adding a *policy rule* for the delegatee and by updating the *delegation level* for the resource (line 12). Before the authorization, the system examines the delegator's repository for a *policy set* defined for the target resource (line 13). If a *policy set* already exists, the new policy is added to the existing *policy set* of the delegator (line 14). Otherwise, a new *policy set* is created for the delegator and the policy is added to it (lines 15-16). Once the delegation policy is created and stored in the repository of delegator (line 17), the delegation workflow is generated by instantiating the corresponding ontological classes (delegator, delegatee, delegated resource and policy set) and the corresponding relationships (line 18).

### 5.3.2 Verification of Authority

The authority verification starts as a result of an *access request* generated by the requester that accesses a resource. As illustrated in Figure 5.1(c), the *access request* contains the identity of the requester, the intended resource and the action that needs to be performed on that resource. Algorithm 2 processes this *access request* by considering these elements in order to verify the delegation authority that granted the access rights (i.e., the intended action) to the requester on the resource.

Due to the fact that policies are managed by the delegators themselves, they can only be retrieved from the database through the delegator's instances. Therefore, (in line 2) the process starts by determining the instances of the delegators (from the ontology-based workflow generated during the delegation process i.e., in Algorithm 1) that have issued policies to the requestor regardless of the requested resource. For example, in Figure 5.4, as a result of an *access request* by the employees working in *dept1*, the instances of *org-1* and *CSP1* will be retrieved. Thus, the interrelations of each delegator's instance are parsed one by one (in line 3) in order to get the appropriate delegator that has issued a policy to the requester on the given resource. To do so, the policy instance that is *written for* the corresponding delegator instance is determined (in line 4), which is also connected to the instances of the target resource and the requester within the workflow (e.g., in Figure 5.4, *P2* connects *org-1* (delegator), *printing service* (resource) and *dept1* (delegatee)). Then, in line 5, the *Policy set* is retrieved from the policy repository through the policy instance that is already determined. The *policy set* repository is managed by the delegator but the relationship of the policy instance with the delegators instance makes it possible to retrieve it from the delegator's repository. From the *policy set*, the policy issued to the requester is searched in line 6. If the policy exists for the requester the rule is matched with the requested action; if not, the same process is repeated for the other delegator instances (lines 7-9). In the next step, the authority of the delegator is verified in case that the requested action (e.g., read request or write request) matches with the delegated access rights stored as a policy rule (line 9-16). Once the authority of the delegator is verified, the decision to grant or deny access to the resource is made (lines 17-19).

**Algorithm 2:** *AccessRequest (subject_requester, object_resource, action)*

```
policyFound = false;
subjectInstances ← get all subject instances connected with the policy instance of the
subject_requester from ontology-based delegation workflow;
while there are subjectInstances to parse and policyFound is false do
    policyInstance ← get policy instance related to object_resource that connects
    subjectInstance and subject_requester through the ontology-based workflow;
    policySet ← get policy set related to object_resource through policyInstance from the
    policy repository;
    policy ← get policy issued to subject_requester;
    if policy is valid then
        rule ← read rule of the policy;
        if rule matches with the action then
            authority = VerifyDelegationAuthority (subjectInstance, object_resource, action);
            if authority is valid then
                policyFound = true;
            end if
        end if
    end if
end while
if policyFound is true and authority is valid then
    grant Access;
else deny access; end if
```

Algorithm-3 shows the process of verifying the delegator's authority. It receives three parameters: (i) *subject* (a delegator that needs to be verified), (ii) *resource* and (iii) *action* (that contains the granted access rights by the delegator). In line 2, the instances of the subject's predecessor that have issued policies related to the target resource are determined (e.g., in Figure 5.4, *org-1* and *CSP1* are the predecessors of *dept1*) and (in line 3) the same process of backward chaining is repeated for the predecessors at each delegation level. This recursive process is repeated to determine the rules of predecessors of each subject until the owner is found (lines 4-7). Besides, the authority of each delegator (at each level of delegation) is verified by consulting their access privileges within the rules of the policies, which are issued to them by their predecessors (lines 8-12). Thus, if the primary subject (i.e., the issuer of the policy to the requester) has a path of delegation interrelations that connects with the owner, the signature of the owner is verified using her public key (lines 13-24). At the end, the decision is made in the form of valid or invalid authority of the primary subject (line 25).

**Algorithm 3:** *VerifyDelegatorAuthority (subject, object_resource, action)*

```
subInstances ← get all subject instances connected to the policy instances of the
object_resource that are connected to the subject from the ontology-based workflow;
while there are subjectInstance to parse do
        policyInstance ← get subject's policy instance connected to the owner's instance from
        the ontology;
        policySet ← get policy set through policyInstance from the database;
        policy ← get policy issued to subjectInstance from the policySet;
        rule ← get rule from the policy;
        if subjectInstance is NOT the owner then
             if rule matches with the action then
                  authority ← VerifyDelegatorAuthority (subjectInstance,object_resource,action);
             else authority = 'invalid';
                  continue with other subjectInstances; end if
        else
           if rule matches with the action then
                  signature ← verify identity of the owner using her public key;
                  if signature is valid then
                     authority = 'valid';
                     end while;
                     return authority;
                  else authority = 'invalid';
                  end if
             else authority = 'invalid';
                  continue with other subjectInstances; end if
        end if
end while
return authority;
```

## 5.3.3 Revocation of Authority

The delegator can revoke the delegated access rights by generating a revoke request. In Algorithm 4, the system handles this request by examining the instances of the delegator, delegatee, and the privilege being revoked on the given resource. To do so, the authority of the delegator is verified in order to ensure that the delegator has issued the policy to the delegatee and that it is capable of performing the operation (lines 1-5). As a result of valid request, the delegated policy is removed from the *policy set* and it is also updated in the repository of policies (lines 6-8 and 10). Moreover, the policy instance is deleted in the case that the *policy set* does not hold any other policy (line 9).

**Algorithm 4:** *RevokeDelegatorAuthority (subject_delegator, subject_delegatee, object_resource, action)*

*authority* ← *VerifyDelegatorAuthority* (*subject_delegator,object_resource,action*);
*policyInstance* ← get *subject_delegator's* policy instance from ontological workflow;
*policySet* ← get policy set through *policyInstance* from database;
*policy* ← get policy issued to *subjectInstance* from the *policySet*;
*rule* ← get rule from the *policy;*
**if** *authority* is valid & *rule* matches with the *action* **then**
    delete policy issued to subject_delegatee from policy set;
    **if** policy set is empty **then**
        delete policyInstance;
    **else** update policy set in database; **end if**
**else** invalid request; **end if**

## 5.3.4 Policy Combining Algorithm

As discussed in section 5.2.4, the policy combining algorithm is invoked if there is a conflict within the policies. In such case, the policies are matched in order to find the precedence in the ontology-based delegation workflow. Algorithm 5 implements this task and receives the instances of the conflicting policies: the subject for whom the conflict occurred (i.e., delegatee) and the conflicting resource. First, the *policy sets* related to the conflicting resource are retrieved from the policy database through the policy instances of the delegator subjects (lines 1-2). From both *policy sets*, the delegation level of the policies is determined and matched in order to decide which policy should be implemented (lines 3-11).

**Algorithm 5:** *PolicyCombiningAlgorithm (policyInstance1, policyInstance2, subject_delegatee, object_resource)*

1: *policySet1* ← get policy set from policy database using *policyInstance1* instance for *object_resource*;
2: *policySet2* ← get policy set from policy database using *policyInstance2* instance for *object_resource*;
3: **if** policySet1 and *policySet2* are valid **then**
4:     *policy1* ← get policy from *policySet1* issued to *subject_delegatee*;
5:     *policy2* ← get policy from *policySet2* issued to *subject_delegatee*;
6:     *DL1* ← get delegation level of *policy1*;
7:     *DL2* ← get delegation level of *policy2*;
8:     **if** *DL1* is greater than *DL2* **then**
9:         **return** *policyInstance1;*
10:     **else return** *policyInstance2;* **end if**
11: **else** invalid policies; **end if**

## 5.4 Performance Evaluation

As discussed in section 5.2, the main practical benefit of our ontology-based delegation workflow is the improvement of the performance of the XACML delegation profile. Contrary to the XACML profile, our approach overcomes the cost of generating the delegation graph on each access request for a resource; instead, a delegation workflow is generated and updated on each delegation of access rights. Moreover, each delegator maintains a separate *policy set* in order to manage access rights for the delegatees on a given resource, which also improves the performance of the policy search during the authority verification process. In order to measure the performance of our system, in this section, we study: (i) the computation complexity of our algorithms; and (ii) the cost of searching policies from the policy sets and generating a graph for performing the verification of authority. We also compare these costs with the cost of the policy management activities of the standard XACML profile.

### 5.4.1 Verification of Authority

In order to process an *access request* in the XACML profile, the following two tasks are performed: (i) a graph of policies is generated; and (ii) the authority of the delegation is verified using that graph. To generate the graph, the policies within the policy set are compared on each request by matching the credentials of the *access request* or the delegators with the elements of the policies. As a result, the corresponding delegated flow is represented as a policy graph. Therefore, in the worst case, $N$ policies are compared on each request to find the requester's policy or the delegator's policy in order to generate the delegation graph. On the other hand, and also in the worst case, it would be required to evaluate $N$ edges (to check whether the policies either permit or deny the delegation) in order to verify the authority. Therefore, the overall cost of this process is $O(2N)$.

Contrary to the XACML profile, our approach does not incur in any graph generation cost at the time of an access request, because the workflow is automatically generated during the access control delegation process by the corresponding authorities. The verification process is divided into two steps, the first step processes the access request and the second one verifies the authority of the delegator. Processing an access request requires finding an appropriate delegator from the already available workflow; this is done by getting all the $n$ nodes connected to the requestor (where $n<N$) and determining the one related to the requested resource. It is important to note that matching nodes in the workflow is also less expensive than searching policies from the repository (as done in the XACML profile). Moreover, $n$ is usually much smaller than $N$, because $N$ is the total number of policies managed by all delegators irrespective of delegatees, and $n$ is the number of delegator nodes that delegated resources only to the intended delegatee. Once the search is completed, the chosen delegator is verified from the delegation workflow, which is achieved by following the workflow that already exists. For this purpose, the delegator and its predecessor delegators are verified from the delegation workflow. The number of these delegator nodes can be considered constant, because they will be lower than the maximum delegation level $l$ set by the owner (i.e., the maximum number of delegations permitted on a given resource). Thus, the computation cost of both of these processes (i.e., searching and verifying) is, in the worst case, $n+l$. As $l$ is a constant and bounded number, the cost of the resulting computation is just $O(n)$, with $n<<N$ in most cases.

In terms of policies, contrary to the XACML profile that requires the evaluation of $N$ policies in the worst case, our system verifies only those policies that are relevant to the delegator and its predecessors. Thus, the number of evaluated policies cannot exceed delegation level $l$ set by the owner of the resource, because a resource cannot be shared more times than specified. Therefore, in the worst case, the number of policies that are compared is constant.

### 5.4.2 Delegation of Authority

The delegation algorithm is divided into three steps: (i) verification of the delegation authority; (ii) instantiation and interlinking of the entities in the delegation workflow; and (iii) storage of the policy in the database. As discussed above, the verification process evaluates $n$ nodes to verify the authority, whereas the instantiation and interlinking of entities evaluate a fixed number of nodes (i.e., the delegatee, the policy set and the resource being delegated) which takes constant time in any case. Similarly, the storage of the policy set in the database is also constant and depends on the processing time of the machine. Therefore, in the worst case, the cost of the delegation algorithm is $O(n)$ for any given delegation request. On the other hand, the XACML profile requires $O(2N)$ for verification (as explained above) and a constant time to store it in the database.

### 5.4.3 Revocation of Authority

The revocation process has two steps: (i) verification of the authority of the issuer of the revocation request; and (ii) retrieval of the related policy and deletion of instances from the delegation workflow. As above, the cost of verification is $O(n)$, whereas accessing the related policy and deleting node takes constant time; as a result, the overall cost remains $O(n)$. On the other hand, the cost of the XACML profile to verify the authority (as above) is $O(2N)$, whereas the retrieval and deletion of policy can be done in constant time. Therefore, the overall cost of the revocation process for the XACML profile is $O(2N)$.

### 5.4.4 Experimental Analysis

In order to evaluate the practical performance of our system, we have implemented our ontology and the proposed algorithms and compared it with the XACML delegation profile within several scenarios and policy loads. The test environment uses Windows 7 professional (64-bit) OS and has the following hardware specifications: Intel core i-5 CPU, 4 GB RAM and 500 GB hard disk drive. We have used Java and the OWL API to create and manipulate the ontology, which is written in the standard Ontology Web Language (OWL).

The main task involved in the verification and revocation processes is to find the target policy. Thus, we considered this task the core of the empirical analysis in both systems (i.e., our ontology-based one and the XACML profile).

Because the XACML delegation profile generates a policy graph modeling the delegation workflow on each access request, it requires repeated accesses to the policy database to create that workflow (i.e., one for each delegation level). As stated above, the cost of searching for a policy is proportional to the size of the database (i.e., the total number of policies *N* stored in it). According to [Turkmen, 2008], in a standard implementation of XACML, the policy decision point, PDP, (i.e., the module that makes access decisions on a given access request) requires 4 seconds in a similar execution environment as ours to search for a target policy from a database containing 10,000 policies; that is, 0.4 milliseconds for a database containing only one policy. Thus, the overall cost of the process associated to an access request is proportional to the cost of the policy search (i.e., *N* x 0.4 ms) multiplied by the total number of delegator nodes *n* to be evaluated. This corresponds to the repeated accesses to the database for each delegated node that is needed to generate the graph. For example, if there are *N*=10 policies stored in the database and the number of delegator nodes related to a request is *n*=5, then the system requires 20 ms (i.e., (10*0.4)*5) to process the request.

On the contrary, our system does not require searching for policies in the database of policies, but just following the delegation workflow that has been already generated by instantiating the ontology. Thus, it does not incur in the cost of searching for policies in a, potentially large, database; that is, its cost is independent of the size of the database (*N*), bust just of the number of delegator nodes (*n*).

Figure 5.6 depicts the actual runtime required to process an access request in both systems for several policy database sizes (*N*), as a function of the number of delegator nodes (*n*). Results show that the runtime of the XACML profile significantly increases as the size of the policy database increases. On the other hand, our ontology-based system is not affected by the size of the database, but just on the number of delegator nodes to be evaluated within the workflow. Thus, the runtime is constant for different database sizes. In practice, and for large databases (i.e., N>1000), the difference in performance between our system and the XACML profile accounts several orders of magnitude, which is illustrated by the $log_{10}$ runtime scale that we used for the Y-axis. Thus, our system provides better scalability in crowded scenarios as the cloud.

**N=10**

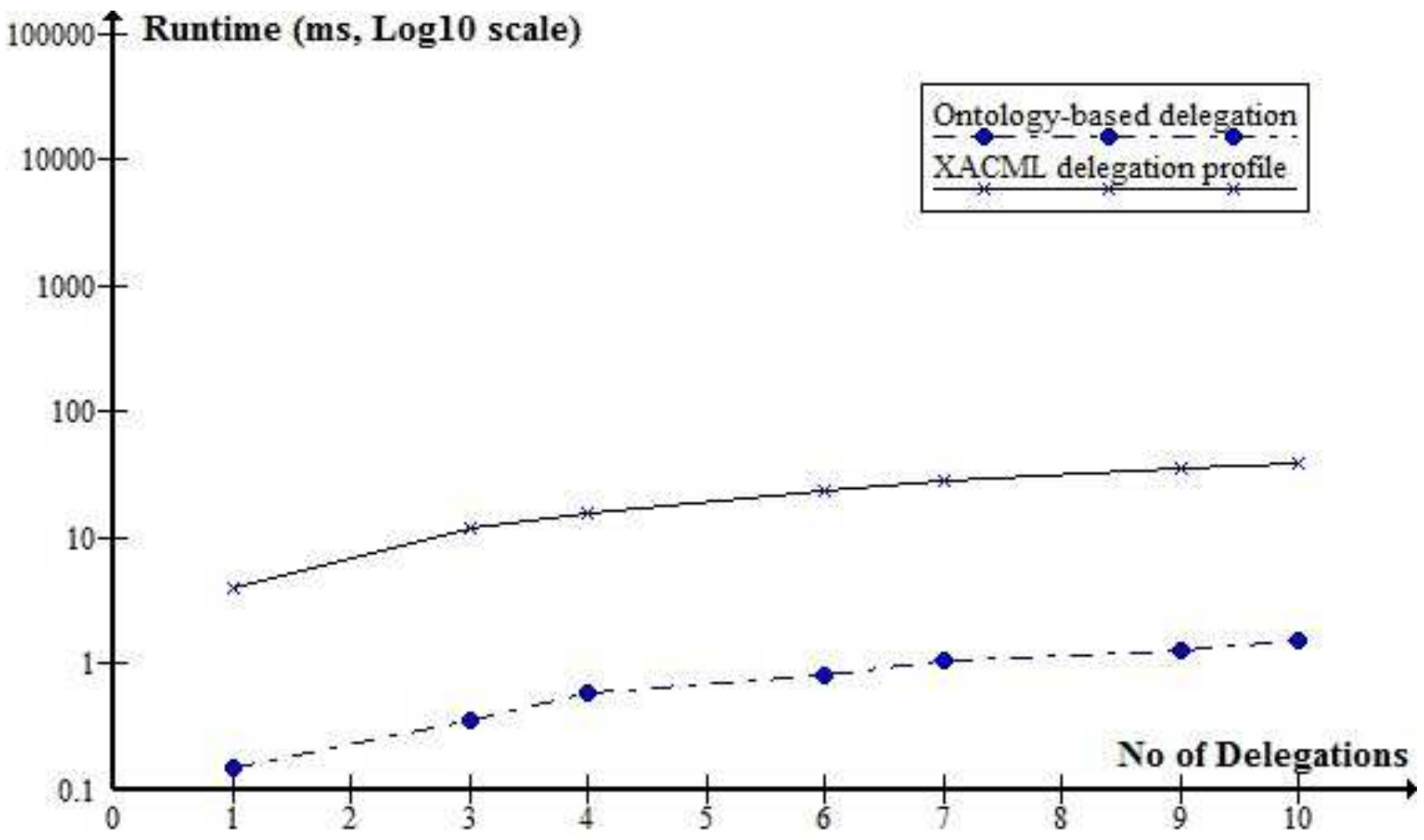

**N=100**

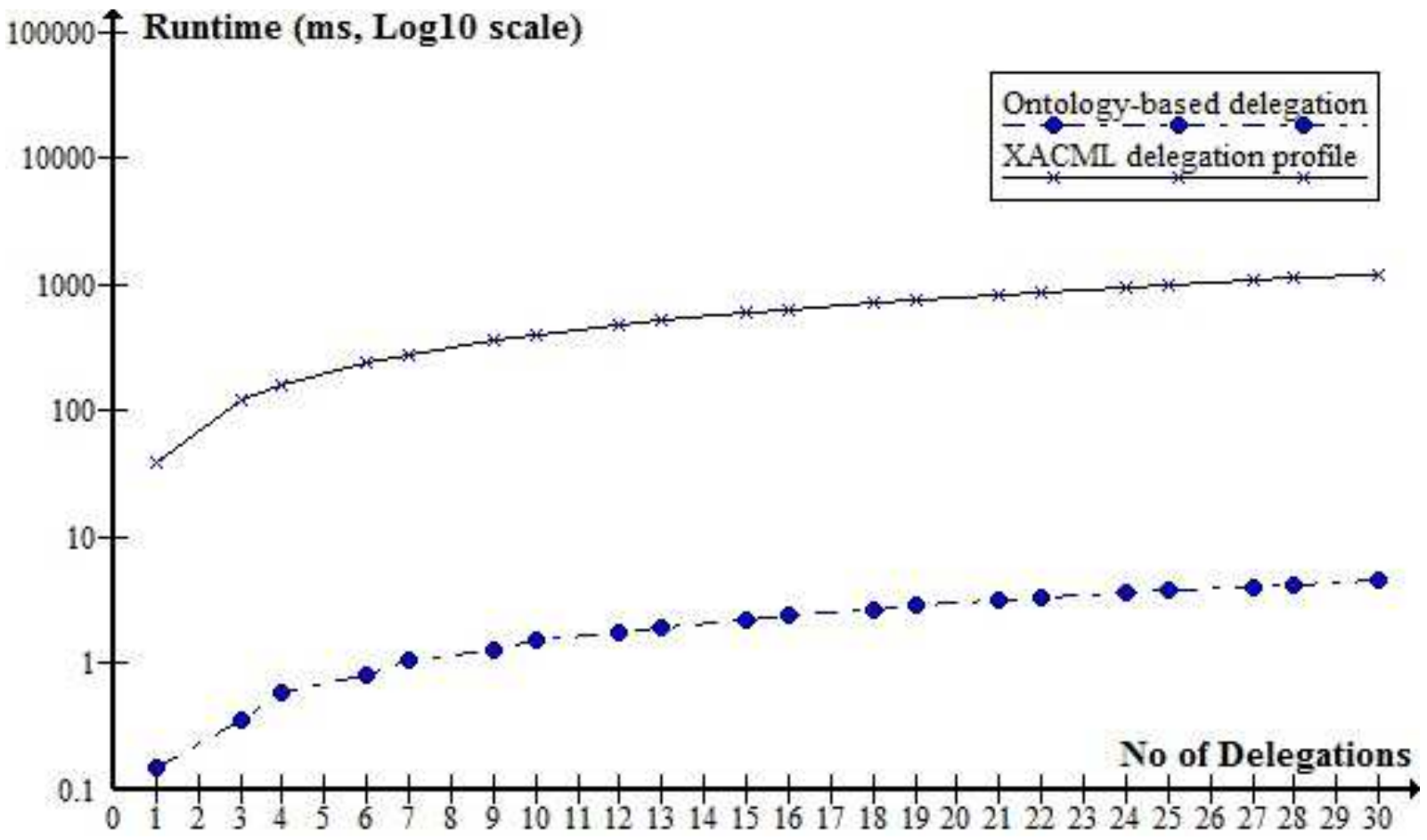

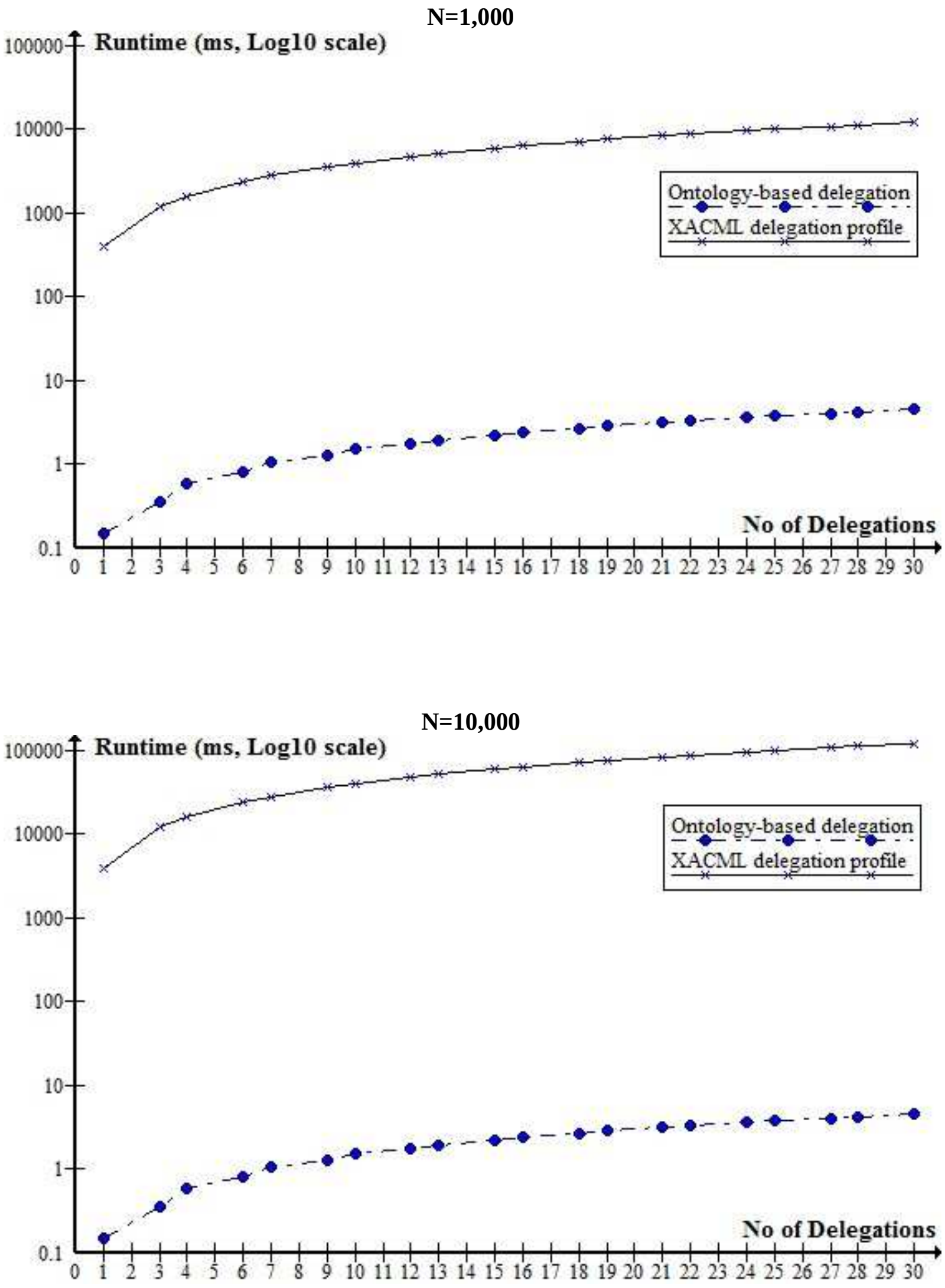

**Figure 5.6** Runtime of the XACML profile and our ontology-based system for several policy database sizes (N)

## 5.5 Conclusions

In this chapter, we proposed a delegation enforcement mechanism that enhances the XACML delegation profile by relying on an ontological modeling of the delegation workflow. As a result, the processes of delegation, verification and revocation are managed in a more scalable way, which is of great importance in crowded environments such as the cloud. Contrary to the XACML profile, our proposal does not require generating a graph on each access request and only those policies that are relevant to the entities of the delegation are evaluated for the verification of authority; this provides a more efficient solution for cloud environments where entities are numerous, heterogeneous and distributed. Moreover, the size of the policies database does not hamper the performance of the proposed system. In addition, it does not require defining special rules (as needed by some related work [Sohr, 2012]) for the enforcement of the delegation policies, since this is automatically achieved by the proposed algorithms, which rely on the coherence of the ontology-based delegation workflow. Thanks to this coherency, the policy conflicts within the policy rules are also automatically handled by an intuitive algorithm. Moreover, we also provide CSPs with the ability to limit the number of delegation levels on a delegated resource; this provides an improved control on the delegation process. Finally, the system is also capable of detecting forged identities by providing a mechanism to digitally sign policies by the owners of the resource.

# Chapter 6 **Privacy-driven Access Control by Means of Ontology-based Semantic Annotation**

## 6.1 Introduction

Social media are open platforms that allow users to interact and exchange information. OSNs have been particularly successfully in the recent years and have attracted many internet users [PewResearchCenter, 2010], who have produced and published vast amounts of (personal) data. As mentioned in chapter 4 (section 4.3), these publically shared data may contain sensitive information, and thus, they require proper access control management in order to guarantee the privacy of the individuals. In the following paragraphs, we discuss the efforts made by the research community to protect the access to sensitive resources made available through social media (and OSNs in particular). Then, we propose an automatic content and privacy-driven solution to manage the access control to these resources that overcome most of the shortcoming of current solutions.

In the last few years, OSNs (such as Facebook) introduced some measures to improve users' privacy by implementing access control features. In order to incorporate such access control, the user profile is broken down into small customizable elements [Aïmeur, 2010]. In order to manage the access to related resources, the information can be classified as "public", "private", "friend" or "friend of friend" [Cheng, 2012a]. According to Aïmeur et al. [Aïmeur, 2010], these features are unreliable or fail to provide desirable results, because they are not fully understood [Eecke, 2010] or it is difficult for the users to manage them correctly [Johnson, 2012]. Furthermore, while configuring privacy settings, users need to perform a tedious job of defining policies for each user, type of resource and to classify those resources.

In order to overcome these shortcomings, the scientific community has proposed some access control solutions (Masoumzadeh et al [Masoumzadeh, 2010b] & Carminati et al. [Carminati, 2009,2011]) that take into consideration the type of resources to be protected (e.g., photos, videos, wall messages, etc.) before allowing/rejecting an access request. These methods rely on ad-hoc structures (i.e., application ontologies) that provide a preliminary modeling of the resources. In order to manage the access control, the users or the OSN administrators need to define access control rules for each resource type. The proposed solutions bear some limitations. On one hand, the classification of resources is coarse grained, fixed and rigid. Similarly, access control policies are applied as a whole on the object or resource, regardless of their actual contents or sensitiveness. As a result, the access to the resource is binary, that is, complete access or complete restriction. For example, if a user declares *WallMessages* as private for a special group of friends, all the published messages will be hidden from that category of friends, regardless the messages contain any sensitive information or not. Furthermore, it is usually difficult for the users to configure the access control policies, since they may not be familiar with such notations and privacy issues.

In order to address the limitations introduced above, in this chapter, we present a new ontology-based scheme [Imran-Daud, 2016b] to enforce access control over content-based resources published in social networks. Ontologies are particularly helpful to analyze and understand the semantics of the published content. Moreover, they are also helpful to identify the sensitive content that may hamper privacy of the publisher. We next summarize the main contributions of our work:

- We propose a transparent, dynamic and privacy-driven access control mechanism. Privacy is ensured by automatically protecting the content of messages to be published according to the privacy requirements of the publishers. The privacy requirements are defined by stating the type of information and the level of detail that is allowed to be accessed by each type of publisher's contact within the OSN. Contrary to access control policies defined over specific resources, such requirements are only defined once in a generic way and can be intuitively stated. Moreover, the user does not need to have a priori privacy notions.

- Contrary to solutions [Masoumzadeh, 2010b, Masoumzadeh, 2010a] [Carminati, 2011], the privacy assessment is performed by semantically analyzing the contents to be published in an automatic way. Moreover, instead of evaluating the privacy for a resource (e.g., a publication) as a whole, our approach examines the privacy risk of each part of the resource individually (i.e., each textual term in a message).

- The semantics that drive the privacy assessment are gathered by means of an automatic semantic annotation process, which relies on available ontological knowledge bases (i.e., DBPedia[5]) and several linguistic tools.
- In contrast to the binary access control policies proposed by other researchers (which just completely allow or deny the access to a resource), our access control enforcement provides each type of reader with a sanitized version of the original publication that is coherent with the privacy requirements specified by the publisher for that type of reader. The different sanitized versions are semantically coherent with regard to the original publication, and are created automatically according to the semantic annotation process and the privacy risk assessment.

The rest of this chapter is organized as follows. In section 6.2 we present our access control mechanism and give a detailed description of its different components and how potential policy conflicts are managed. Section 6.3 illustrates the feasibility of the proposal through a real example. In section 6.4, we evaluate the system, under the perspectives of feasibility, scalability and accuracy of the privacy protection. Finally, in section 6.5 we provide conclusions of this work.

*The content of this chapter has been published in:*

- *Imran-Daud M, Sánchez D, Viejo A, (2016), Privacy-driven access control in social networks by means of automatic semantic annotation, Computer Communications, Volume 76, 15 February 2016, Pages 12-25, ISSN 0140-3664, http://dx.doi.org/10.1016/j.comcom.2016.01.001.*

## 6.2 Our Proposal

As shown in figure 6.1, the actors involved in our system are the *publisher* of a message, the *reader* of that message and the *social network*, which provides the framework. The *publisher* is responsible for specifying her privacy requirements and to publish data in the *social network*. The *reader* is the one who initiates a request for accessing to the published content; as a result, he gets a sanitized version of the publication in coherence with the *publisher's* privacy requirements with regard to him. The *social network* is in charge of controlling the whole process by (i) semantically annotating the messages submitted for publication and (ii) enforcing *publisher*'s privacy by creating semantically-coherent *sanitized versions* of the published content according to the type of *reader*. To tackle these tasks, two components are incorporated, respectively, in the *social*

[5] http://dbpedia.org/About

*network*: the *annotator* and the *monitor*. The following workflow of the system is explained with regard to figure 6.1.

In order to define her privacy requirements, the *publisher* specifies the level of content disclosure allowed for each type of contact in the *social network* in a generic way (e.g., only *family contacts* can know her *sexual orientation*). These requirements are stored as privacy rules (e.g., $Pr_1$, $Pr_2$…….$Pr_n$), which are associated to each *publisher* in the privacy rule database that is managed by the *social network*. The specification of the privacy requirements of each potential *publisher* is a process that is performed once, at the deployment stage of the system; afterwards, the system applies them for all of her future publications. The details of this process are explained in section 6.2.2.1.

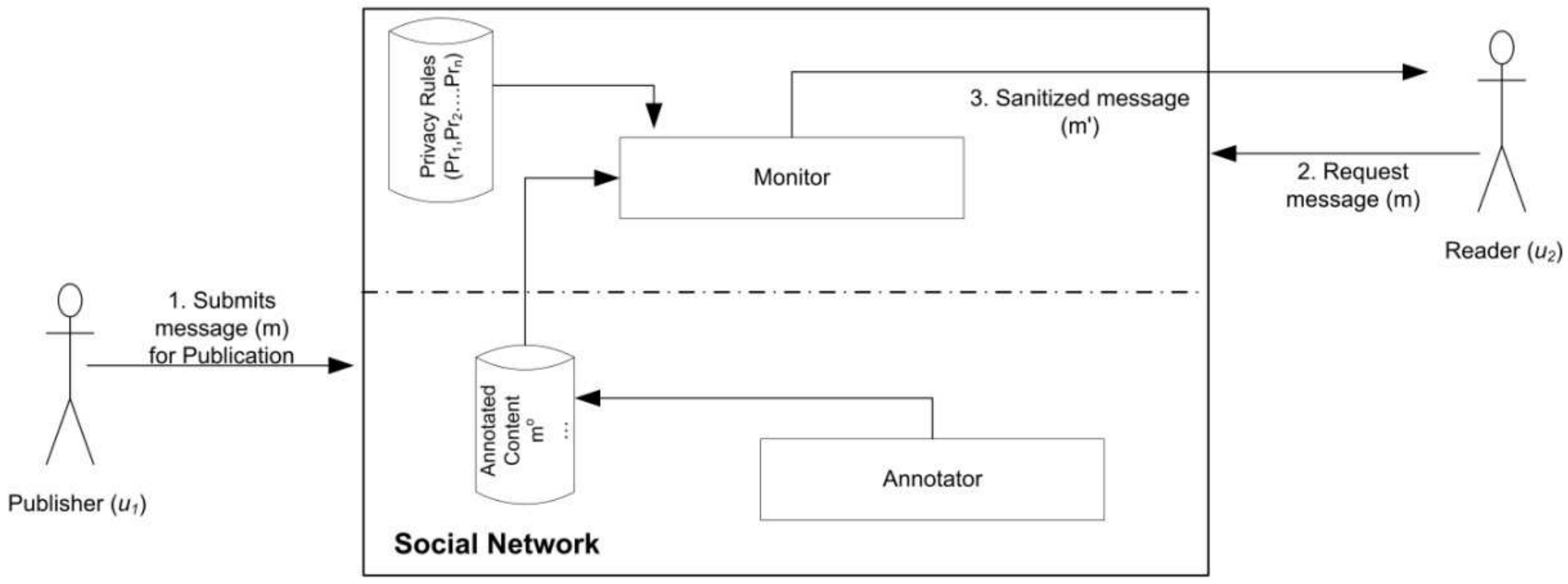

**Figure 6.1** System architecture

Once the privacy requirements are specified, the subsequent workflow of the proposed system is as follows. For a message *m* to be published by the *publisher* $u_1$, the *annotator* module analyses the text of the message by performing syntactic and semantic analyses in order to identify and semantically annotate the content. The resulting annotated message $m^o$ is stored in the annotated content database within the *social network*. The annotation methodology of the *annotator* module is elaborated in section 6.2.1. After that, when a *reader* $u_2$ requests a message *m* of the *publisher* $u_1$, the request is evaluated by the *monitor* module. The *monitor* assesses the content sensitiveness of the annotated version of the requested message ($m^o$), with respect to the privacy requirements defined by the *publisher* $u_1$, which are retrieved from the privacy rules database. As a result, the *monitor* sanitizes the sensitive content of a message according to the level of disclosure allowed by the *publisher* ($u_1$) for the type of *reader's* contact (i.e., $u_2$) with respect to the *publisher*. The resulting sanitized message (m′) is finally forwarded to the *reader* $u_2$. The details of the *monitor* module are explained in section 6.2.2.

### 6.2.1 Submitting a Message for Publication

The *annotator* module is invoked whenever the *publisher* sends a message *m* to be published by the social network. The message contents are processed by the *annotator* module in order to perform the semantic annotation process. The formal semantics associated to the message content during the annotation are used in a later step to evaluate the sensitiveness of a publication, because our sensitiveness assessment is driven by the content of the message.

Within a discourse, nouns are the part of speech that provides the richest semantics and usually carry the sensitive content [Sánchez, 2013a]. Therefore, the *annotator* module identifies the terms that are nouns from a given message, and then derives their semantics by associating them with a formal conceptualization. Since words may have multiple senses, there is a need to resolve the ambiguity by choosing the appropriate word sense that corresponds to the actual semantics of a message. Consequently, the *annotator* module also performs a word sense disambiguation to select the most appropriate sense in order to semantically annotate nouns. The activity diagram in figure 6.2 depicts the workflow of the *annotator* module that is explained in the following paragraphs.

Two different types of nouns that potentially carry sensitive data are distinguished: (i) proper nouns (usually referred as *named entities* (NEs)), which are instances of concepts (e.g., person names, locations, etc.); and (ii) common nouns that refer to concepts (e.g., a sensitive disease, a sexual orientation, etc.). The former are especially relevant from the privacy-preserving perspective, because their specificity and the fact that they usually identify an individual (e.g., a person) may produce a privacy leak. In order to identify them, we rely on a Named Entity Recognition package [Finkel, 2005], which is able to identify named entities and classify them into seven categories: *Time, Location, Organization, Person, Money, Percent* and *Date*. Since this package relies on the lexical regularities of named entities (e.g., they are usually expressed with capital letters), rather than on the syntactical structure of a sentence, no priori analysis is needed to detect them. As a result of this process, the *annotator* module stores the tagged NEs and passes the remaining text of the message for further processing.

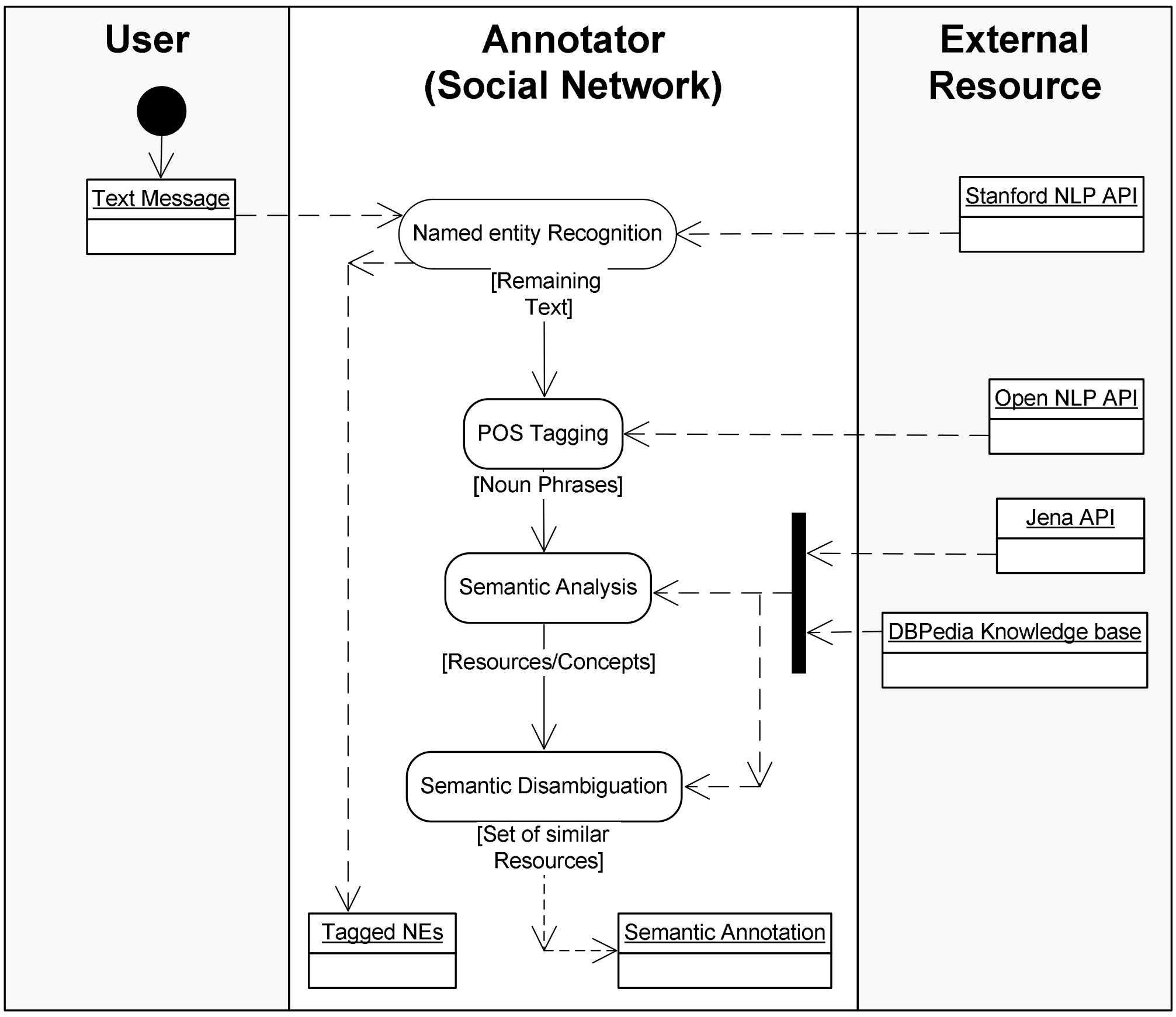

**Figure 6.2** Semantic annotation activity diagram

The second step of the annotation process consists on a *Part-of-Speech (POS) tagging* of the remaining text, which aims at identifying the common nouns potentially referring to sensitive topics (e.g., sensitive diseases). For this purpose, our system relies on a set of natural language processing libraries [OpenNLP, 2010], which perform sentence detection, tokenization of terms, POS tagging and chunking. As a result of this step, noun phrases are identified.

In the next step, a *Semantic Analysis* is performed over the set of noun phrases provided by the *POS tagging* module. This consists in deriving the meanings of noun phrases by associating them with the concepts to which they refer in the message, which will serve as semantic annotation tags. To do so, our system relies on an ontological knowledge base DBPedia [Lehmann, 2014], which provides a taxonomically structured representation of Wikipedia articles (called resources). Within DBPedia, the resources are classified into several ontologies such as the Wikipedia categories, YAGO and WordNet [Lehmann, 2014]. By exploring the classifications associated to a resource, we are able to associate a conceptualization to the noun phrases identified in the message that match with such resources.

In order to explore these classifications, we exploit the ontological properties of DBPedia resources that map them with each other based on their common categories. To do so, we use

SPARQL [W3C-SPARQL, 2013] as a query language and the Semantic Web API [Jena, 2014] for its implementation. Accordingly, the SPARQL query is customized in the following order:

i. First it determines the DBPedia resources that contain a noun phrase identified in the message as a substring in its title (e.g., for a noun phrase like *Apple* we can have DBPedia resources such as *Apple, Apple_inc, Apple_iOS,* etc).

ii. Then, for each resource determined in step one; it derives a list of other resources that are linked with them on the basis of their relational properties (e.g., *Apple* ingredient_of *Fruit, Apple_inc* developer *Mac_OS,* where *ingredient_of* and *developer* are the properties and their ranges are other resources). As a result, an extended list of resources is gathered, which includes the resources that are semantically related to those retrieved in step-i by simple keyword matching. With this step, we aim to extend the number of possible conceptualizations of the noun phrase, which will be useful to derive the semantics of the noun phrase in the message and, thus, to annotate it.

iii. Finally, the system determines the Wikipedia categories of the resources gathered in step-ii (e.g., Wikipedia categories for *Apple* are: *Apples, Malus, Plants with sequenced genomes and Honey plants,* whereas, categories for *Apple_Inc* are: *Apple Inc, Computer companies of the United States, Computer hardware companies, Electronics companies, Steve Jobs and others*).

As a result of the *semantic analysis,* a set of possible conceptualizations (i.e., each one representing a word sense) and their taxonomic categories are retrieved for each noun phrase. If several conceptualizations have been retrieved, the most appropriate one (according to the sense) to which the noun phrase is referring within the message should be determined (e.g., for an ambiguous noun phrase like *Apple* we get conceptualizations like *Fruit* and *Apple_inc*). To perform the *Semantic Disambiguation,* our system uses the other noun phrases in the message as contextual information. Specifically, it calculates the semantic similarity of all senses retrieved for all noun phrases, and selects the combination of senses (one for each noun phrase) that are, in aggregate, the most semantically similar with each other. This strategy relies on the hypothesis that, to be semantically coherent, the senses of the noun phrases appearing in a sentence should refer to a common topic (i.e., they should be semantically similar). This is in fact the premise of most semantic disambiguation approaches proposed in the literature (developed around the senseval initiative [Kilgarriff, 2000]), which exploit the context of a term to identify its appropriate sense.

To compute the semantic similarity between senses, our system relies on the taxonomic structure of DBPedia. Because the semantic similarity between two concepts is understood as their degree of taxonomic resemblance [Batet, 2014] (e.g., *flu* and *pneumonia* are similar because both are respiratory disorders), semantic similarity is usually computed as a function of the commonalities and/or differences between their taxonomic ancestors [Harispe, 2014]. Specifically, in Sánchez et al. [Sánchez, 2012] the authors propose to measure the semantic distance (i.e., the inverse to similarity) between two concepts *a* and *b*, as the ratio between the number of taxonomic ancestors (*T*(*a*) and *T*(*b*)) that they do not share (as an indication of *distance*), divided by the union of both ancestors' sets (to normalize values to the size of the sets of ancestors). The logarithmic function is used as non-linear smoothing of the differences between the compared concepts (which better correlates with human assessment of similarities/distances), and the (1+) factor is added within the expression to avoid Log(0) calculations (i.e., in case of synonyms with identical sets of ancestors) and to ensure that the distance is within the [0..1] range [Sánchez, 2012]:

$$dist(a,b) = \log_2\left(1 + \frac{|T(a) \cup T(b) - T(a) \cap T(b)|}{|T(a) \cup T(b)|}\right) \tag{6.1}$$

By applying this measure (eq. 2.1) to all pairs of senses amongst all the possible combinations of senses retrieved from DBPedia for the noun phrases identified in the message, we can identify/disambiguate the most appropriate combination of senses as the one that, in aggregate, results in the smallest semantic distance (i.e., the highest similarity). As a result, each noun phrase is semantically annotated with the conceptualization associated to the disambiguated sense. The result of the semantic annotation is stored by the *annotation* module in the *annotated content database*. This semantic information will be the base for assessing term sensitiveness and performing the privacy driven access control in the next stages.

### 6.2.2 Accessing a Message

In this section, we propose an access control system for messages published in the OSN, where the policies are seamlessly defined by the owners of the resources, who are in charge of categorizing their OSN friends and specifying their allowed level of access to sensitive information. Note that the use of user categories is analogous to the use of roles in the well-known role-based access control (RBAC); therefore, our proposal can be considered to be inspired by the RBAC model but working in a discretional way [Sandhu, 1996].

Moreover, our system is flexible enough to be integrated in any OSN that implements publications, and it relies on existing OSN procedures in which an access request for a resource is monitored for authorization. To do so, our system requires a *monitor* module to be deployed in the OSN, which is responsible for the authorization of *reader's* access request for any given message. The *monitor* processes each access request based on the following three inputs: (i) the annotated message of the publisher that a reader desires to access; (ii) the reader classification with regards to the publisher; and (iii) the publisher's privacy requirements, which are defined by her to manage the access to her publications.

In order to process an access request, the *monitor* module retrieves the requested annotated message from the *annotated content* database. This message is annotated with three tags (i) the publisher of a message; (ii) the co-publisher of a message (i.e., the users who are tagged by the publisher in the message because the message content may refer to them); and (iii) the semantic annotations automatically defined by the *annotator* module (discussed in section 6.2.1). The publisher and the co-publisher tags are provided by a user who publishes the message in the social network (these tags are currently supported and managed by most OSNs). On the other hand, the semantic elements are calculated and stored in the *annotated content* database by the *annotator* module.

The *monitor* processes an access request by annotating the reader according to the type of contact that he represents for the publisher. This contact type represents the nature of the relationship between the publisher and her contacts (e.g., *close friends*, *family friends*, etc.), which are defined by the publisher by following existing OSN settings and can be assigned while accepting a user's request for friendship. This categorization of friends reduces the administrative efforts of the users that define the privacy requirements for once on the whole group of friends, and the access is automatically managed for all future publications. Based on this reader annotation, the *monitor* assesses and manages an access request by determining the reader's type of contact with the publisher, and by applying an appropriate privacy rule that is defined by the publisher for that type of contact. These rules are defined according to the publisher's privacy requirements, which define the level of disclosure of information for each type of contact. According to such requirements, the system automatically generates related rules and stores them in a local repository of the social network. The definition of rules is a onetime process, that is, at the time of creation of an account, and the system automatically manages all future access requests to the publisher's messages according to these rules. The process required to retrieve the rules according to the privacy settings is detailed in the next section.

#### 6.2.2.1 Defining Access Rules According to Users' Privacy Requirements

An important goal of our system is to ensure the user's privacy while minimizing her administrative efforts to manage access to her publications. To do so, our system facilitates the users to configure their privacy requirements at the time of creation of OSN account and, then, it automatically and seamlessly manages the access to their publications according to these requirements. By means of these requirements, the system defines a list of rules that contains the access levels of disclosure to sensitive data for the publishers' types of contacts in the OSN. The following paragraphs elaborate the process of rules specification and management.

The rules are defined according to three types of elements: (i) the sensitive topics (*ST*), that is, the type of data that are sensitive according to, for example, privacy regulations [EU, 1995]; (ii) the contact categories (*CC*), which are defined by the publisher; and (iii) the access level (*AL*), that is, the level of sensitive information disclosure allowed for a contact type. In our approach, the user can choose the sensitive topics that she wants to protect from others (e.g., religion, race, health, etc.) from a list provided by the system. On the other hand, the user manages her OSN contacts by classifying them into distinct categories (as discussed in section 6.2.2). This categorization of friends is based on the level of trust. After that, the user defines access levels by providing a list of terms that represent the maximum degree of information disclosure for each sensitive topic and each contact category (*CC*). By means of that, each user can control the access to her sensitive data by restricting the level of information detail that will be provided to each type of contact. The following tuple represents an access rule.

$\text{rule}_i \equiv < st_i, cc_i, al_i >$

Where, a $rule_i \in Rules$ has the following elements: (i) a *sensitive_topic* ($st_i$), (ii) a *contact_category* ($cc_i$), and (iii) an *access_level* ($al_i$). Sensitive topics $ST=\{st_1,st_2,..., st_n\}$ could be any topic that is considered sensitive. Contact categories $CC=\{cc_1, cc_2, ..., cc_m\}$ are those implemented by the OSN. Access levels $AL= \{al_1, al_2,..., al_m\}$ are defined by terms representing the maximum level of information disclosure for each element in *CC*. In the following paragraphs we discuss each element in detail.

**Definition 1:** *Sensitive topics (ST)*: A set of topics that are provided by the system and may be sensitive (e.g., according to privacy legislations) because it portrays the information about a user that can be misused if disclosed to others.

For example, according to the *European parliament and the council of the EU* [EU] and the U.S. laws on medical data privacy [DoHNY, 2013], sensitive individual’s data is such that is related to *medical health, religion, race, politics* and *sexuality*. In view of that, the users can define their privacy rules related to these topics that are considered as sensitive topics by our system.

**Definition 2:** *Contact categories* (*CC*): The users classify their contacts into distinct categories based on the level of trust. Examples of *CC* can be *close friends, friends, family*, etc., as defined by the OSN.

This categorization is helpful in order to reduce the administrative efforts of the users, because, by defining a rule for each category of friends, the system can automatically manage the access to new members of this group or to users that can change between categories (e.g., a *friend* becomes a *close friend*). However, the possibility to define rules for specific individuals is also supported by the system.

**Definition 3:** *Access levels (AL)*: The publisher can restrict the access to the contents of her publications by defining the allowed level of disclosure of information that is related to a specific *ST*, and by assigning it to each type of contact in *CC*. Thus, terms in *AL* define the maximum level of information that a reader of a certain *CC* type can access in any message of the publisher.

*Example 1*: A user B*ob* configures his privacy settings related to *medical health* (that is *ST*), defines access levels *AL* by specifying terms *AL*={*Disease, Hepatitis*}, and classifies her OSN friends into the following OSN contact categories *CC*={*close friends, family friends*}. Hence, the access level *AL* assigned to *close friends* is *disease,* whereas, the level of access for *family friends* is *hepatitis*. As a result, any publication of *Bob* that contains information about *hepatitis* will not be completely disclosed to *close friends* (in fact, they will get a sanitized version of the message, as it will be explained in section 6.2.2.2). On the other hand, *family friends* will get the information about *hepatitis* but not more specific details (e.g., types of *hepatitis B, hepatitis C*). The following rules are generated as a result of these settings.

*rule*$_1$=<*medical health, close friends, 'diseases'*>
*rule*$_2$=<*medical health, family friends, 'hepatitis'*>

In the following examples, we illustrate the process of rules definition and their enforcement by the system.

*rule*$_3$ ≡ < *religion*, *friends*, *'religion'*>
*rule*$_4$ ≡ < *religion*, *family friends*, *'Muslim'*>

*Rule*$_3$ restricts *friends* contacts to get any details of the publisher's *religion* (e.g., the publishers belief, sect, others), whereas, *rule*$_4$ permits the *family friends* to know that the publisher is *Muslim*, but anything more specific will be sanitized.

*rule*$_5$ ≡ < *sexuality*, *friends*, *null*>
*rule*$_6$ ≡ < *sexuality*, *family friends*, *'homosexual'*>

In *rule*$_5$, the level of disclosure for *friends* is *null,* that is, *friends* contacts cannot get any information related to the sexual life of the publisher. As a result, any information related to the publisher's sexual life will be sanitized from any publication accessed by *friends*. In contrast, the *rule*$_6$ permits *family friends* to know that the publisher is *homosexual* but nothing more detailed.

The previous rules are defined at a conceptual level, as a function of the semantic annotation performed by the *annotator* module. Since the annotation process is also able to detect and classify Named Entities (NEs), rules can be also defined in order to protect specific types of NEs (e.g., *persons*, *locations*, *organizations*, etc.) that, due to their specificity, may reveal sensitive information.

$rule_7$ ≡ < *NE_person*, *strangers*, *null*>
$rule_8$ ≡ < *NE_person*, *family friends*, *person_name*>
$rule_9$ ≡ < *NE_location*, *family friends*, *location_name*>
$rule_{10}$ ≡ < *NE_organization*, *family friends*, *organization_name*>

$Rule_7$ restricts *strangers* contacts to get any information that refers to *person names*. However, $rule_8$, $rule_9$ and $rule_{10}$ permits only *family friends* contacts to access the specific name of a *person*, a *location* or an *organization* mentioned in the publications.

Notice that, according to the nature of the sensitive topics considered in the requirements, the protection will focus on confidential data (e.g., sensitive diseases, sexuality, etc.), thus protecting against *attribute disclosure*, or on identifying data (e.g., person names, locations etc.), thus protecting against *identity disclosure*.

#### 6.2.2.2 Enforcing Flexible Access Control

As mentioned in section 6.2.2.1, publishers manage their privacy requirements by assigning a level of information disclosure to each contact type. To enforce the appropriate access to sensitive data according to such requirements, the publisher's messages are assessed for sensitiveness according to the type of reader that is accessing it. This sensitiveness is determined according to the following elements: (i) the contact type of the reader; (ii) their allowed level of disclosure, as defined in the rules; (iii) the taxonomy associated to the access level (*AL*) term and (iv) the semantic annotations of the message. The contact type of the reader is evaluated in coherence with the privacy requirements of the publisher, in which she categorized her friends in to distinct contact categories (*CC*). According to the contact type of the reader, the rule assigned to them is retrieved from the *privacy rule* repository (which is managed by the social network). As a result, the access level (*AL*) assigned to this contact type is determined from the rule assigned to him. The taxonomy associated to access level terms is retrieved from DBPedia. Finally, the annotated message of the publisher is retrieved from the *annotated content* database.

The system measures the sensitiveness of each term in the message by comparing their semantic annotations with the assigned level of access allowed for the reader. To do so, the taxonomic branch of assigned access level term is retrieved from DBPedia, in which the top level node is the access level (*AL*) term. As a result, any content that lies under the access level node is considered as sensitive for the reader and it is sanitized (i.e., replaced) with the term that is defined in *AL*, which defines the maximum level of disclosure for that type of reader.

*Example 2*: By considering $rule_1$ and $rule_2$ defined in *Example 1* (section 6.2.2.1), *Bob* publishes a certain text that contains the term *hepatitis*. Then, a contact named *Alice*, who is classified as a *close friend* of *Bob* tries to access that message. The *monitor* intercepts the request and it checks the rule assigned to the contact type *close friend* (i.e., $rule_1$) in order to determine the level of access for Alice (which is *diseases*). Then, it retrieves the taxonomy branch related to *diseases* (shown in figure 6.3) from DBPedia and checks that the term *hepatitis* lies under the *AL* (i.e., *diseases*). Finally, the monitor sanitizes (i.e., replaces) the term *hepatitis* by the *AL* (i.e., *disease*) allowed to *Alice* and provides this sanitized message to her.

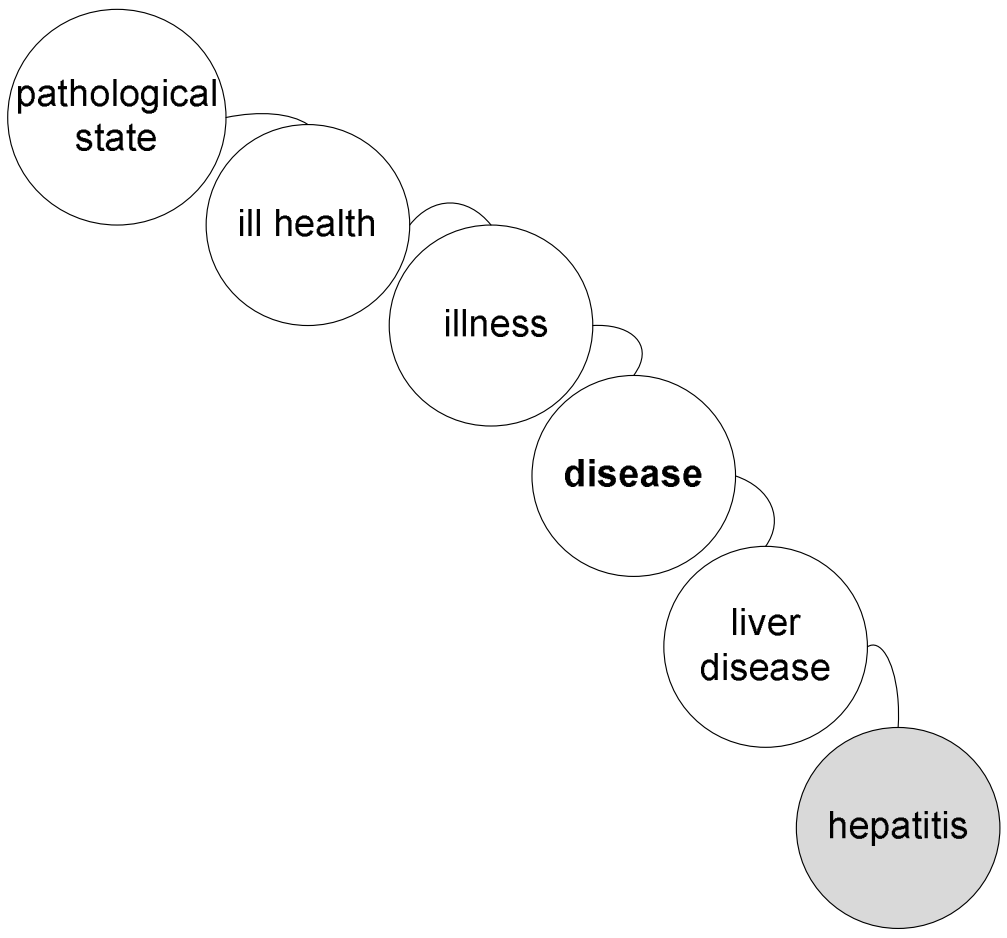

**Figure 6.3** Taxonomic generalizations of Hepatitis

#### 6.2.2.3 Policy Conflicts

In addition to the sanitization of messages, the *monitor* also handles potential *policy conflicts* within the users. A conflict may appear when the publisher posts a message on another user's timeline or tags another user in her publications, because the message content may refer to the latter. To handle the privacy requirements that may apply in such cases, in our system, the users (other than the publisher) who are associated with the content of a publication become the co-publishers of the publication and their privacy rules are also taken into consideration. However, the fact that several rules are associated to a message may cause a *policy conflict* between the publisher and the co-publisher(s), because the publisher and co-publisher(s) may define different access levels for their types of contacts (e.g., *close friends*, *friends*, *strangers*, etc.). As a result, there can be conflicting policies for their contact types. In this situation, in order to fulfill all the privacy requirements, the strictest rule amongst those of the publisher and the co-publisher(s) is the one that will be applied for their respective contact types. In practice, this means that the access level of disclosure that is higher in the taxonomic tree (i.e., the more generic in terms of semantics and, thus, the one that imposes the strictest restriction with regard to term sanitization) is the one that will be considered to sanitize the message contents.

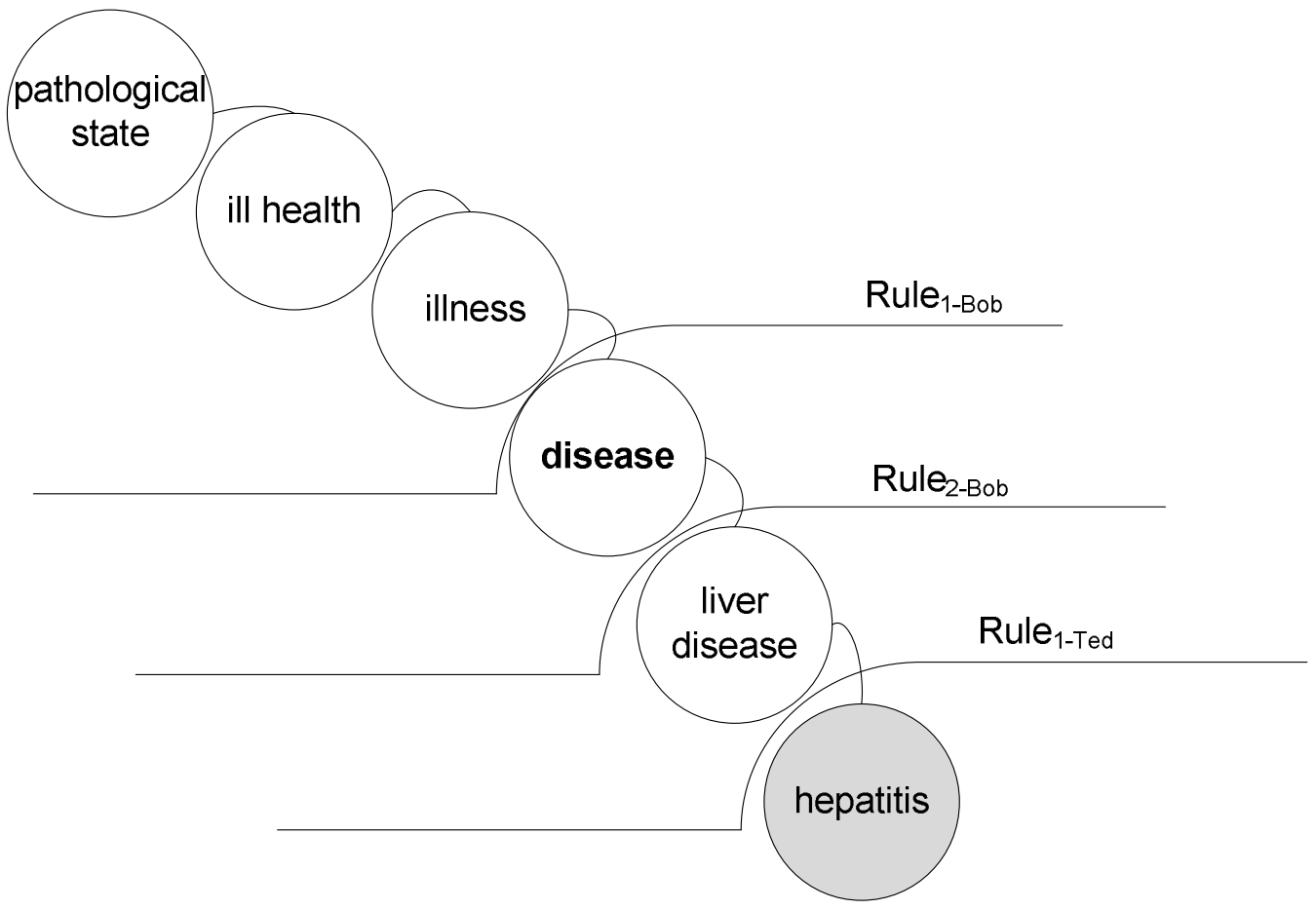

**Figure 6.4** Access levels defined by Bob and Ted

*Example 3*: Extending *Example 1* and *Example 2* (section 6.2.2.1), *Bob* tags her friend *Ted* in a publication. As a result, *Ted* becomes co-publisher of the message. *Alice* can be a common friend or a stranger for *Bob* or *Ted*. The access levels defined by *Bob* and *Ted* are depicted in figure 6.4.

Accordingly, the following rules are generated as a result of privacy requirements of both users.

$\text{Rule}_{1\text{-Bob}}$ ≡ < *medical health, strangers, 'illness'*>
$\text{Rule}_{2\text{-Bob}}$ ≡ < *medical health, close friends, 'disease'*>
$\text{Rule}_{1\text{-Ted}}$ ≡ < *medical health, close friends, 'liver disease'*>

In the following cases, we discuss the automatic resolution strategy for the policy conflicts:

**Case 1:** *Alice* is a *close friend* of *Bob* and *Ted*

The access request by *Alice* to *Bob*'s publication, with *Ted* as a co-publisher, produces a *policy conflict*; that is, there are two different levels of disclosure for *Alice* by the publisher and the co-publisher of the message (i.e., $\text{Rule}_{2\text{-Bob}}$ and $\text{Rule}_{1\text{-Ted}}$). In order to resolve the conflict so that the privacy requirements of both *Bob* and *Ted* are fulfilled, the *monitor* retrieves the taxonomic branches (show in figure 6.4) of both *ALs* defined by the publisher and co-publisher (i.e., *disease* and *liver disease*, respectively) and selects the level that is more general, that is, higher in the taxonomic tree. As a result, the level of disclosure *disease* is chosen by the *monitor* in order to sanitize the message content for *Alice*.

**Case 2:** *Alice* is a stranger to *Bob* and a *close friend* of *Ted*

In this case, the level of access for *Alice* defined by *Bob* is *illness*, whereas by *Ted* it is *liver disease*. By applying the same resolution strategy as above, the sanitization of the content of the message will be performed by using *illness* as access level.

# 6.3 Feasibility study

In this section, we illustrate the practical feasibility of our system by analyzing the scalability of its different modules, and show its behavior with an example framed within social network specialized in healthcare.

As already discussed in section 6.2.2.1, it is also important to recall, that the specification of the privacy requirements (that would be done just once, during the initialization of their user accounts), is the only interaction that the system requires from the user. The rules generated as a result of these requirements and the automatic assessment of sensitive information that is driven by the semantic annotation will provide the means to enforce a transparent and automatic access control over all the subsequent publications. These privacy requirements can be based on the sensitive topics (*ST*) defined in current legislations on data privacy, such as the EU Data Protection Act [EU, 1995] (i.e., *medical health, religion, race, politics and sexuality*), U.S. laws on medical data privacy [DoHNY, 2013] (which define lists of sensitive diseases such as *HIV*, *hepatitis*, *sexually transmitted diseases*, etc.) or the Health Insurance Portability and Accountability Act (HIPAA) [HIPAA, 1996] (which specifies the protection identifying census features such as *names*, *locations*, etc.). Furthermore, the topics can be chosen according to the thematic scope of the OSN. Finally, the access levels (*AL*) to be defined for each topic would match the number of contact categories (*CC*) in the OSN that is 3, in average, according to [Carminati, 2009,2011]. Thus, the definition of privacy requirements requires minimal manual efforts by the users.

To illustrate this, let us compare this with the configuration burden of a standard approach in which the users would need to specify the access permission for each publication and contact type. To enforce the same level of access control that our system provides in a standard OSN, the user would need to: (i) assess the sensitiveness of the contents to be published for *every* new message; (ii) for each message with sensitive contents, create as many sanitized versions of the message as contact types with different privacy requirements; and (iii) define the appropriate access control rules, so that only allowed contact types can access to the corresponding message. According to the Statistics Brain Research Institute [BRI, 2015] and kissmetrics [Kissmetrics, 2015] on average, a Facebook user publishes 90 pieces of content per month; from these, around a 58% of the publications require privacy-conscious settings [Liu, 2011]. Thus, users should manually evaluate those 90 pieces in order to identify which of them may cause privacy risks and protect the 52 (90*0.58) pieces that, in average, are sensitive. If we consider an average of 3 contact types, then the user would need to create 3*52 message versions and define 3*52 access rules, in the worst case. In comparison, in our approach the user only needs to specify as many access levels (*AL*) as contact types (*CC*) per sensitive topic (*ST*). If we consider a generic implementation with 6 sensitive topics (those defined in the EU Data Protection Act plus census-related features), we have that the user only needs to define 6*3 *AL* just once, during the initial configuration step.

Let us now illustrate the whole process within the context of the sample social network. Because of the medical scope of this social network, the privacy protection can be restricted to *health*. Therefore, privacy rules can be configured so that they are related to the *health* topic. Let us also assume that contacts are categorized into the following three groups: *Clinicians/Researchers*, *Followers* and *Registered users*. Thus, for privacy rules, the access levels (*AL*) for the contact categories (*CC*) are related to the different levels of disclosure that may be allowed for the medical condition of the user of the SN (which is the sensitive topic (*ST*). The following sets show the customized access levels (*AL*) and the contact categories (*CC*) for this social network:

*AL* = {(HIV/AIDS/Hepatitis/STDs), Infections, ill health, Condition/State}
*CC* = {(Clinicians/Researcher), Followers, Registered users}

The first elements of *AL* (i.e., HIV/AIDS/Hepatitis/STDs) correspond to the diseases that are considered sensitive by the U.S. federal laws on data privacy [DoHNY, 2013], whereas, the rest of the elements are semantically coherent generalizations of the former according to the taxonomic structure of DBPedia.

Once the *AL* and *CC* are defined, the user can configure her privacy requirements by assigning a specific *AL* to each element in *CC*. To do so, the system provides an intuitive interface to define her privacy requirements in the form of questions related to the sensitive topics with a list of contact categories and a predefined set of taxonomically coherent access levels. Figure 6.5 shows an example of the list of *AL* that can be assigned to each element of *CC*.

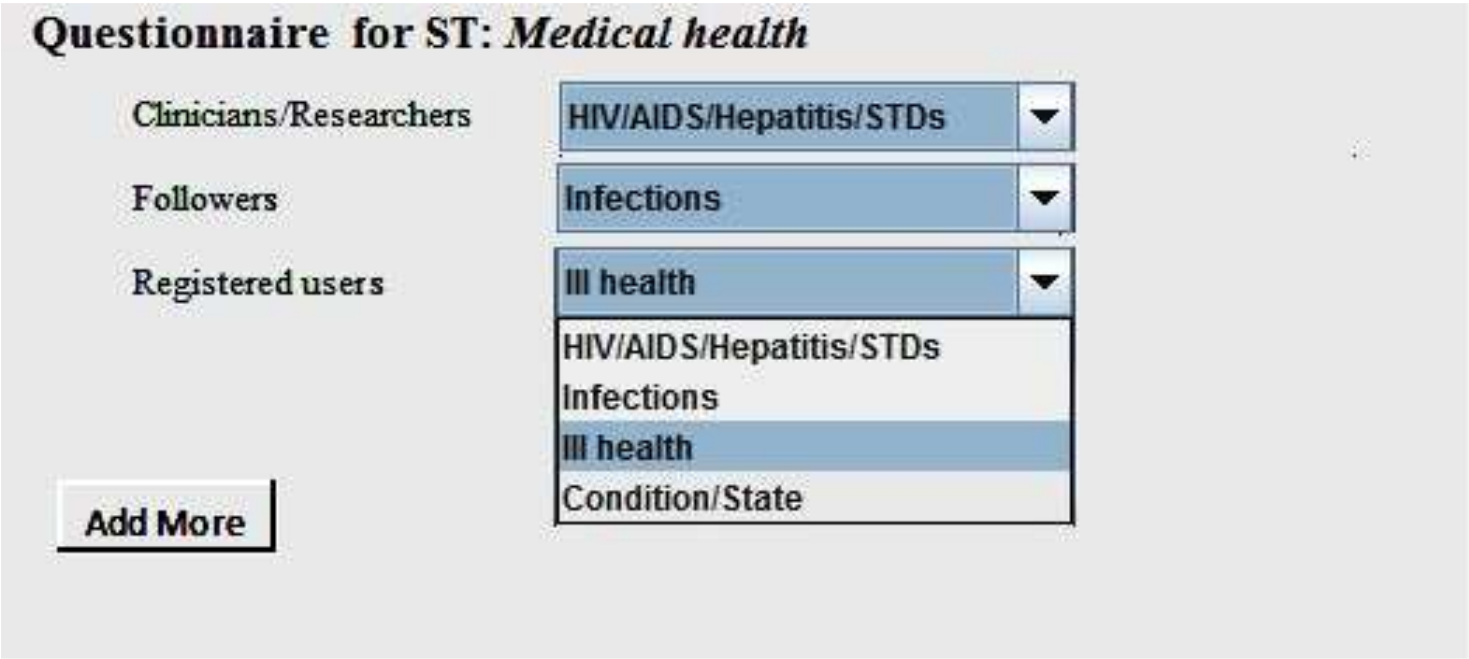

**Figure 6.5** Privacy requirements of a user in sample social network for the *medical health* topic

In this example, the access level for the *Clinicians/Researchers* group is *HIV/AIDS/Hepatitis/STDs*, whereas, the maximum allowed disclosure for *Followers* regarding these diseases is *infections* (but not specific diseases) and the *Registered users* are only allowed to know the general notion of *ill health*, but nothing more specific other than *ill health*. The formalization of these privacy requirements as rule tuples ($rule_i \equiv < st_i, cc_i, al_i >$) are as follows:

rule$_1$ = < *Medical health*, *Clinicians/Researchers*, *HIV* >
rule$_2$ = < *Medical health*, *Clinicians/Researchers*, *AIDS* >
rule$_3$ = < *Medical health*, *Clinicians/Researchers*, *Hepatitis* >
rule$_4$ = < *Medical health*, *Clinicians/Researchers*, *STDs* >
rule$_5$ = < *Medical health*, *Followers*, *Infections* >
rule$_6$ = < *Medical health*, *Registered users*, *ill health* >

In order to show how these privacy rules are applied in practice, we use the sample message shown in figure 6.6, in which a patient shares her personal feelings about her disease (HIV). As per privacy requirements, this message is related to the *medical health* topic (i.e., *ST*) and is sensitive because it contains information about the disease of the publisher.

**Dealing with Hiv and then being told that you suffer from AIDS is almost the hardest thing to face with in life. The hardest thing is dealing with the virus because there are people that just do not understand and think that you are a leper.**

**Figure 6.6** Sample message to be published

In order to process the publisher's messages, the *annotator* module performs the semantic annotation of the messages' content, as detailed in section 6.2.1. First, the message is syntactically analyzed to detect POS (see the output in figure 2.7 that corresponds to the sample message).

**Dealing with Hiv[NNP] and then being told that you suffer from AIDS[NNP] is almost the hardest thing[NN] to face with in life[NN]. The hardest thing[NN] is dealing with the virus[NN] because there are people[NNS] that just do not understand and think that you are a leper[NN].**

**Figure 6.7** POS tagging of the sample message

Then, the semantic annotation is performed. Given the number of messages that an OSN is expected to receive on daily basis, the scalability of the annotation process is crucial. On the one hand, we can consider that the average length of publications in current social networks is relatively short and, in some cases, restricted by the number of characters (e.g., Twitter allows only 140 characters). On the other hand, our semantic analysis focuses on the noun phrases of the publication (marked as NN (singular noun), NNP (proper noun) and NNS (plural noun) in figure 6.7) and hence, it scales according to the cost of analyzing them. In order to perform the annotation, our system derives the semantics of noun phrases by finding their potential conceptualizations in DBPedia. For this purpose, SPARQL queries are performed to retrieve the possible senses and their corresponding taxonomic structures. Given that, the DBPedia queries are the most costly part of the annotation process, the cost of analyzing a message is proportional to the time ($T\acute{s}$) required to get the number of senses ($n_{senses}$) for each noun phrase within a given publication, which is formalized as follows.

$$Cost \propto \left\{ \sum_{i=1}^{n_{NP}} T\acute{s}\left(n_{senses}\right) \right\} \qquad (6.2)$$

A single query is executed for each noun phrase to retrieve any number of senses from the knowledge base. Table 6.1 summarizes the actual runtime required to execute SPARQL queries with respect to the number of senses of different noun phrases of the message (as shown in figure 6.6).

**Table 6.1** Time required by a query to get senses of noun phrases

| Noun Phrase | No of senses | Time (ms) |
|---|---|---|
| HIV | 5 | 0.24 |
| AIDS | 12 | 0.21 |
| Thing | 54 | 0.24 |
| Life | 86 | 0.33 |
| Virus | 100 | 0.19 |
| People | 12 | 0.21 |
| Leper | 86 | 0.33 |

We can see that the runtime of a SPARQL query is independent of the number of senses of each noun phrase, and that the average cost per noun phrase is 0.25ms. In consequence, as we can see in figure 6.8 that the runtime of the annotation is linear with regard to the number of noun phrases, that is, it scales linearly according to the number and length of the messages to be analyzed.

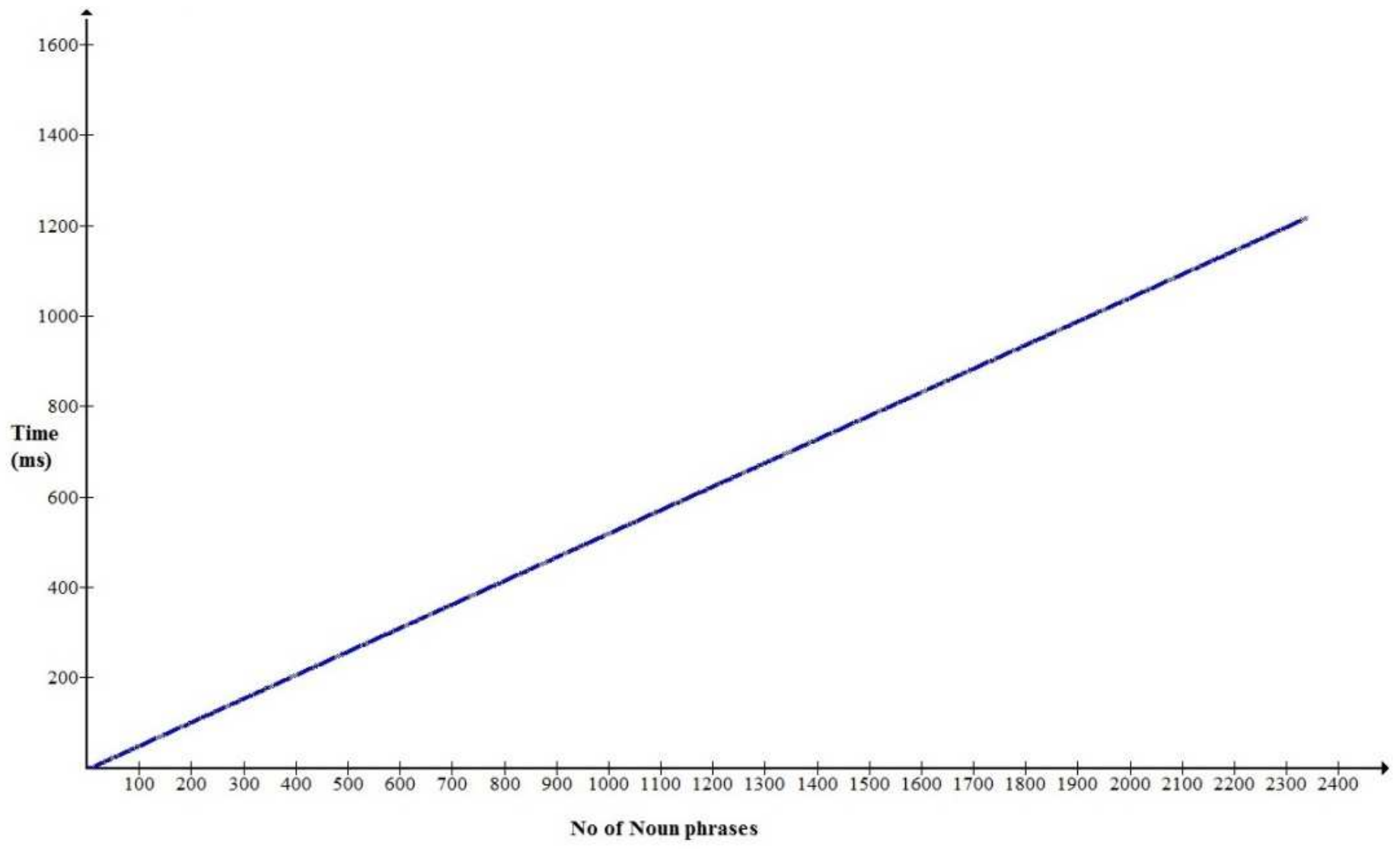

**Figure 6.8** Runtime of the annotation process w.r.t. the number of noun phrases to analyze

As a result of SPARQL queries, a number of potential senses are retrieved for each noun phrase. Table 6.2 shows a list of senses of the noun phrases of the sample message.

**Table 6.2** Senses of the noun phrases in the sample message

| HIV | LIFE | VIRUS | AIDS | PEOPLE | LEPER |
|---|---|---|---|---|---|
| Sexually_transmitted_diseases (STDs) | Biological_science | Viruses | HIV/AIDS | Humans | Tropical_diseases |
| HIV | Systems | Virology | Pandemics | People_(magazine) | Leprosy |
| Lentiviruses | Biology | Pediatrics | Health_disasters | | Bacterial_diseases |
| | Life | Organism | Syndromes | | Neglected_diseases |
| | | Others | | | |

As detailed in section 6.2.1, senses need to be semantically disambiguated in order to get the appropriate set of senses to get actual semantics of the message. To do so, our system calculates the semantic distance of each sense with respect to senses of the other noun phrases according to their taxonomic structure in DBpedia (that is retrieved as a result of the previous SPARQL queries). For example, let us consider *HIV* and *LIFE* nouns appearing in the sample message. The semantic distances between their senses are shown in table 6.3 that are calculated according to the taxonomies retrieved from DBPedia (which are shown in figure 6.9).

**Table 6.3** Semantic distances between the senses of *HIV* and *LIFE*

| | | LIFE | | | |
|---|---|---|---|---|---|
| | | **Life** | **System** | **Biology** | **Biological_science** |
| **HIV** | **STDs** | 0.888 | 0.937 | 0.94 | 0.93 |
| | **HIV** | 0.857 | 0.928 | 0.93 | 0.92 |

By performing the same process for all the senses of all the noun phrases, we obtain the set of most suitable senses, which are shown in table 6.4. Notice that the disambiguation process does not require any additional SPARQL query, but just the pairwise evaluation of already retrieved taxonomies.

**Table 6.4** Set of most suitable senses/annotation for the sample message

| HIV | LIFE | VIRUS | AIDS | PEOPLE | LEPER |
|---|---|---|---|---|---|
| *HIV* | *life* | *Viruses* | *AIDS* | *Humans* | *Leprosy* |

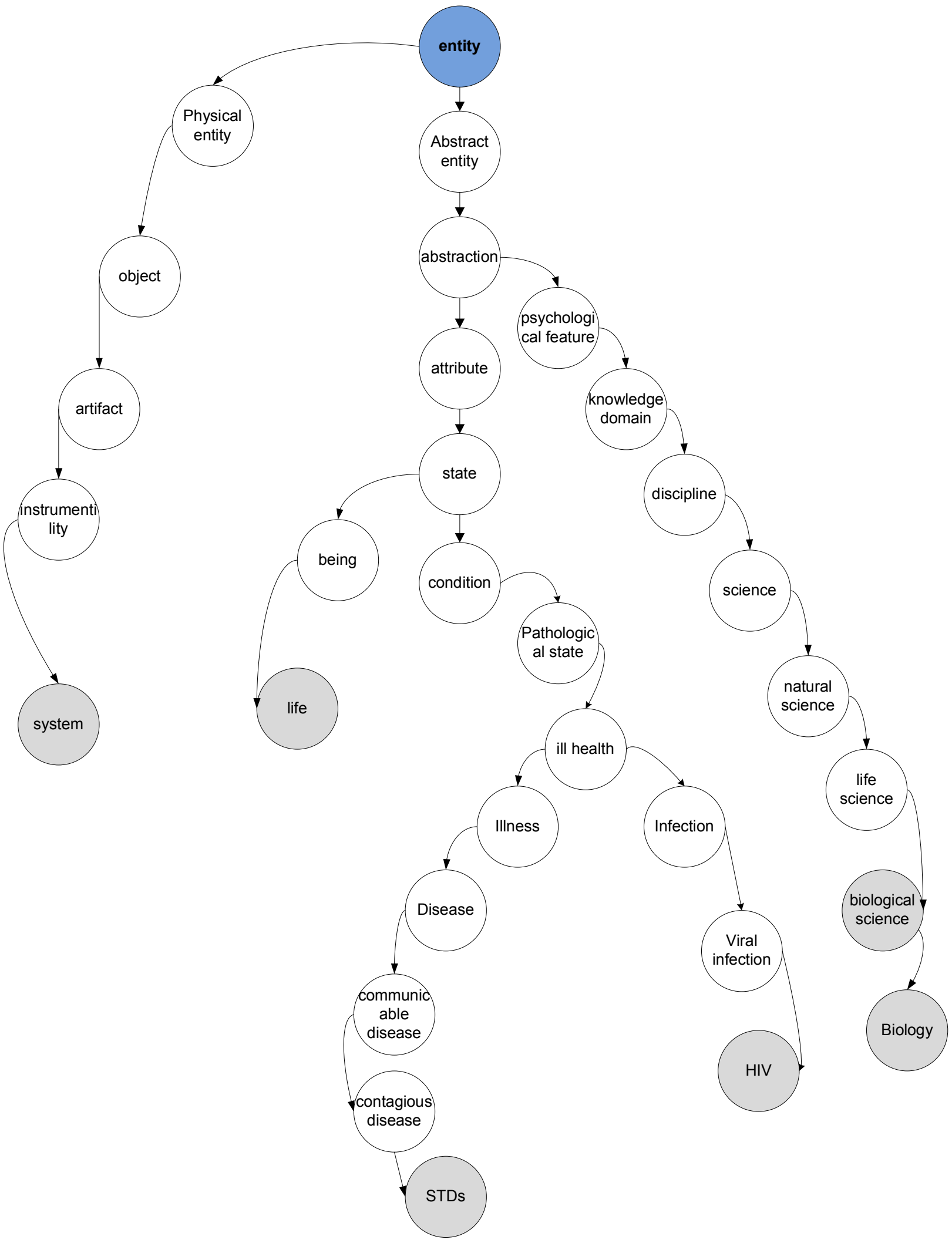

**Figure 6.9** Taxonomic tree of senses for the sample message

Once the message is annotated and stored, the *monitor* processes access requests of the readers and assesses the sensitiveness of the messages according to the privacy requirements of the publisher (shown in figure 6.5). Eventually, any sense (shown in table 6.4) that lies within the branch of access level (that is, a level of disclosure for the contact type of the reader) is considered sensitive, and the corresponding noun phrase is sanitized accordingly. In this case, the computational cost for the sensitiveness assessment of a message for a specific contact type is proportional to the product of number of senses/annotations ($n_{senses}$) and the number of nodes ($r$) in a taxonomic branch of the access level corresponding to the contact type:

$$Cost \propto n_{senses} \times r \quad (6.3)$$

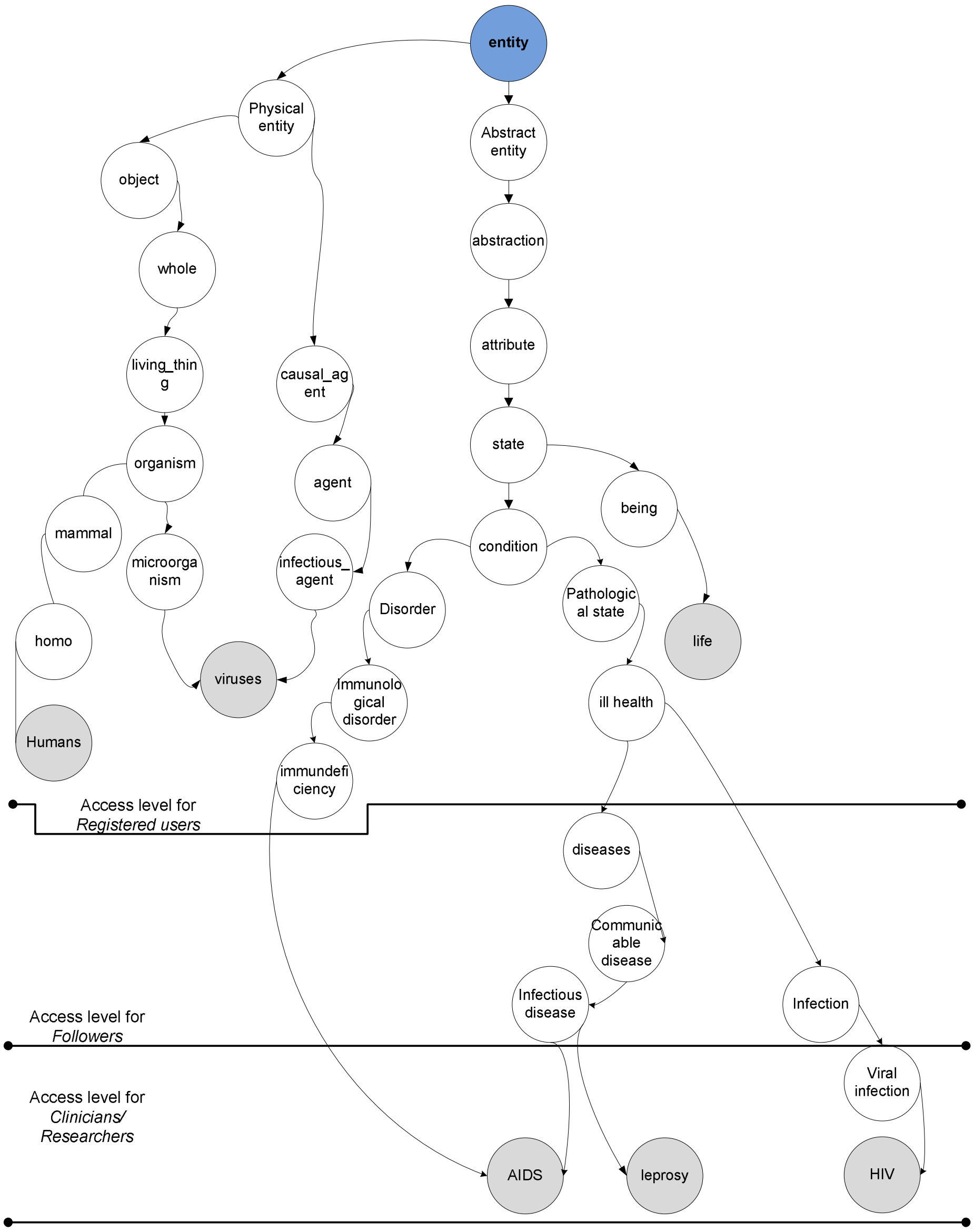

**Figure 6.10** DBPedia taxonomies and access levels of the noun phrases in the sample message

Given the short length of typical messages, the limited amount of contact types and the fact that the taxonomies corresponding to the noun phrases have been already retrieved during the annotation stage, the sanitization process is highly scalable. Moreover, the numbers of distinct sanitized versions of a message are also limited to the types of contacts and access levels defined by the user and, thus, once they are created, they can be cached for further access requests by the readers of same contact type.

Let us illustrate this process for the sample message with respect to the privacy requirements (defined in figure 6.5). The sanitization process will be based on the taxonomies retrieved (listed in table 6.4) from DBPedia for each annotated noun phrase and the access levels defined for each contact type, which are shown in figure 6.10.

According to $rule_5$, the access level for *followers* is *infections*. Therefore, the concepts below *infection* and *infectious disease* in the taxonomy shown in figure 6.10 (e.g., *AIDS*, *Leprosy* and *HIV*) are considered sensitive for *followers*. Therefore, the corresponding noun phrases need to be sanitized in order to fulfill the privacy requirements of the publisher. For this purpose, the sensitive noun phrases are replaced with the terms of access level nodes (i.e., *infection* and *infectious disease*). As a result, the *monitor* of our system will prepare a sanitized version of a message (shown in figure 6.11) for *followers*, which hides the specific details of the publisher's condition.

Dealing with **infection** and then being told that you suffer from **infectious disease** is almost the hardest thing to face with in life. The hardest thing is dealing with the virus because there are people that just do not understand and think that you are a **infectious disease**.

**Figure 6.11** Sanitized message for *followers*

Likewise, according to $rule_6$, the access level for *registered users* is *ill health*, thus producing the sanitized message shown in figure 6.12.

Dealing with **ill health** and then being told that you suffer from **ill health** is almost the hardest thing to face with in life. The hardest thing is dealing with the virus because there are people that just do not understand and think that you are a **ill health**.

**Figure 6.12** Sanitized message for *registered users* group

Finally, according to the first four rules (i.e., $rule_1$ to $rule_4$), the *clinicians/researchers* can access all the details of diseases of the publisher. Consequently, there are no sensitive terms that fall under the access level defined for this type of group. Therefore, the information will not be sanitized and the members of this group will get the plain text as shown in figure 6.6.

## 6.4 Evaluation

To complement the feasibility study that mainly considered the scalability of the system from both manual and algorithmic sides, in this section we evaluate the accuracy of the semantic annotation and the subsequent sensitivity assessment and privacy protection. To do so, we measure and evaluate (i) the accuracy of the detection of sensitive terms; and (ii) the accuracy of the semantic disambiguation process.

As evaluation data, we considered a set of entities related to the sensitive topics which are covered by current legislations on data privacy (i.e., healthcare data, religion, sexuality and location data); we simulated a set of messages to be published (and protected) referring to those entities by taking their descriptions in their corresponding Wikipedia articles. Note that, due to the highly informative nature of Wikipedia articles describing the entities to be protected, using this text as messages to be published represents a very challenging test bed from the perspective of document sanitization [Sánchez, 2013b,2014b].

As a benchmark for assessing the accuracy of our proposal regarding both sensitive term detection and term disambiguation, the collaboration of a human expert was required. More specifically, for the first operation, the human expert was asked to manually identify the textual terms in the simulated messages that, according to her opinion, unequivocally disclosed the underlying entity to be protected with respect to the level of disclosure specified by the user (i.e., *AL*); regarding the second operation, the human expert was asked to manually validate the terms that were correctly disambiguated by the system with respect to the senses available in WordNet for such terms. According to that judgment, the accuracy of the process for detecting sensitive term was quantified in terms of *precision*, *recall* and *f-measure*; while the accuracy of the semantic disambiguation process was quantified just in terms of *precision*.

*Precision*, in the first operation, it measures the percentage of automatically identified terms in the message (*S*) that are truly sensitive according to the expert's opinion (*H*). A high precision is desirable, because it indicates that the system has incurred in a low number of false positives, which may unnecessarily hamper the utility and readability of the protected messages. See equation (2.4) for a formal representation of *precision*:

$$Precision = \frac{|S \cap H|}{|S|} \times 100 \tag{6.4}$$

In the second operation, *precision* is just the percentage of properly disambiguated terms, according to the expert's opinion. In this case, a high *precision* is desirable in order to replace sensitive terms by semantically coherent generalizations.

On the other hand, *recall*, which only applies to the first operation, measures the percentage of sensitive terms correctly detected by the system ($S \cap H$) from the total number of the terms detected by the human expert (*H*). A high *recall* is desirable because it indicates that the protected message fulfils with the privacy requirements of the user. See equation (2.5) for a formal representation of *recall*:

$$Recall = \frac{|S \cap H|}{|H|} \times 100 \qquad (6.5)$$

Finally, *f-measure* provides the harmonic mean of *precision* and *recall* and, thus, summarizes the accuracy of the process in charge of detecting sensitive terms in the messages to be published. See equation (2.6) for formal representation of *f-measure*:

$$F - measure = \frac{2 \times Recall \times Precision}{Recall + Precision} \qquad (6.6)$$

Evaluation results that show the accuracy of the sensitive term detection process are depicted in table 6.5. To evaluate the effect of the configuration of the privacy requirements, each entity has been protected and evaluated (by the human expert) for two access levels (*AL*) with different degrees of generality.

**Table 6.5** Evaluation results for the process in charge of detecting sensitive terms

| **Entity/ Wikipedia article** | **Related *ST*** | **# words in text** | **# noun phrases** | **Access Level** | ***H*** | ***S*** | **Recall** | **Precision** | **F-measure** |
|---|---|---|---|---|---|---|---|---|---|
| HIV | Health | 49 | 20 | Condition | 9 | 7 | 77.77 % | 100 % | 87 % |
| | | | | Infection | 4 | 3 | 75 % | 100 % | 85 % |
| Christianity | Religion | 66 | 22 | Belief | 10 | 9 | 90 % | 100 % | 94 % |
| | | | | Religion | 7 | 5 | 71 % | 100 % | 83 % |
| Homosexu-ality | Sexual orientation | 78 | 26 | Process | 6 | 6 | 100 % | 100 % | 100 % |
| | | | | Sexual activity | 5 | 4 | 80 % | 100 % | 88 % |
| Berlin | Census data | 105 | 31 | Location | 10 | 8 | 80 % | 100 % | 88 % |
| | | | | City | 3 | 3 | 100 % | 100 % | 100 % |

In all cases, the system achieves perfect *precision* because, as stated in the privacy requirements, it sanitizes terms that are semantic specializations of the entities defined as access levels; thus, by definition, all the detected terms that may disclose the entity must be protected. On the other hand, *recall* figures fluctuated between 71-100%, showing that there is still room for improvement. Indeed, according to the expert assessment, some *combinations* of terms that are not actual specializations of the entity to be protected, but that can be related on some way with it, may also enable disclosure and should be adequately protected. For example, an informed attacker may infer that a publisher suffers from a certain sensitive disease from the fact that specific treatments or symptoms are mentioned in a discourse, despite of the fact that the disease has been already sanitized in the published message and that those terms are not specializations of the former. We are currently working on this issue and we provide some insights on how to tackle it in the next section. Finally, we can also see that the recall (i.e., the accuracy of the privacy protection) tends to increase as more general terms are defined as *ALs*. Indeed, a more general *AL* will impose a stronger restriction and force the system to sanitize more terms and, thus, the outcome would tend to offer a more robust protection.

Regarding the accuracy achieved by the semantic disambiguation process, table 6.6 shows the evaluation results that have been obtained. As it has been previously explained, in this case the human expert just validates the terms that have been correctly disambiguated by the proposed system according to the senses available for such terms in WordNet. Results reflect that, on average, the scheme disambiguated 66% of the terms correctly; even though this value may seem on the low side, it is coherent with the state of the art in semantic disambiguation [Kilgarriff, 2000], which rarely achieves very high precision figures. Moreover, improperly disambiguated terms would only affect the semantic coherence of the protected message, but not the privacy of the user, which is our main goal.

**Table 6.6** Evaluation results for the process in charge of disambiguating terms

| Entity/ Wikipedia article | Related *ST* | Access Level | *H* | *S* | Precision |
|---|---|---|---|---|---|
| HIV | Health | Condition | 7 | 4 | 57 % |
| | | Infection | 3 | 2 | 66 % |
| Christianity | Religion | Belief | 9 | 5 | 55 % |
| | | Religion | 5 | 3 | 60 % |
| Homosexuality | Sexual orientation | Process | 6 | 4 | 66 % |
| | | Sexual activity | 4 | 2 | 50 % |
| Berlin | Census data | Location | 8 | 6 | 75 % |
| | | City | 3 | 3 | 100 % |

## 6.5 Conclusions

In this chapter, we proposed a privacy-preserving content-driven access control mechanism for textual publications in the context of social media. Contrary to existing solutions [Masoumzadeh, 2010b] [Carminati, 2009,2011], the proposal is content driven in the sense that the semantics of the messages are automatically assessed by relying on an ontological knowledge base in order to detect the sensitive information they contain according to the privacy requirements of the publishers. These requirements are defined in general (i.e., an allowed level of disclosure is defined for the different contact types defined in a given environment), and the publications whose contents are related to these requirements are automatically protected. To do so, the sensitive information is sanitized and different versions of the publication are generated according to the access level of the readers. Thus, the privacy enforcement is transparent both to the publishers and readers, thus requiring no administrative efforts at the publication time, contrary to most related works [Carminati, 2009,2011]. In addition, the proposed mechanism is flexible enough to be incorporated in any scenario that publishes messages and classifies contacts into categories.

# Chapter 7 **Conclusion and Future work**

## 7.1 Summary

The main goal of this thesis was to tackle the privacy issues inherent to the information exchange in open and distributed environments, by proposing privacy-preserving access control solutions that could alleviate the limitations of the existing solutions.

To achieve this goal, we faced a number of challenges. First, we need to deal with large number of heterogeneous entities involved in open scenarios and their diverse (and dynamically changing) privacy requirements. Moreover, in order to propose a feasible and scalable solution, we need to get rid of manual management of rules/constraints (in which most available solutions rely) that constitutes a serious burden for the users and the administrators. Finally, access control management should be intuitive for the end users, who usually lack technical expertise, and they may find access control mechanism more difficult to understand and rigid to apply due to its complex configuration settings. We tackled these challenges by means of the following contributions:

- Most of the existing works propose scenario-specific solutions aiming to alleviate the limitations of standard access control models. In contrast, we proposed a generic ontology-based solution that can potentially model entities and policies of any scenario. Our ontology-based solution is inspired in the ABAC paradigm and, at its backbone, it models the generic *subject*, *object* and *policy* entities, which can be easily extended to accommodate the specificities of concrete scenarios. Moreover, the generic design of our ontology allows defining access rules at different levels of abstraction; that is, from concrete instances or classes to the most general/abstract classes in the ontology. In the latter case, and thanks to the ontological inference enabled by the taxonomic structure of entities modeled in the ontology, the system can automatically enforce rules for the more specific entities. The applicability and extensibility of our ontology has been illustrated in two paradigmatic open scenarios: OSNs and the cloud.

- We further extended our generic ontology to propose a delegation enforcement mechanism by enhancing the capabilities of XACML delegation profile (in terms of scalability and performance) by means of an ontology-based modeling of the delegation workflow. As a result, the delegation, verification and revocation processes are managed in a way that they can be easily adapted to the large scale environments (e.g., cloud) where entities are numerous, heterogeneous and distributed. With our approach, policies are modeled in the ontology and linked to the entities to which they refer. In this way, we do not incur in the overhead of searching for policy/entity matching in a potentially large database; as shown both theoretically and in practice, this significantly increases the performance of the system. Moreover, the proposed algorithms automatically enforce and resolve policy conflicts through the delegation workflow, without requiring the definition of special rules. The proposed system is also capable of detecting forge identities by relying on policies digitally signed by the owners of the resources.

- We also proposed a privacy-driven access control mechanism for textual resources that is able to enforce access control rules over the resources according to the sensitiveness of their contents. By relying on the formal domain knowledge modeled in ontological knowledge bases, our proposed system is able to automatically detect pieces of information that are sensitive according to the privacy requirements of the user. Moreover, the user may define different access/disclosure level on her own sensitive data according to the trust she has on the others. According to this, the system sanitizes the sensitive content and prepares different versions of textual resources according to the access levels of the readers. This offers a much more flexible (and transparent) mechanism than the manually defined and binary access control models usually implemented in environments such as OSNs. Due to the automatic management and enforcement of privacy configurations, our system requires no administrative efforts by the users at the time of publishing/resealing sensitive contents.

## 7.2 Publications

The contributions of this thesis have been published or submitted for publication to the following venues:

**Conferences:**

- Imran-Daud M, Sánchez D, Viejo A, (2016), Ontology-based Access Control Management: Two Use Cases, in: Proceedings of the 8th International Conference on Agents and Artificial Intelligence, Rome, Italy, pp. 244-249. **Core Rank: C**

- Imran-Daud M, Sánchez D, Viejo A, (2015), Ontology-Based Delegation of Access Control: An Enhancement to the XACML Delegation Profile, in: Proceedings of the 12th International Conference, TrustBus 2015, Valencia, Spain, September 1-2, 2015, Proceedings, Springer International Publishing, pp. 18-29. **Core Rank: B**

**JCR Indexed Journals:**

- Imran-Daud M, Sánchez D, Viejo A, (2016), Privacy-driven access control in social networks by means of automatic semantic annotation, in: Computer Communications, 76, 2016, pp. 12-25. **(Published) Impact Factor-2014: 1.695 (1st Quartile, Computer Science-Information Systems)**

- Imran-Daud M, Sánchez D, Viejo A, Ontology-based Access Control Delegation Enforcement for the Cloud, in: Security and Communication Networks. ***(Under Review)*** **Impact Factor-2014: 0.72 (2nd Quartile, Computer Networks and Communications)**.

## 7.3 Future Work

In this section, we give some lines for future research that derive from the work developed in this thesis, and also provide some ideas to tackle them. Regarding thegenericaccess control ontology, following lines can be considered:

- We plan to study the interoperability issues that arise in access control between heterogeneous systems and evaluate whether our ontology-based mechanism (with its common ontological backbone, which offer a common modeling for the entities involved in different systems) may provide a suitable solution to interoperate between rules and entities of different scenarios.

- The backbone of our generic ontology is inspired in the ABAC paradigm, thus, access control is managed based on the attributes of the entities. However, ABAC does not support constraints specification [Coyne, 2013]; for example, when we want to constrain delegation privileges with date and time or with a valid registration. Such constrains are supported in the RBAC paradigm through the roles (which contain role names, constrains and permissions). As future work, we can extend our ontology to support constraints specification by incorporating role as an attribute. However, ABAC can only constrain role names (which are stored as an attribute) but not the role with the associated permissions. For this purpose, we need to find a solution to manage the roles modeled in the ontology and also devise methods to constrain the specifications of attributes. Existing solutions [Damiani, 2005, Bijon, 2013] proposed for open environments can be considered to incorporate constraints in our ontology-based mechanism.

- The delegation model presented in chapter 5 is an enhancement of the XACML profile, which is able to improve the performance of the policy search for each access request by relying on the ontology-based workflow. To apply this improvement in practice we require modifying the architecture of XACML (shown in figure 2.1) in order to support our ontological framework. To do so, the XACML architecture can be extended with the modules (in addition to the actors explained in section 2.3.1) that are responsible for (i) retrieving information of entities from the ontology and (ii) verifying the delegation from the certification authority. In addition, the procedures of the XACML actors need to be redefined in order synchronize them with ontological procedures and the proposed algorithms.

Regarding the content-driven privacy-preserving access control mechanism, the following lines of research can be defined for further improvement:

- In order to alleviate users from completely specifying their privacy requirements, we will consider the automatic inference of access control rules according to the social relationships implemented in the social web (e.g., In OSNs, the privacy rules for friends could be same for the friends of friend). At this respect, a machine learning approach [Bilogrevic, 2013] can also be considered to semi-automatize the configuration of privacy rules.

- The user’s privacy can also be compromised by the (co-)occurrence of information that is correlated to the sensitive topic to be protected. For example, even though a user may publish a message that does not directly references the disease she suffer from, the message may contain combinations of terms (such as treatment, symptoms or drugs) that may enable to univocally inferring such sensitive disease. We are currently working on automatic solutions to address this issue that, in a nutshell, would assess the disclosure that potentially correlated terms (e.g., symptoms) may produce for a sensitive one (e.g., a sensitive disease) according to their mutual information, which is computed from the information distribution of data in large corpora [Sánchez, 2013b,2014b,a, Sánchez, 2016a]. We plan to incorporate them to the developed in order to improve the assessment of privacy risks by detecting correlated terms or term aggregations that may disclose more information about a sensitive topic than the one specified in the privacy rules.